\documentclass[printer]{aa}
\usepackage{multicol}
\usepackage{graphicx}
\usepackage{url}
\usepackage{subcaption}
\usepackage{psfrag}
\usepackage{hyperref}
\DeclareGraphicsExtensions{.pdf,.png,.jpg,.ps,.eps}
\usepackage{natbib}
\usepackage{amsmath}
\usepackage[varg]{txfonts}
\usepackage{booktabs}
\usepackage{amssymb}
\usepackage{wasysym}
\usepackage{longtable}
\bibpunct{(}{)}{;}{a}{}{,} 

\usepackage[normalem]{ulem}
\usepackage{color}

\usepackage{rotating}
\usepackage{longtable}
\usepackage{booktabs}
\usepackage{graphicx} 

\begin{document}

\title{Spatial and spectral constraints on resolved mass-loss of the massive Post-RSG star IRAS 17163–3907 and its Fried Egg Nebula}

\author{E.~Koumpia\inst{\ref{inst1},\ref{inst2}}\fnmsep\thanks{ekoumpia@eso.org}, A.~Cikota\inst{\ref{inst3}}, W.-J.~de Wit\inst{\ref{inst2}},
G. Muñoz-Sanchez\inst{\ref{inst11}},
T.~Kim\inst{\ref{inst5}},
A.~Corporaal\inst{\ref{inst2}}, 
R.~D.~Oudmaijer\inst{\ref{inst4},\ref{inst12}}, S.~Muller\inst{\ref{inst8}}, J.~S.~Vink\inst{\ref{inst6}},  L.~Cerrigone\inst{\ref{inst1},\ref{inst16}}, A.~Zijlstra\inst{\ref{inst9}}, R.~Szczerba\inst{\ref{inst16},\ref{inst17},\ref{inst18}}, Y.~Asaki\inst{\ref{inst1},\ref{inst13},\ref{inst14}}, E.~Lagadec\inst{\ref{inst7}}, F.~Millour\inst{\ref{inst7}}}

\institute{Joint ALMA Observatory, Alonso de Córdova 3107, Vitacura, Santiago, Chile\label{inst1}
\and 
ESO, Alonso de Córdova 3107, Vitacura, Casilla, 19001, Santiago, Chile\label{inst2}
\and
Gemini Observatory/NSF's NOIRLab, Casilla 603, La Serena, Chile\label{inst3} 
\and Royal Observatory of Belgium, Ringlaan 3, 1180 Brussels, Belgium\label{inst4} 
\and
Department of Astrophysical Sciences, Princeton University, Princeton, NJ, 08544, USA\label{inst5}
\and
Armagh Observatory and Planetarium, College Hill, Armagh, BT61 9DG, UK\label{inst6}
\and
Observatoire de la C\^ote d'Azur Nice, 96 Boulevard de l'Observatoire, 06300, Nice, France\label{inst7}
\and
Department of Space, Earth, and Environment, Chalmers University of Technology, Onsala Space Observatory, 43992, Onsala, Sweden\label{inst8}
\and
Jodrell Bank Centre for Astrophysics, Alan Turing Building, The University of Manchester, Oxford Road, Manchester, M139PL, UK\label{inst9}
\and IAASARS, National Observatory of Athens, I. Metaxa \& Vas. Pavlou St., 15236, Penteli, Athens, Greece\label{inst11}
\and
School of Physics \& Astronomy, University of Leeds, Woodhouse Lane, LS2 9JT, Leeds, UK\label{inst12}
\and 
National Astronomical Observatory of Japan, Los Abedules 3085 Oficina 701, Vitacura, Santiago, 763 0000, Chile\label{inst13}
\and The Graduate University for Advanced Studies, SOKENDAI, Osawa 2-21-1, Mitaka, Tokyo, 181-8588, Japan\label{inst14}
\and National Radio Astronomy Observatory, 520 Edgemont Road, Charlottesville, VA, 22903, USA\label{inst15}
\and Xinjiang Astronomical Observatory, Chinese Academy of Sciences, 150 Science 1-Street, Urumqi, Xinjiang 830011, China\label{inst16}
\and Xinjiang Key Laboratory of Radio Astrophysics, 150 Science 1-Street, Urumqi, Xinjiang 830011, China\label{inst17}
\and 
Nicolaus Copernicus Astronomical Centre, PAS, ul. Rabiańska 8, 87-100, Toruń, Poland\label{inst18}
}

\date{Received date/Accepted date}

\abstract
{The fate of massive stars during the latest stages of their evolution is highly dependent on their mass-loss rate and geometry. The geometry of the mass-loss process can be inferred from the shape of the circumstellar material, having a significant influence on the evolution of massive stars (between 25 and 40 $M_{\astrosun}$), i.e., type II SN progenitors. In this context, post-Red
Supergiants (post-RSGs) offer an excellent opportunity to study mass-loss events.}{We aim to investigate the mass-loss history, geometry, and physical conditions of the yellow hypergiant in a post-RSG stage, IRAS~17163$-$3907 (IRAS~17163, aka the Fried Egg Nebula). We place it in context with another famous evolved massive star, the yellow hypergiant IRC+10420.}{We combine M-band spectra of the source using high-resolution CRIRES+ spectroscopy, with VLTI/MATISSE mid-infrared interferometry in the L-band, and FORS2 optical spectropolarimetry to probe both the small-scale circumstellar structure and the large-scale dusty environment of IRAS~17163. The interferometric observables were analysed with simple geometric fitting and a more advanced parametric modelling using \texttt{PMOIRED} to extract the morphology of the hot inner shell that was previously reported via radiative transfer modelling.}{The CRIRES+ spectrum provides the first M-band coverage of IRAS~17163, revealing prominent low-excitation metal lines and hydrogen recombination features, but lacking the pronounced CO absorption seen in IRC+10420. The MATISSE observations reveal the first high angular scales of the source in the L-band and spatially resolve the Br$\alpha$ line-emitting region, which is a factor of two more extended than the continuum emission and hint to a marginally asymmetric and variable ionised wind. FORS2 spectropolarimetry shows intrinsic continuum polarisation and line effects in Stokes Q parameter across H$\alpha$, pointing to deviations from perfect spherical symmetry also on larger scales. The interferometry reveals no evidence for a binary companion within the explored parameter space, indicating that the observed clumpy and time-variable mass loss is likely intrinsic to the star rather than companion-driven.}{Our results demonstrate that IRAS~17163 hosts a dense, structured, and time-variable wind, coexisting with extended dusty shells. The comparison with IRC+10420 highlights diversity among post-RSG YHGs, with IRAS~17163 showing an ionised environment without apparent molecular signatures. These findings emphasise the role of clumpy and near-symmetric mass-loss in shaping the circumstellar medium of evolved massive stars, with implications for their subsequent evolution and core-collapse supernova progenitor properties.}


\titlerunning{Resolved mass-loss of the massive Post-RSG star IRAS 17163–3907 and its Fried Egg Nebula.} 
\authorrunning{Koumpia et al.} 
 \maketitle

\vspace{0.2cm}

\section{Introduction}

Massive stars ($>8M_{\astrosun}$) are among the most influential objects in the universe: they provide the primary energy budget in galaxies, they enrich the chemical composition of the interstellar medium by producing heavy elements and even trigger star formation in their immediate surroundings. Many important open issues related to the final stages of stellar evolution can only be addressed once we comprehend the geometry of circumstellar material close to the star. For massive stars, which eventually explode as core-collapse supernovae (SNe), the problem of understanding mass-loss rate and mass-loss geometry is particularly acute \citep[e.g.][]{Heger1998,Heger2003,Langer2012,Smith2014}. Mass-loss events impact angular momentum evolution and final mass, and thus the fate of the individual massive
star. In addition, these mass-loss events create the circumstellar environment with which SN ejecta will interact. Following an SN explosion, the mass-loss history of the progenitor is reflected in the light-curve morphology and likely also on observed aspherical SN remnants \citep[][]{Moriya2014}.

However, the physical mechanisms that govern mass-loss and shape the ejecta of evolved massive stars are still poorly constrained. Proposed drivers include pulsational instabilities \citep[e.g.][]{Lobel1994,deJager1998} and radiation pressure acting on spectral lines. Line-driven winds are subject to changes in radiative acceleration associated with the bi-stability jumps, which can strongly affect wind velocity and mass-loss rates. In particular, the second bi-stability jump occurring at T $\sim$ 8800~K due to the recombination of Fe {\sc iii} to Fe {\sc ii} is the most relevant in the context of yellow hypergiants and post-RSG objects \citep{Vink1999,Vink2000,Vink2001,Petrov2016}. Yet distinguishing between these mechanisms observationally remains challenging, as accurate constraints on the dynamics, geometry, and timescales of mass-loss are scarce.

In recent years, binarity has emerged as an increasingly important ingredient in shaping the circumstellar environments of evolved massive stars. Binary interaction can influence mass-loss rates, angular-momentum evolution, and outflow geometry, leading to asymmetric nebulae, cavities, and complex shell morphologies that may not be easily explained by single-star wind models alone. High angular resolution observations have recently begun to resolve such systems directly, revealing the impact of binarity on mass loss and nebular structure in post-RSG objects \citep[e.g. AFGL 4106,][]{Tomassini2026}. Motivated by these results, assessing the presence and potential role of companions has become an integral part of interpreting the circumstellar geometry of evolved massive stars.

Post-red supergiants (post-RSGs) are rare but crucial laboratories for investigating these processes. Among them, the A–K type yellow hypergiants (YHGs) are particularly important, as they exhibit high mass-loss rates and dusty circumstellar envelopes, making them excellent candidates to study episodic ejections and the transition toward the blue supergiant phase. Examples include IRC+10420, HD~179821, and IRAS~17163$-$3907. For reviews on the matter, see \citet{Jones2025}, \citet{2019Gordon}, \citet{2009Oudmaijer} and \citet{1998deJager}.

In this work, we focus on IRAS~17163$-$3907 (WRAY~15-1676, IRAS~17163 hereafter), also known as the “Fried Egg Nebula” due to its multiple detached shell morphology \citep{Lagadec2011}. At an estimated initial mass of $25$–$40\,M_\odot$ and a distance of about 1.2 kpc \citep[see Sect.~\ref{distance_sec}, and][]{Koumpia2020}, IRAS~17163 is one of the brightest infrared sources in the sky and one of the best-studied post-RSG candidates. Its A6\,Ia optical spectrum closely resembles that of IRC+10420, suggesting that it is evolving toward the blue part of the Hertzsprung–Russell diagram \citep[e.g.][]{Meynet2015}. We note that the designation of IRAS~17163 as a post-red supergiant reflects an evolutionary interpretation rather than a direct observational classification; however, its high luminosity, extended detached dusty shells, chemical properties, and similarity to archetypal objects such as IRC+10420 strongly support an evolutionary scenario in which the star has already passed through a red supergiant phase \citep[e.g.][]{Lagadec2011,Koumpia2020}.

The detached shells appear circularly symmetric to first order in far-infrared \textit{Herschel} and VLT/VISIR images, but substructure is also clearly present, with far-IR dust clumps co-spatial with enhanced CO emission \citep{Lagadec2011, Hutsemekers2013, Wallstrom2015}. ALMA Compact Array (ACA) observations of the Fried Egg Nebula have revealed a red-shifted spur extending to $\sim$20\arcsec, suggestive of a past, possibly unidirectional mass-ejection event \citep{Wallstrom2017}. These properties provide evidence for episodic and possibly asymmetric mass-loss. While the detached shells of IRAS~17163 trace the circumstellar environment at large distances from the star ($>1''$, $\gtrsim1200$~au at 1.2~kpc), the innermost regions remain comparatively unconstrained \citep{Koumpia2020}. 

In this work, we combine high-angular-resolution mid-infrared interferometry, high-resolution mid-infrared spectroscopy, and optical spectropolarimetry to probe both the small-scale ionised and dusty environment and the larger-scale circumstellar morphology of this key post-RSG object. We, therefore, aim to obtain a coherent view of mass-loss across several orders of magnitude in radius. To place IRAS~17163 into the broader context of yellow hypergiant evolution, we additionally present CRIRES+ $M$-band data of the archetypal post-RSG \object{IRC+10420}. Its rapid spectral changes and resolved ionised and dusty structures have made it a key reference object \citep[e.g.][]{Oudmaijer1994, Oudmaijer1996, Humphreys1997, Humphreys2002, Koumpia2022}. We directly contrast its spectral properties with those of IRAS~17163, thereby probing the diversity of mass-loss behaviour across this rare evolutionary phase.

This paper is structured as follows. Section~2 describes the observations and data reduction procedures for the CRIRES+, VLTI/MATISSE, and FORS2 data sets. Section~3 presents the observational results, while Section~4 focuses on geometric and parametric modelling of the interferometric data. In Section~5, we discuss the implications for the mass-loss geometry and evolutionary status of IRAS~17163, and in Section~6 we summarise our conclusions.

\section{Observations and data reduction}
\label{obs} 

\subsection{Spectroscopy - CRIRES+}
\label{sect:crires}

Mid-infrared spectroscopic observations of both IRAS~17163 and IRC+10420 
were carried out on 15 September 2021 using the CRyogenic InfraRed Echelle Spectrograph (CRIRES) Upgrade, CRIRES+ \citep{Dorn2014, Dorn2023} mounted on Unit Telescope 3 (UT3) of the Very Large Telescope (VLT) at ESO’s Paranal Observatory, Chile (Program ID: 107.22TZ.001). CRIRES+ provides high-resolution spectroscopy in the infrared domain, and the observations were conducted in nodding mode, which enables effective background subtraction.

The targets were observed at global central wavelengths of 4265.706 nm, 4318.114 nm, 4368.182 nm, and 4415.885 nm, covering distinct spectral windows within the range of 3412 nm - 5359 nm, using the four spectral settings, M4266, M4318, M4368, and M4416\footnote{\url{https://www.eso.org/sci/facilities/paranal/instruments/crires/doc/ESO-323064_2_CRIRES+UserManual.pdf}}, respectively. CRIRES+ is equipped with a detector mosaic comprising an array of three Hawaii 2RG detectors, or "chips", with a read-out window size of 6144 $\times$ 2048 pixels, and an 18 microns pixel size. The CRIRES+ detector records multiple orders simultaneously. In this case, six orders are covered per spectral set-up, with a wavelength bandwidth between 175~nm and 340 nm (increasing with wavelength). The spectrograph was configured in high-resolution spectroscopy mode. The slit width was set to 0.2\arcsec, corresponding to a resolving power of R $\sim$ 80,000, with a slit length of 10\arcsec. The total exposure time per frame was 1.427 s.

The observations were conducted under adaptive optics (AO) correction with the system operating in closed-loop mode. The guide star used for AO corrections was located at RA = 17:19:17.2, DEC = -39:14:10.5 with an {\it R-}band magnitude of 9.185. The atmospheric conditions measured at the time of observation indicated a mean Strehl ratio of 0.569, a seeing corrected by airmass of 1 arcsec, a Fried parameter (r0) of 1.910 m, and a correlation time ($\tau$0) of 33.575 s. These conditions allowed for optimal correction of wavefront distortions, ensuring high spectral resolution.

Standard calibration frames, including darks and flats, were obtained as part of the standard ESO calibration plan. Data were reduced using the CR2RES pipeline recipes (version~1.1.7\footnote{\url{https://www.eso.org/sci/software/pipelines/cr2res/cr2res-pipe-recipes.html}}), within the ESO Reflex environment, which includes corrections for bad pixels, background subtraction, and spectral extraction. Post-processing of the spectra included telluric correction and a refinement of the wavelength calibration using atmospheric absorption features. The default pipeline wavelength solution (obtained without the use of a gas cell) has an absolute accuracy of approximately three detector pixels. At the spectral resolving power of CRIRES+ ($R\sim8\times10^{4}$--$10^{5}$), this corresponds to roughly one resolution element, i.e.\ an absolute velocity uncertainty of a $\sim$ 4 km\,s$^{-1}$. The subsequent alignment to telluric features improves the relative wavelength scale within each spectrum.

\subsection{Interferometry - MATISSE}
\label{sect:matisse}

Interferometric observations of \object{IRAS~17163} were carried out using the Multi AperTure mid-Infrared SpectroScopic Experiment \citep[][MATISSE]{Lopez2022, Woillez2024} on the four Auxiliary Telescopes (ATs) of the Very Large Telescope Interferometer (VLTI) at Paranal Observatory. Observations were conducted over multiple nights between April 2022 and March 2023 in both standalone and imaging/GRA4MAT modes (fringe stabilisation was achieved using the GRAVITY fringe tracker), utilising the following AT configurations: A0-B2-D0-C1, A0–G1–J2–J3, A0–G2–J2–J3, A0–B2–D0–J3, K0–G2–D0–J3, and A0–G1–D0–J3, under ESO program IDs 109.22VF.001 (small), 109.22VF.002 (medium), 109.22VF.003 (imaging in all three configurations), and 109.22VF.005 (large). A summary of the interferometric data finally used in this work is provided in Table~\ref{tab:visibilities_two_column}. Given that the observations span several months, we note that potential time variability on sub-au to few-au scales cannot be excluded a priori; any evidence for or implications of such variability are addressed in the analysis and discussion sections. 

MATISSE operates in the $L$, $M$, and $N$ bands. The focus of this study is the $L$ band ($\lambda \approx 3$--$4.2\,\mu$m), with a spectral resolving power of $R \approx 35$ (LOW) for standalone observations and $R \approx 3300$ (HIGH$+$) for imaging. The L-band detector integration time (DIT) was set to 0.111\,s for standalone L-band observations and 10\,s for the GRA4MAT imaging mode. The projected baselines ranged from 11\,m to 130\,m, corresponding to angular resolutions of approximately 65\,mas to 5.5\,mas for a representative wavelength of 3.5 $\mu$m in the $L$ band, and position angles from 0$^{\circ}$ to 170$^{\circ}$ (Table~\ref{tab:visibilities_two_column}). At the estimated distance of 1.2\,kpc to \object{IRAS~17163}, these baselines allow us to probe spatial scales of down to $\sim$ 7\,au.

The observing sequence followed the standard MATISSE calibration plan, alternating between the science target and a calibrator to account for instrumental and atmospheric effects. In particular, the standalone observations followed the standard CAL-SCI sequence, while during the imaging observations, each science observation was bracketed by calibrator measurements (CAL-SCI-CAL). Two calibrators were used during the observations: HD~156216, a standard calibrator listed in the JMMC catalog with flux densities of about 20\,Jy in the L band, 17\,Jy in the M band, and 3.3\,Jy in the N band; and HD~127755, with L- and N-band fluxes of 34\,Jy and 2.2\,Jy, respectively, retrieved from a SIMBAD-based catalogue \citep[II/361/mdfc-v10;][]{Cruzalebes2019}. Since HD~127755 was not listed in the JMMC catalog it was manually added as a calibrator FITS file during the reduction. The K-band magnitudes of the calibrators are 2.5\,mag for HD~156216 and 3.2\,mag for HD~127755. The science target, \object{IRAS~17163}, has L- and N-band fluxes of approximately 213\,Jy and 400\,Jy, respectively.

The observations were carried out under generally good atmospheric conditions. The optical seeing during science exposures ranged from 0.4$\arcsec$ to 0.88\arcsec, with a median value of approximately 0.6\arcsec. The atmospheric coherence time ($\tau_0$) varied between 4.1 and 8.4 ms across observing nights, typically remaining above 5 ms. These conditions ensured stable fringe tracking and meaningful interferometric data, although some observations were affected by transient instabilities or partial data loss, which were excluded from the final dataset.

The data were reduced using the standard MATISSE data reduction pipeline recipes (version 2.0.2\footnote{\url{https://www.eso.org/sci/software/pipelines/matisse/matisse-pipe-recipes.html}}), provided by ESO. 
The reduction process involved transforming raw interferometric frames into calibrated observables (squared visibilities, closure phases, and differential phases) using the \texttt{mat\_raw\_estimates} and \texttt{mat\_cal\_oifits} recipes. These steps included background subtraction, flat-fielding, bad pixel correction, optical distortion correction, and the estimation of interferometric observables through Fourier analysis. The final calibrated observables were exported in OIFITS format for further analysis.

\subsection{Spectropolarimetry - FORS2}

Observations of IRAS 17163 were conducted using the Focal Reducer and Low Dispersion Spectrograph (FORS) in spectropolarimetric (PMOS) mode, mounted at the UT1 Cassegrain focus of the VLT \citep{1998Msngr..94....1A}. Data were collected over two epochs, on July 4 and 8, 2023, as part of ESO program ID 111.24P0.004. Seeing conditions ranged between 0.9\arcsec \ and 1.2$\arcsec$. The sky was clear on July 4, while thick cirrus clouds were present during the observations on July 8.

FORS2 contains a Wollaston prism that splits the incoming light into two orthogonally polarised components: the ordinary (o) and extraordinary (e) beams. 
IRAS 17163 was observed at each epoch through the 300V grism ($\rm \lambda_c$ = 590 nm, $\rm \lambda / \Delta \lambda$ = 440) with and without the GG435 order-separating filter. 
The GG435 filter, which has a cut-off wavelength around 435 nm, suppresses second-order spectral contamination. While second-order effects are usually minimal in spectropolarimetric data, they can become non-negligible for particularly blue sources (see 
\citealt{2010A&A...510A.108P}, appendix).

Each polarisation sequence involved positioning the half-wave retarder plate at four distinct angles: 0, 22.5, 45, and 67.5 degrees. 
The raw data were bias-subtracted but not flat-fielded; however, the impact of detector inhomogeneities is mitigated by using multiple retarder plate positions \citep{2006PASP..118..146P}. Cosmic rays were removed using L.A.Cosmic  \citep{2001PASP..113.1420V}.

The o and e beams were extracted using standard procedures implemented in IRAF, as described in \citet{Cikota2017}. 
Wavelength calibration was performed using exposures of He–Ne–Ar arc lamps, achieving a typical root-mean-square (rms) accuracy of about 0.3 \AA. The Stokes Q and U parameters, the polarisation degree P, and the polarisation angle $\theta$ were calculated following the FORS2 User Manual, and as described in e.g. \citet{Cikota2017}. The flux spectrum was calculated by combining all the ordinary and extraordinary beams. The flux calibration is based on a sensitivity curve template and is therefore suitable only for qualitative comparison with the polarisation data.


\section{Observational results}

\subsection{M-band spectrum} 

\begin{figure}[ht]
    \centering
    \includegraphics[width=\columnwidth]{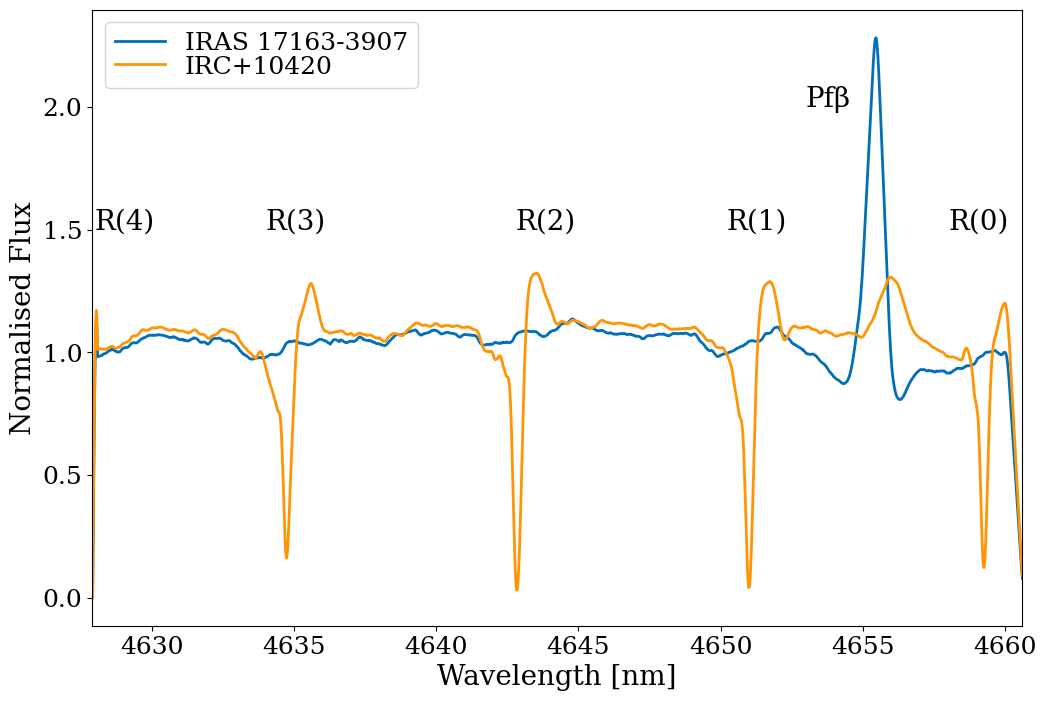}
    \caption{CRIRES+ spectra comparison of \object{IRAS~17163} (blue) and \object{IRC+10420} (yellow), showing the Pfund$\beta$ line and CO R-branch transitions. The fluxes are normalised, and a telluric correction has been applied.}
    \label{fig:pfbeta_comparison}
\end{figure}

\begin{figure}[ht]
    \centering
    \includegraphics[width=\columnwidth]{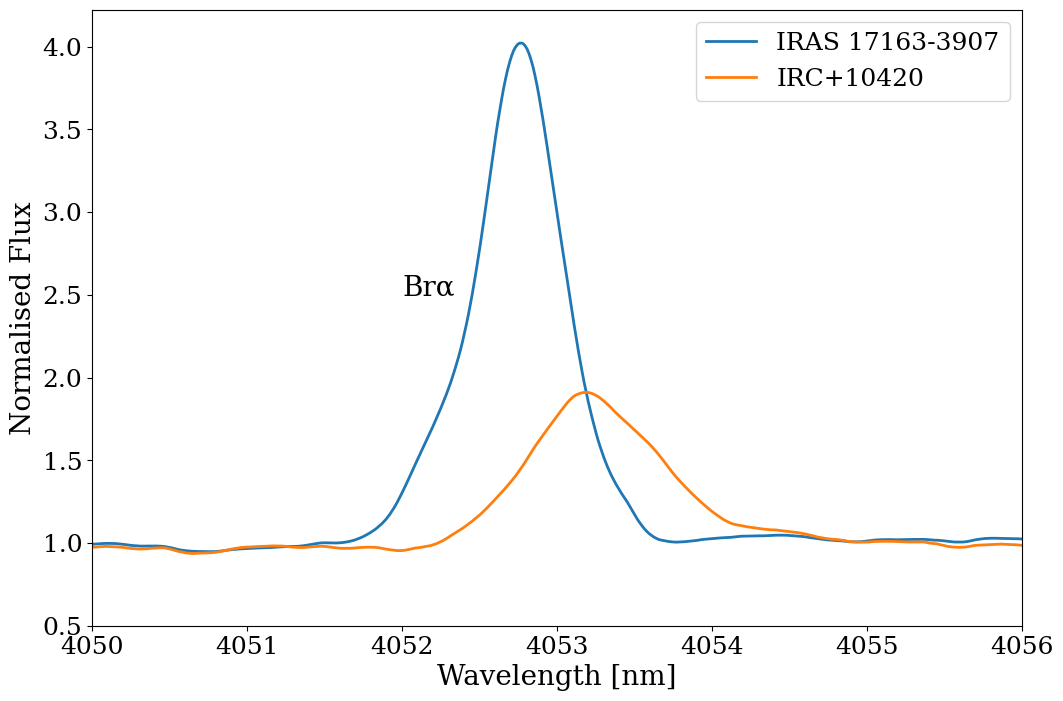}
    \caption{Normalised and telluric corrected CRIRES+ spectra of \object{IRAS\,17163} (blue) and \object{IRC+10420} (yellow) around the Brackett~$\alpha$ line at $\sim$4.052\,$\mu \textrm{m}$. The emission is significantly stronger in IRAS\,17163 compared to IRC+10420, indicating a higher density of ionised gas in the inner wind region.}
    \label{fig:br_alpha}
\end{figure}

\begin{figure}[ht]
    \centering
    \includegraphics[width=\columnwidth]{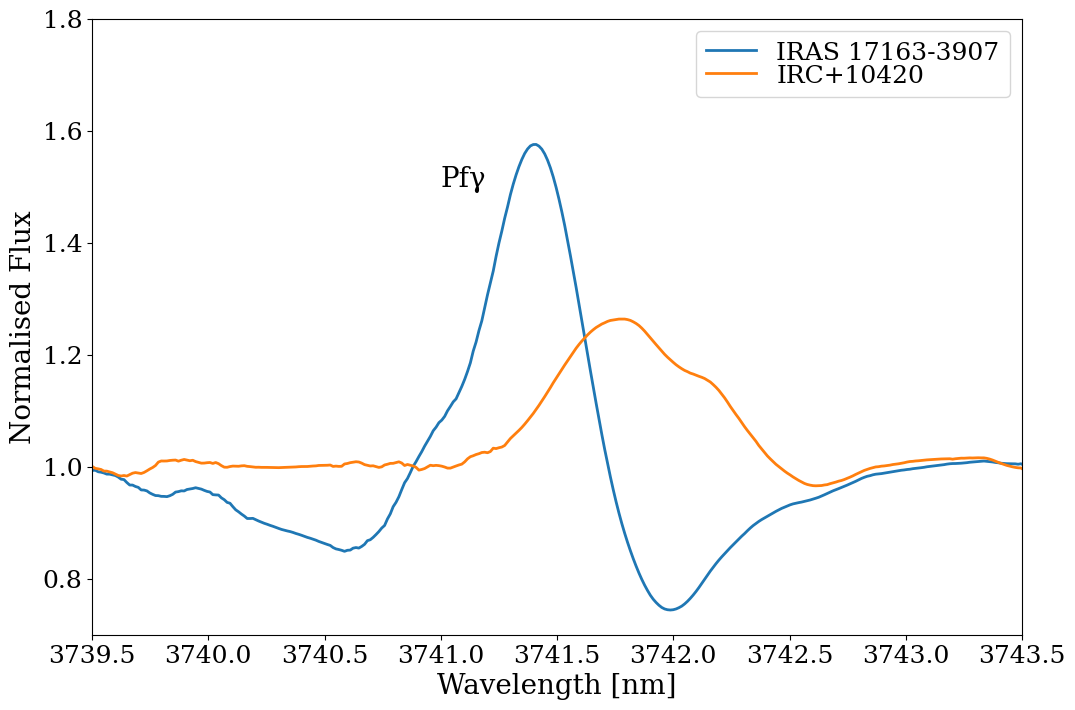}
    \includegraphics[width=\columnwidth]{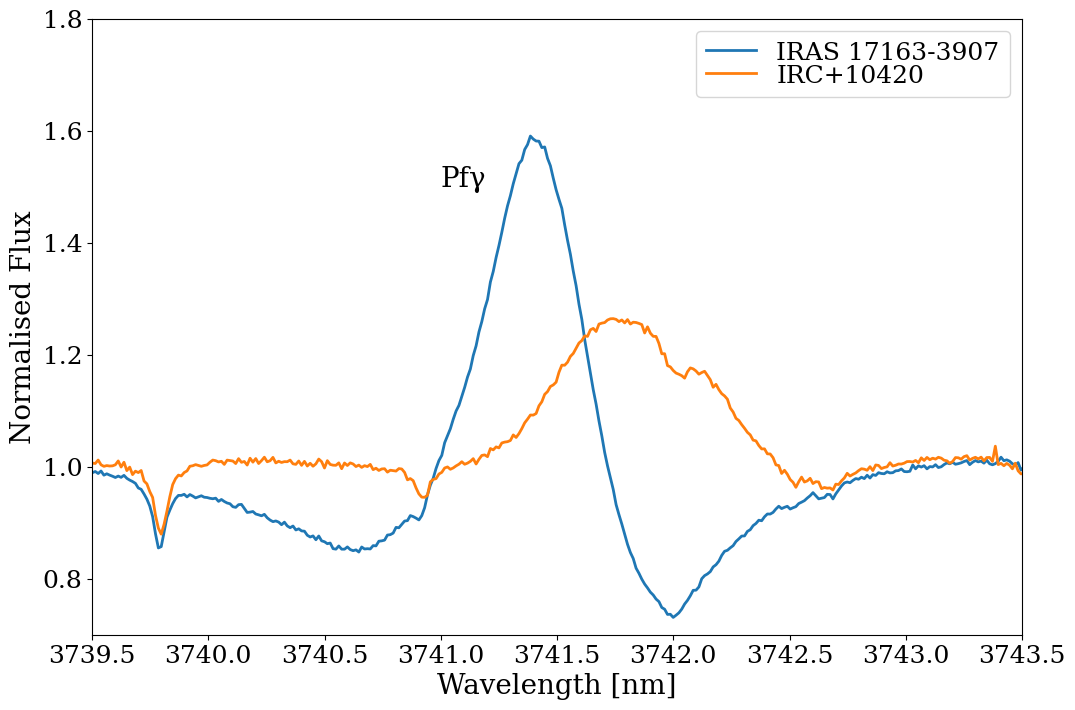}
    \caption{Comparison of the Pfund~$\gamma$ line profiles in \object{IRAS\,17163} and \object{IRC+10420} before (top) and after (bottom) telluric correction. The uncorrected profile shows residual atmospheric features affecting both the continuum and line wings, while the corrected version provides a cleaner view of the intrinsic emission and absorption structure.}
    \label{fig:pfund_gamma_comparison}
\end{figure}

We present the $M$-band spectrum of the yellow hypergiant \object{IRAS\,17163}, obtained with CRIRES+, covering hydrogen recombination lines (Brackett, Pfund, and Humphreys series), CO rovibrational transitions (v\,=\,1\,$\rightarrow$\,0), and selected atomic and molecular features that trace the innermost circumstellar environment. To aid the interpretation, we also obtained and analysed similar CRIRES+ observations of the well-studied post-RSG yellow hypergiant \object{IRC+10420}, allowing for a direct comparison between two sources at a similar evolutionary stage. Spectral features from \object{IRC+10420} are overplotted in all relevant figures for reference (see Fig.\ref{fig:pfbeta_comparison}-Fig.\ref{fig:pfund_gamma_comparison},Fig.~\ref{fig:crires_overview}-\ref{fig:crires_select},  and Table~\ref{crires_comparison}). The full spectral coverage was systematically examined to identify secure line detections, tentative features, and non-detections. An overview of all CRIRES+ spectral windows with the line identifications of the most prominent transitions is presented in Figure~\ref{fig:crires_overview}.

CO first overtone transitions ($\nu$\,=\,1\,$\rightarrow$\,0), particularly in the R- and P-branches, exhibit stark contrasts between the sources (e.g., Fig.~\ref{fig:pfbeta_comparison}). \object{IRC+10420} displays well-defined absorption features across multiple transitions (e.g., R(0)--R(6), P(1)--P(3)), consistent with classical P~Cygni morphologies and the presence of a molecular envelope (Fig.~\ref{fig:crires_overview}). Conversely, \object{IRAS\,17163} shows little to no absorption in the CO lines (Fig.~\ref{fig:crires_overview}). A potential weak absorption seen is more difficult to characterise, given the strong atmospheric contributions that make the detection less confident. For a more direct comparison, the fitted parameters for the line profiles of both sources, including peak velocities, FWHM, and equivalent widths, are listed in Table~\ref{crires_comparison}. Tentative features were also noted among the higher-order P-branch transitions (e.g., upwards of P(3)), though several are potentially affected by telluric residuals or insufficient signal-to-noise (SNR, Fig.~\ref{fig:crires_overview}). 

Hydrogen recombination lines including Br\,$\alpha$, Pf\,$\beta$, Pf\,$\gamma$, Hu\,$\delta$, and Hu\,$\epsilon$ are clearly detected in \object{IRAS\,17163}, typically displaying asymmetric emission profiles (e.g., Fig.~\ref{fig:br_alpha}, Fig.~\ref{fig:pfund_gamma_comparison}). Notably, Pf\,$\beta$ and Pf\,$\gamma$ exhibit broad wings and P~Cygni profiles. These lines are also present in \object{IRC+10420}. The Br\,$\alpha$ line shows strong emission in both stars, with a red-shifted peak in \object{IRC+10420} consistent with its higher systemic velocity relative to IRAS\,17163. Several Humphreys transitions (Hu~$\delta$, Hu~$\epsilon$, Hu~$\theta$), and higher excitation lines (e.g., H\,20--7), are present in \object{IRAS\,17163}. These lines are either absent or significantly weaker in \object{IRC+10420}. The H$_2$ S($n$) transitions were searched for in both targets but remained undetected at the 3$\sigma$ level, consistent with its expected weakness and susceptibility to atmospheric interference.

Several metal lines were examined, including Fe\,\textsc{ii}, 
Mg\,\textsc{i}, and Ti\,\textsc{i}. The C\,\textsc{i}/Fe\,\textsc{ii} $\lambda$4061.77\,nm line is reliably detected in \object{IRAS\,17163} based on peak strength and red-shift, but remains uncertain in \object{IRC+10420}. Although other potential features have been detected (see Fig. \ref{fig:crires_select}),
these require further verification through line-fitting and comparison with spectral atlases.



This comparison reveals distinct kinematic and excitation properties between the two yellow hypergiants. The observed differences may reflect different mass-loss histories or viewing geometries between the two sources, and we return to their interpretation in Sect.\ref{discussion}.


\subsection{Interferometric observables}

\subsubsection{Visibilities and phases}

\begin{figure}[htbp]
    \centering
    \includegraphics[width=0.5\textwidth]{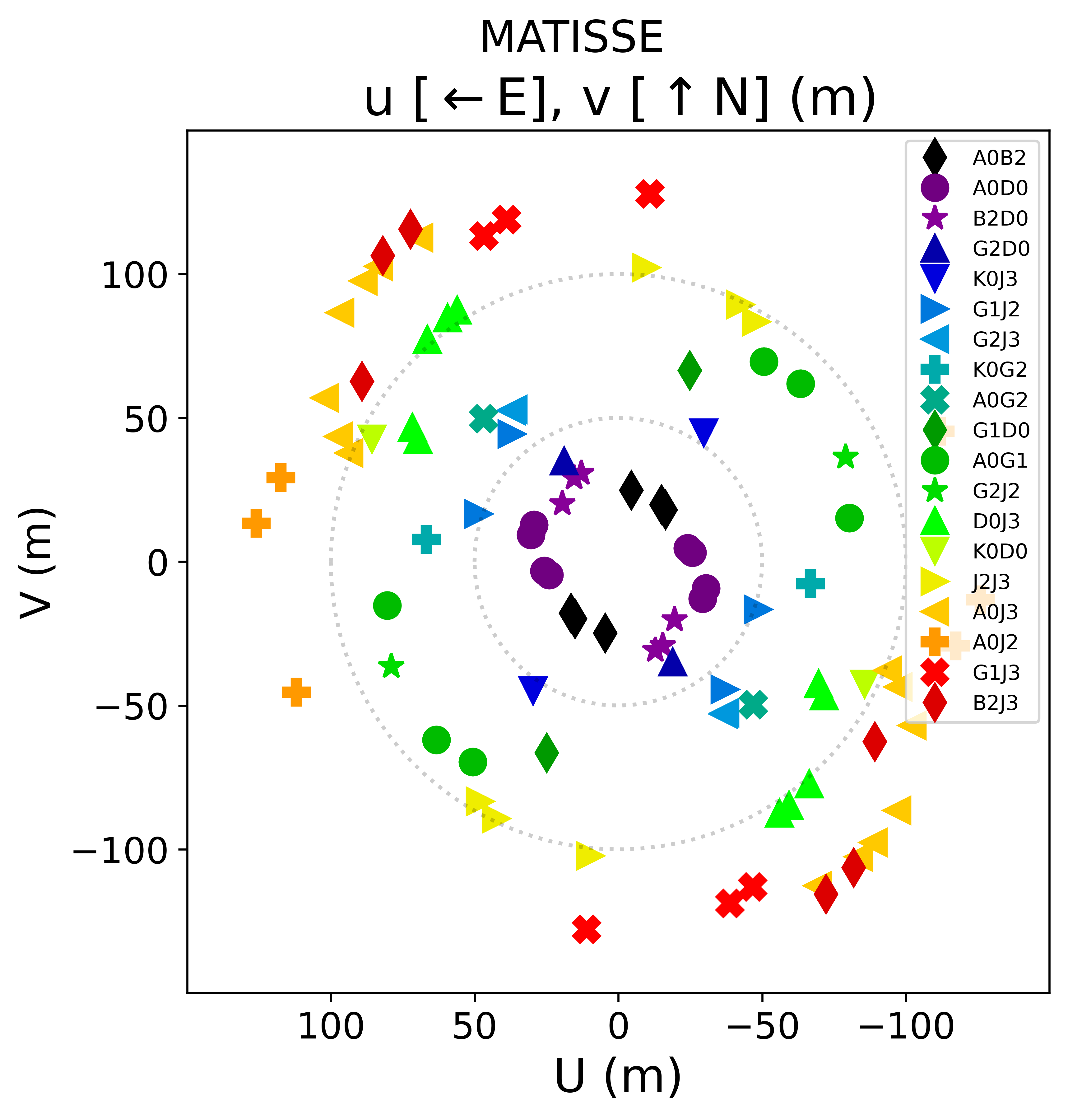} 
    \caption{
        uv coverage of the MATISSE observations. Each point corresponds to a projected baseline at a given wavelength. 
        The plot includes all configurations used, illustrating the sampling of the (u,v) plane 
        across different hour angles and baseline lengths. This coverage enables detailed image reconstruction and model fitting.
    }
    \label{fig:uvplane}
\end{figure}

The VLTI/MATISSE observations of IRAS 17163 were obtained in the $L$-band, covering the Br$\alpha$ transition at 4.05~$\mu$m. The wide range of baselines (Fig.~\ref{fig:uvplane}) allowed us to probe both compact and extended spatial scales, from the warm dust close to the sublimation rim to the more diffuse ionised gas.

In the continuum, the source is only moderately resolved: the calibrated visibilities remain high (0.8--0.95) at short and intermediate baselines, and decrease to $\sim$0.75 at the longest baselines (Table~\ref{tab:visibilities_two_column}). This behaviour is consistent with a compact continuum-emitting region of a few milliarcseconds in size.

In contrast, the Br$\alpha$ visibilities display a stronger dependence on baseline length. They range from near unity on short baselines, to 0.6--0.8 at intermediate baselines, and fall to as low as 0.2--0.4 on the longest baselines. This pronounced drop demonstrates that the Br$\alpha$-emitting region is substantially more extended than the continuum emitting region.

The differential phases exhibit marginal excursions across Br$\alpha$, with perhaps subtle S-shaped profiles (Sec.~\ref{PMOIRED_sec} and figures therein), that could hint to photocentre shifts of a few tenths of a milliarcsecond along the projected baselines. The signal-to-noise ratio (SNR) is not sufficient for robust detections on those signals. The closure phases do not show any significant variation along the line emission or the continuum. We conclude that the Br$\alpha$ emission originates from an extended and close to symmetric structure, yet distinct from the more compact L-band continuum emission.

\subsubsection{Spectra}

\begin{figure}[htbp]
    \centering
    \includegraphics[width=\linewidth]{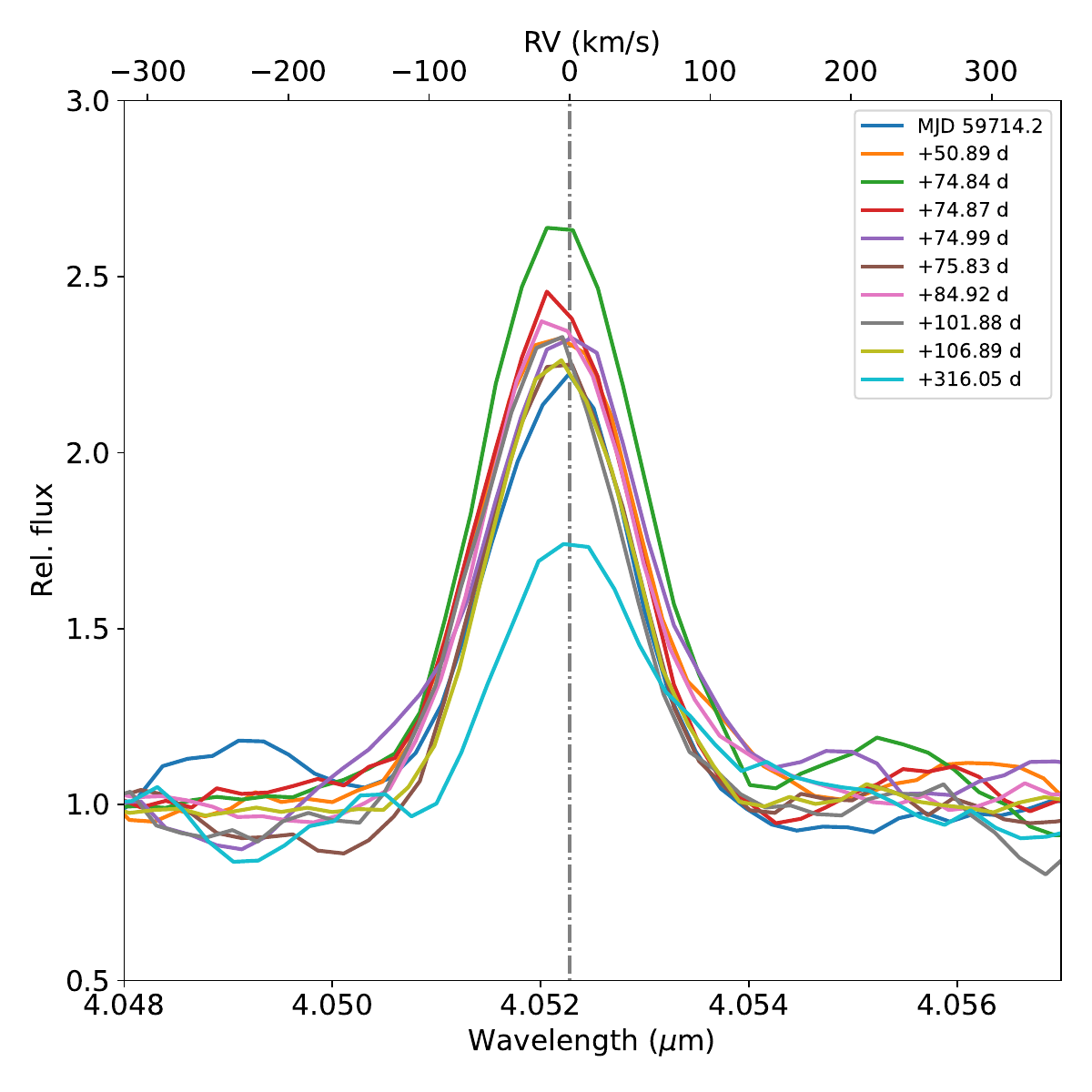}
    \caption{
Br$\alpha$ line profiles observed with MATISSE, corrected for heliocentric velocity shifts, and overplotted for all available epochs. This comparison highlights the temporal variability in the line emission, revealing changes in both the line shape and intensity. The profiles have been normalised to the continuum and aligned in velocity space to facilitate direct comparison.}
\label{fig:bralpha_profiles}
\end{figure}

Figure~\ref{fig:bralpha_profiles} shows the normalised Br$\alpha$ emission profiles extracted from the MATISSE $L$-band spectra over multiple epochs. The line exhibits a consistently strong emission peak centred at $\lambda = 4.052~\mu$m, corresponding to the systemic velocity within the instrumental resolution. The profiles are broad, with total velocity extents of $\sim$200--250~km\,s$^{-1}$, corresponding to Gaussian FWHM values of $\sim$110--130~km\,s$^{-1}$, and show relatively symmetric wings without evidence of significant absorption or P~Cygni-type features.

While the overall line shape remains stable, the peak intensity varies between epochs, with flux changes of up to $\sim$30\%. Such variability on timescales of weeks to months indicates temporal changes in the ionised gas emission, plausibly linked to variations in the mass-loss rate or the excitation conditions of the wind. At later epochs ($+316$~d), the line appears systematically weaker, suggesting a long-term decline in the emission strength. The consistency of the line centroid across epochs confirms that the variability is intrinsic to the emitting region rather than caused by large-scale radial velocity shifts. 

Each Br$\alpha$ spectrum was fitted with a Gaussian profile around the line centre. From these fits, we extracted the centroid velocity, FWHM, and equivalent width, with uncertainties derived from the covariance matrix. The resulting parameters are listed in Table~\ref{tab:bralpha_gauss_fits}.
Figure~\ref{fig:three_plots_Bralpha} shows the temporal evolution of the Br$\alpha$ equivalent width, FWHM, and centroid velocity. The relative stability of the line centroid, combined with variability in line strength and width, points to intrinsic changes in the ionised wind rather than large-scale dynamical shifts.

Overall, the Br$\alpha$ line parameters remain broadly stable throughout the monitoring period, with modest but significant variations. The centroid velocity ranges between $-14.5$ and $+4.9$~km\,s$^{-1}$, while the FWHM spans $110$–$129$~km\,s$^{-1}$. We note, however, that at the spectral resolving power of these observations ($R \sim 3300$), one resolution element corresponds to $\sim 90$~km\,s$^{-1}$. The observed $\sim 20$~km\,s$^{-1}$ centroid excursions are therefore well below the instrumental velocity resolution and cannot be interpreted as genuine radial-velocity variations, but instead reflect small profile or sampling differences within the unresolved line. The equivalent width varies from $-1.12$ to $-2.67$~nm, indicating changes in line strength, while no systematic drift in the centroid velocity is observed. Each Br$\alpha$ spectrum was measured independently for the four telescopes used and subsequently averaged. Because photometric calibration uncertainties are largely uncorrelated between individual telescopes, this averaging significantly suppresses instrumental systematics while preserving intrinsic spectral variability. The observed changes in equivalent width and line width are consistently present across all telescopes and exceed the expected residual calibration uncertainties, supporting their interpretation as genuine source variability rather than interferometric calibration effects.

A comparable pattern of line variability was previously observed in IRAS\,17163 by \citet{Koumpia2020}, who presented multi-epoch GRAVITY spectra over five months showing a gradual decrease of up to $\sim$67\% in the Br$\gamma$ peak intensity, while the Na\,\textsc{i} doublet remained constant within uncertainties of $\sim$13\%. That study demonstrated that the continuum and Na\,\textsc{i} emission are stable on monthly timescales, with the variability originating exclusively in the hydrogen recombination lines. The Br$\alpha$ behaviour reported here follows this same trend: the centroid velocity remains stable across epochs, whereas the line flux varies. 

The measured Br$\alpha$ line widths can be placed in a broader kinematic context by comparison with previous optical and millimetre observations of IRAS~17163. At larger scales, ACA observations reveal CO(2--1) and CO(3--2) emission associated with a fast molecular outflow, with expansion velocities up to $\sim$100~km\,s$^{-1}$, and a FWHM of $\sim$ 60~km\,s$^{-1}$ \citep{Wallstrom2017}. The Br$\alpha$ widths measured here (FWHM $\sim$110--130~km\,s$^{-1}$) are comparable to, or slightly exceed, these molecular expansion velocities, consistent with Br$\alpha$ tracing dense ionised gas in an accelerating inner wind that connects the compact recombination-line region to the larger-scale molecular outflow. Optical H$\alpha$ profiles \citep{Wallstrom2015} show a pronounced P-Cygni morphology with extended wings; however, their apparent emission-core width is strongly affected by self-absorption and optical-depth effects (and potentially electron scattering), and is therefore not directly comparable to the Br$\alpha$ FWHM.

The centroid velocities measured in the CRIRES$+$ spectra further support this picture: they remain remarkably stable across all hydrogen recombination lines, with variations of only a few km\,s$^{-1}$ (Table~\ref{crires_comparison}). This argues against variability driven by orbital motion of a companion or by continuum fluctuations, and instead supports an origin intrinsic to the ionised wind (see \ref{discussion}). 




\subsection{Spectropolarimetry}

\subsubsection{Continuum polarisation}

\begin{figure}[h!]
  \centering
  \includegraphics[width=0.48\textwidth, trim=0cm 0cm 0cm 0cm, clip]{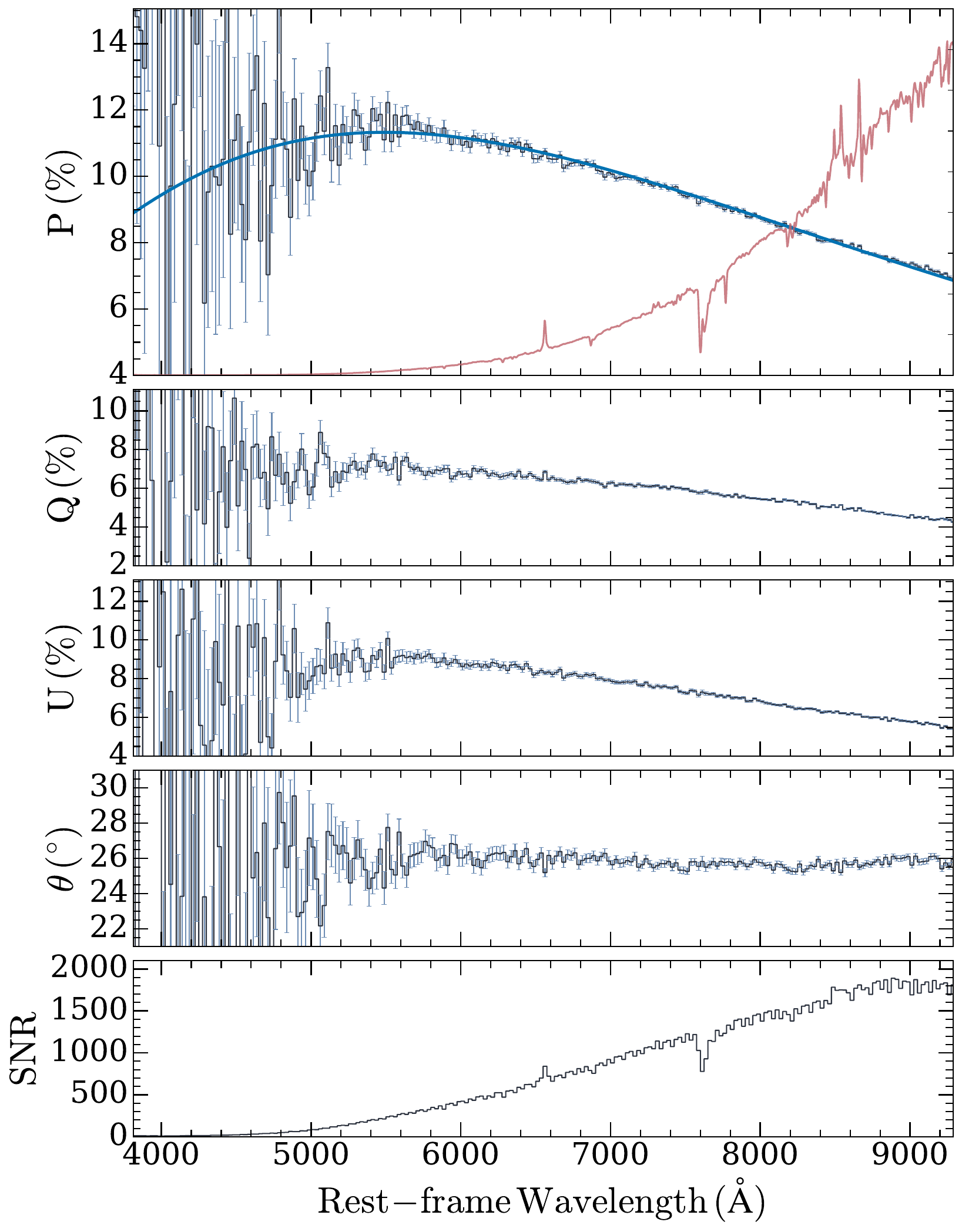}
    \caption{Combined FORS2 spectropolarimetric observations of IRAS~17163 obtained on July 4 and 8, 2023. The blue line in the top panel shows the Serkowski fit to the polarisation spectrum, while the red line represents the flux spectrum for comparison.}
    \label{fig:IRAS17163_specpol}
\end{figure}

We examined the individual spectropolarimetric sequences of IRAS~17163 obtained on July 4 and 8, 2023, and compared them with each other. The data taken on July 8 through the GG435 order-separating filter were saturated and had to be excluded. However, the two epochs without the GG435 filter and the one epoch with the filter are all consistent with each other within the noise. Figure~\ref{fig:IRAS17163_specpol} shows the combined spectropolarimetric observations of IRAS~17163 without the GG435 filter, obtained on July 4 and 8, 2023.

The polarisation curve can be described with the Serkowski law
\begin{equation}
\rm  P(\lambda) = P_{\max} \, \exp \left[ -K \, \ln^{2} \left( \frac{\lambda_{\max}}{\lambda} \right) \right],
\end{equation}
which empirically characterises the wavelength dependence of interstellar linear polarisation \citep{Serkowski1975}, 
where $\lambda_{\max}$ is the wavelength at which the polarisation reaches its maximum, 
P$_{\max}$ is the maximum polarisation, and K describes the width of the polarisation curve. 
The best-fit parameters are $\rm \lambda_{\max} = 5491.8 \pm 76.7$\,\AA, 
$\rm  P_{\max} = 11.33 \pm 0.08$\%, and $\rm  K = 1.81 \pm 0.14$.

The interstellar reddening at the position of IRAS~17163 is $\rm E(B-V) = 9.80 \pm 3.04$\,mag \citep{Schlafly2011}, and the relatively high maximum polarisation $\rm P_{max} \simeq$ 11.3 \% is below the empirical upper limit between interstellar polarisation and reddening
\begin{equation}
\rm P_{\max} \lesssim 9 \times  E(B-V) \quad [\%] , 
\end{equation}
found by \citet{Serkowski1975}.

\citet{Whittet1978} found an empirical relation between the wavelength of maximum polarisation and the ratio of total-to-selective extinction:  
\begin{equation}
R_V \simeq (5.6 \pm 0.3)\,\lambda_{\max} \,[\mu\mathrm{m}],
\end{equation}
where $\lambda_{\max}$ is expressed in microns. A typical Galactic value is 
$\lambda_{\max} \sim 0.55\,\mu$m, which corresponds to $R_V \sim 3.1$, the standard Milky Way extinction curve. 
For IRAS~17163 we measure $\lambda_{\max} = 5491.8 \pm 76.7$\,\AA\, which yields 
$R_V = 3.07 \pm 0.05$, fully consistent with the standard Galactic extinction law.

For the width of the polarisation curve, we found $K = 1.81 \pm 0.14$, which is slightly higher compared to the empirical $\lambda_{max}$ - $K$ relation and found by \citet{Whittet1992}:
\begin{equation}
K = (1.66 \pm 0.09)\,\lambda_{\max}[\mu\mathrm{m}] + (0.01 \pm 0.05).
\end{equation}

For a typical Galactic value of $\lambda_{\max} \sim 0.55\,\mu\mathrm{m}$, this gives $K \sim 0.9$.

However, we note that, because it is a red object, the SNR of the spectropolarimetric data drops below $\lesssim$ 100 at wavelengths below 5000 \AA\ (see Fig.~\ref{fig:IRAS17163_specpol}), and the high K value may be an artefact due to the short wavelength coverage of the data. 
%
%
%
\citet{Bagnulo2017} has shown using Monte Carlo simulations that the scatter in K, and the $\lambda_{max} - K$ correlation depend on the wavelength range in which the observations are carried out. Furthermore, \citet{Cikota2017} found a systematic and statistically significant deviation of the K values in \citet{Whittet1992} compared to FORS data. 


Thus, the Serkowski curve parameters are consistent with normal Milky Way dust, implying that the polarisation of IRAS~17163 is produced by the Galactic interstellar medium along the line of sight. Moreover, IRAS~17163 displays a constant wavelength-independent polarisation angle of 25.8 $\pm$ 0.3 degrees (measured between 6000-9000 \AA), which corresponds to 80.9 $\pm$ 0.3 degrees in Galactic coordinates at a position of \textit{l} = 348.5108102 deg and \textit{b} = -1.1201040 deg.

The wavelength-independent polarisation angle is generally consistent with the polarisation being produced by dust within a single dominant cloud structure along the line of sight with a relatively uniform projected magnetic field orientation (see, e.g., \citealt{1997ApJ...487..314M}).

We calculated bandpass-integrated polarization by integrating the ordinary and extraordinary flux beams over the Bessell $V$, $R$, and $I$ transmission functions and subsequently deriving the polarization. 
The resulting polarization values are $P_V = 11.22 \pm 0.20\%$, $P_R = 9.83 \pm 0.02\%$, and $P_I = 8.44 \pm 0.01\%$. 
The polarization angle remains nearly constant across the three bands, with $\theta_V = 24.92 \pm 0.53^\circ$, $\theta_R = 25.33 \pm 0.04^\circ$, and $\theta_I = 24.47 \pm 0.02^\circ$.

Interestingly, our recent FORS2 spectropolarimetry of 2023 is fully consistent with the 2015 dataset presented in \citet{Koumpia2020}, both in polarisation degree and position angle, suggesting a stable interstellar origin. In contrast, earlier measurements by \citet{LeBertre89} report a higher polarisation level and a wavelength dependent polarisation angle. They reported polarisation degrees from 12.7\% in $V$ to 12.1\% in $R$ and 11.5\% in $I$ (see their Table 5), while the polarisation angle changes from 27$^\circ$ ($V$) to 26$^\circ$ ($R$) and 24$^\circ$ ($I$).
The broader wavelength coverage of the 2023 dataset confirms this difference and highlights that part of the 1988 signal may have been intrinsic, as already speculated by \citet{Koumpia2020}.

\subsubsection{H$\alpha$ and Ca\,{\sc ii} line polarisation}

We investigated the polarisation behaviour across the H$\alpha$ line to probe the geometry of the circumstellar environment. Variations in polarisation across emission lines, which are intrinsically unpolarised, can provide information on scattering and spatial distribution of ionised material close to the star \citep[e.g.][]{Oudmaijer1999}. Detecting changes of polarisation between the continuum and the line emission offers direct evidence for the presence of an asymmetric structure in the innermost regions of the star \citep[see also  Sect. 3.3 in][]{Koumpia2020}.

\begin{figure}[h!]
  \centering
    \includegraphics[width=0.48\textwidth, trim=0cm 0cm 0cm 0cm, clip]{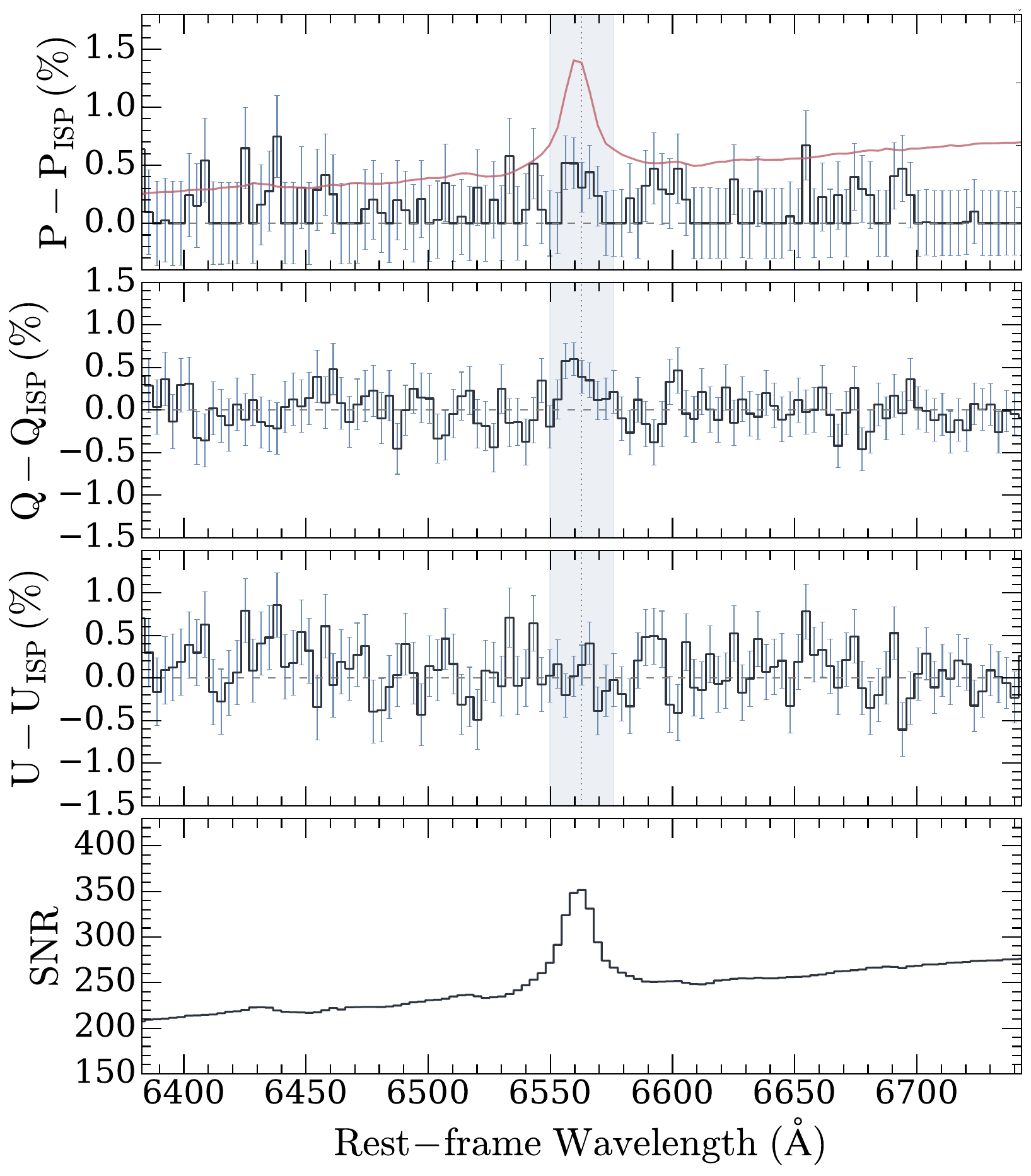}
  \includegraphics[width=0.48\textwidth, trim=0cm 0cm 0cm 0cm, clip]{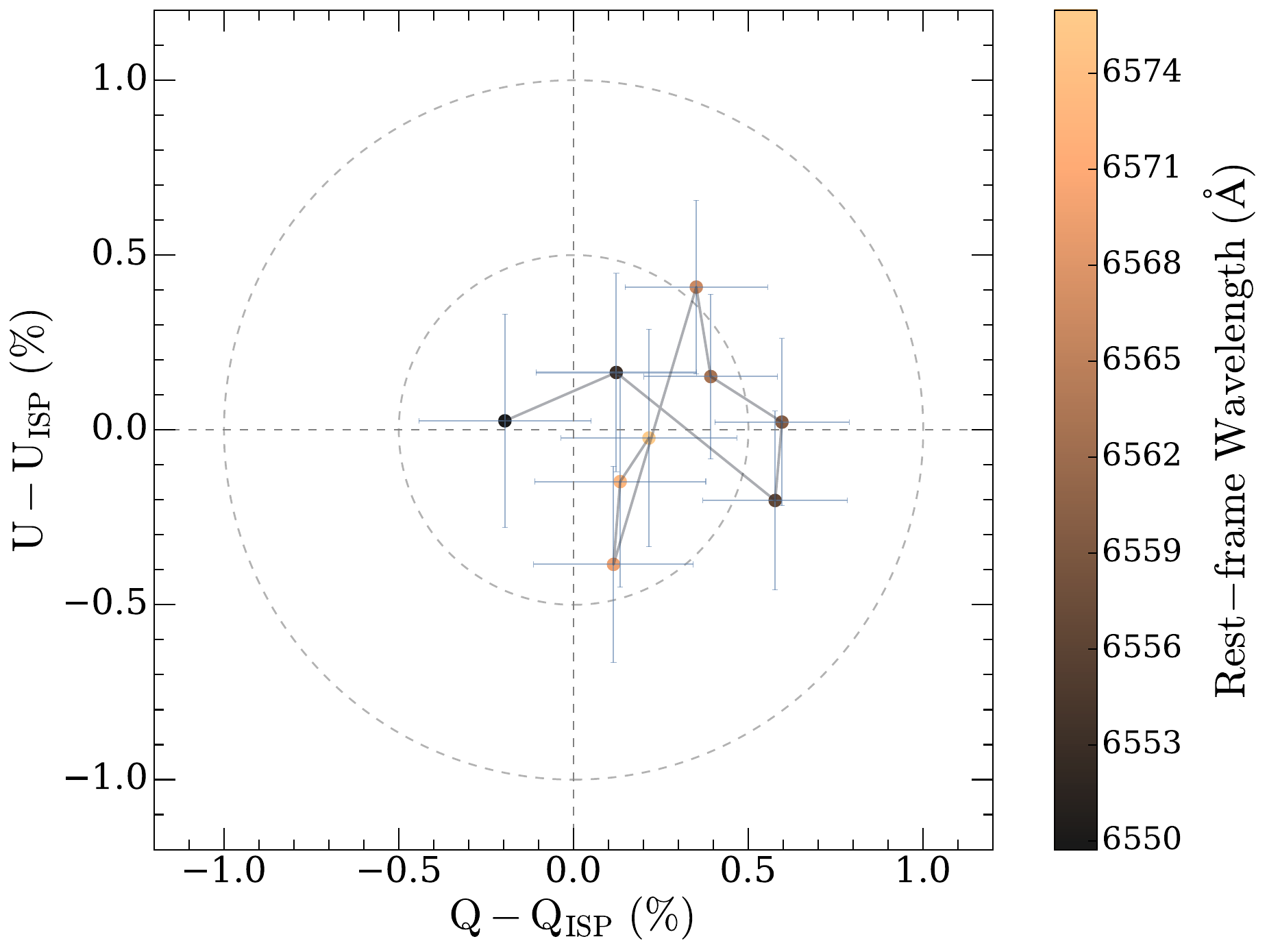}\\
\caption{Combined FORS2 spectropolarimetric observations of the H$\alpha$ line in IRAS~17163 obtained on July 4 and 8, 2023. The data are unbinned and continuum-subtracted. The polarisation spectrum (top panel) has been corrected for polarisation bias. The red line in the top panel represents the flux spectrum. The light-blue shading indicates the wavelength range 6550--6576~\AA\ used for plotting the H$\alpha$ line in the $Q$--$U$ plane (bottom panel).}
    \label{fig:IRAS17163_specpol_Ha}
\end{figure}

We subtracted the continuum polarisation using wavelet decomposition, following the method described by \citet[][see Sect.~4.2]{Cikota2019MNRAS.490..578C}, and present the unbinned and interstellar polarisation subtracted spectrum in Figure~\ref{fig:IRAS17163_specpol_Ha}. 
We confirm a $\sim$0.5\% ($\sim$2-3$\sigma$) increase in polarisation across H$\alpha$ in Stokes $Q$, as previously detected at higher significance in the higher spectral resolution data in \citet{Koumpia2020}, while Stokes $U$ shows no significant variation. Both the detection of the polarisation in $Q$ and the non-detection in $U$ are consistent with the results reported by \citet{Koumpia2020}.
Furthermore, the data does not show obvious lines or loops in the Stokes $Q$-$U$ plane. 

\begin{figure}[h!]
  \centering
    \includegraphics[width=0.48\textwidth, trim=0cm 0cm 0cm 0cm, clip]{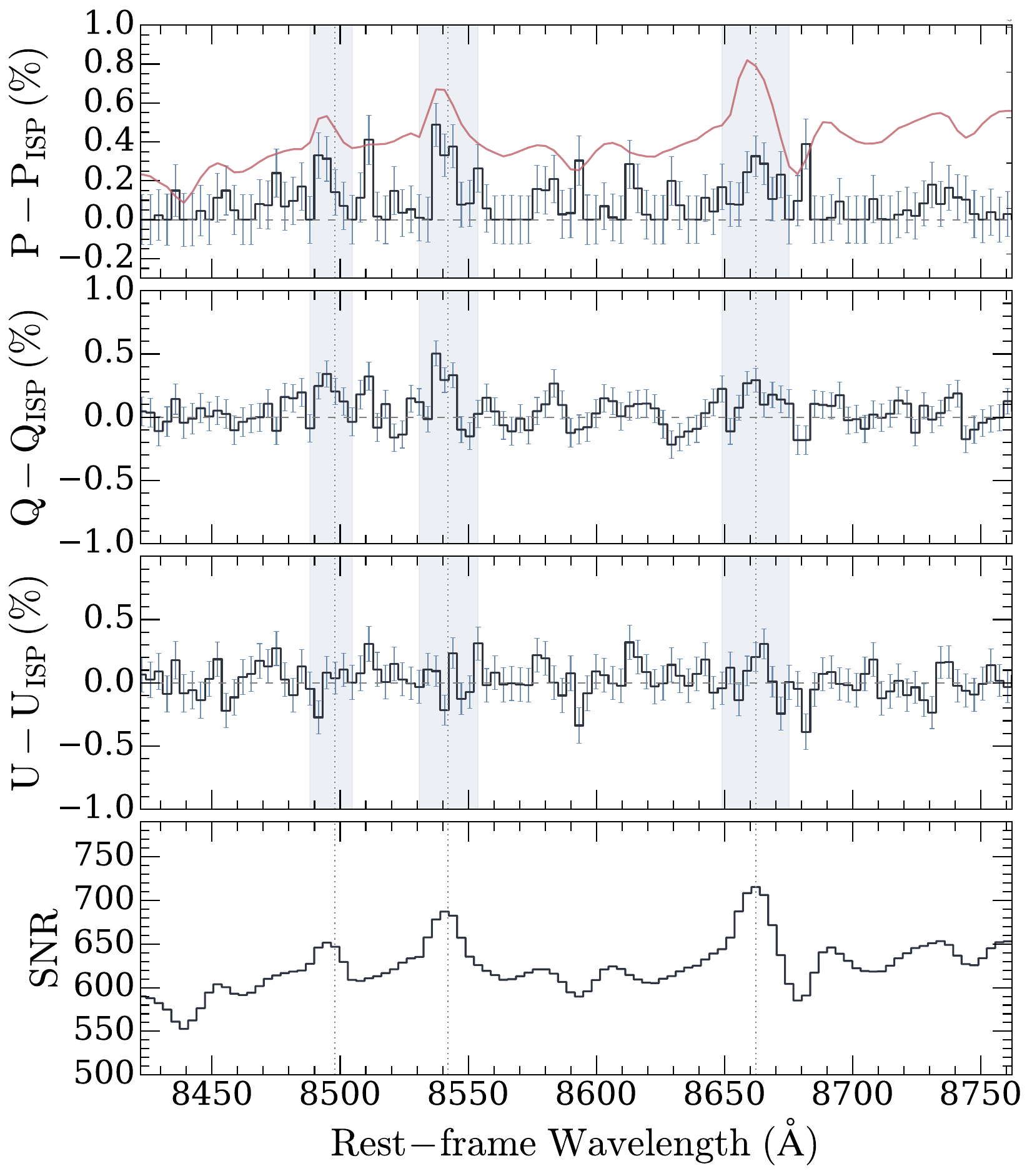}
    \caption{Combined FORS2 spectropolarimetric observations of the Ca\,{\sc ii} triplet lines in IRAS~17163 obtained on July 4 and 8, 2023. The data is unbinned and continuum-subtracted. The polarisation spectrum (top panel) has been corrected due to polarisation bias. The red line in the top panel represents the flux spectrum. The light-blue shading indicates the wavelength ranges used for plotting the Ca\,{\sc ii} line in the $Q$–$U$ plane (Fig~\ref{fig:IRAS17163_specpol_CaII_QU}).}
    \label{fig:IRAS17163_specpol_CaII}
\end{figure}

The present spectropolarimetric data have a wider wavelength coverage than those in \citet{Koumpia2020} allowing for the study of the full spectrum, and in particular the strong Ca\,{\sc ii} infrared triplet lines at 8498.018, 8542.089, and 8662.140\,\AA\ (Fig.~\ref{fig:IRAS17163_specpol_CaII}). It would appear that all three lines show intrinsic line polarisation, which is more pronounced in Stokes~$Q$ than in Stokes~$U$. It could be argued that the excursions are below 3$\sigma$ level; however, the line effects are observed in all three lines with similar Stokes $Q$ and $U$ properties as H$\alpha$, and may thus be real. In fact, line depolarisation has rarely been seen in other lines than H$\alpha$. \citet{Ababakr2016} were among the first to detect a line-effect in the Calcium triplet, which they explain as the Calcium lines arising from outside the (electron-scattering) ionised region.
Generally, the detection of a change in polarisation across emission lines may indicate that the electron-scattering region is not spherical (i.e., there is axisymmetry, see e.g. \citealt{2008MNRAS.385..967P}).

\begin{figure*}[h!]
  \centering
    \includegraphics[width=0.33\textwidth, trim=0cm 0cm 0cm 0cm, clip]{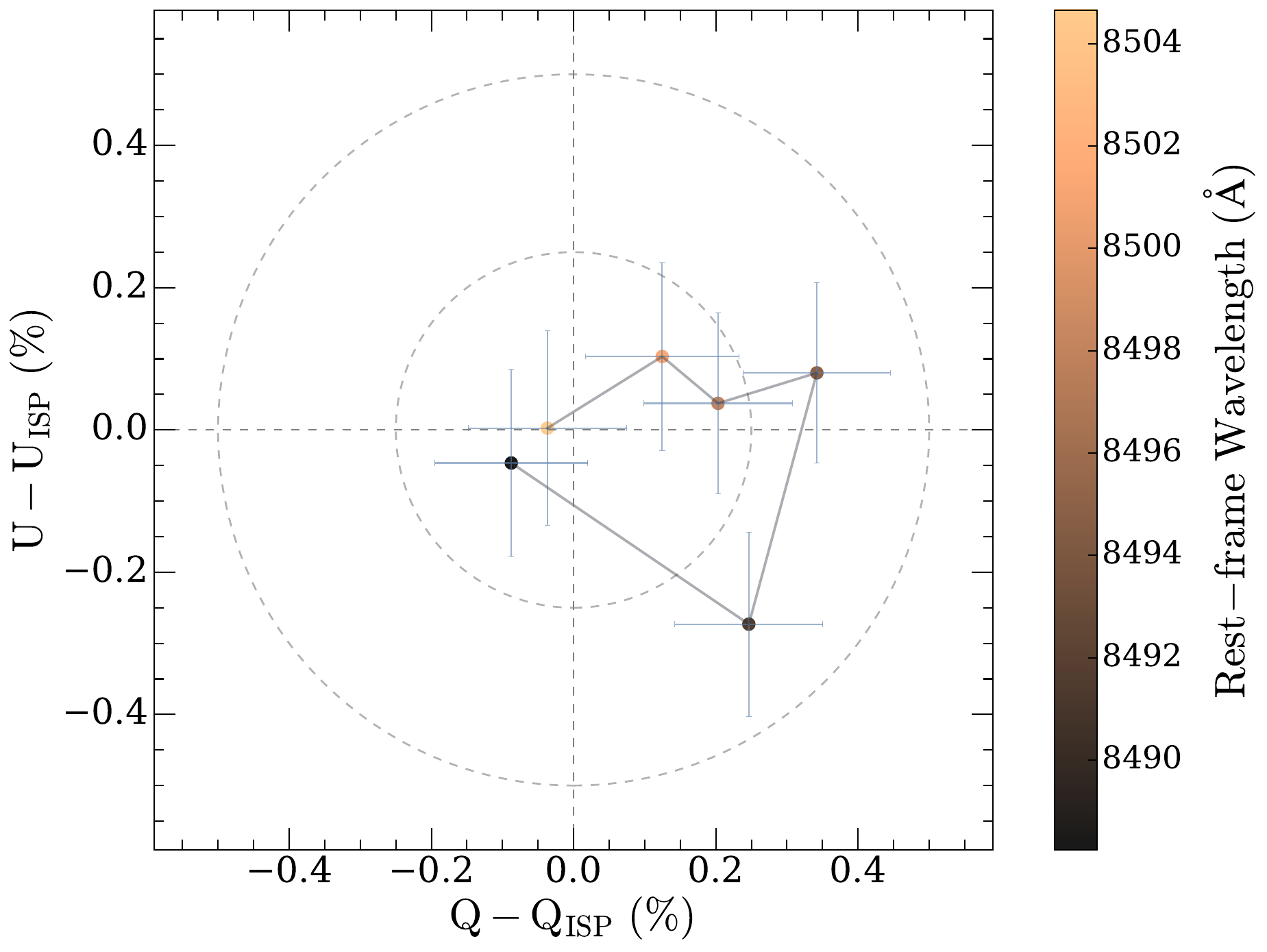}
    \includegraphics[width=0.33\textwidth, trim=0cm 0cm 0cm 0cm, clip]{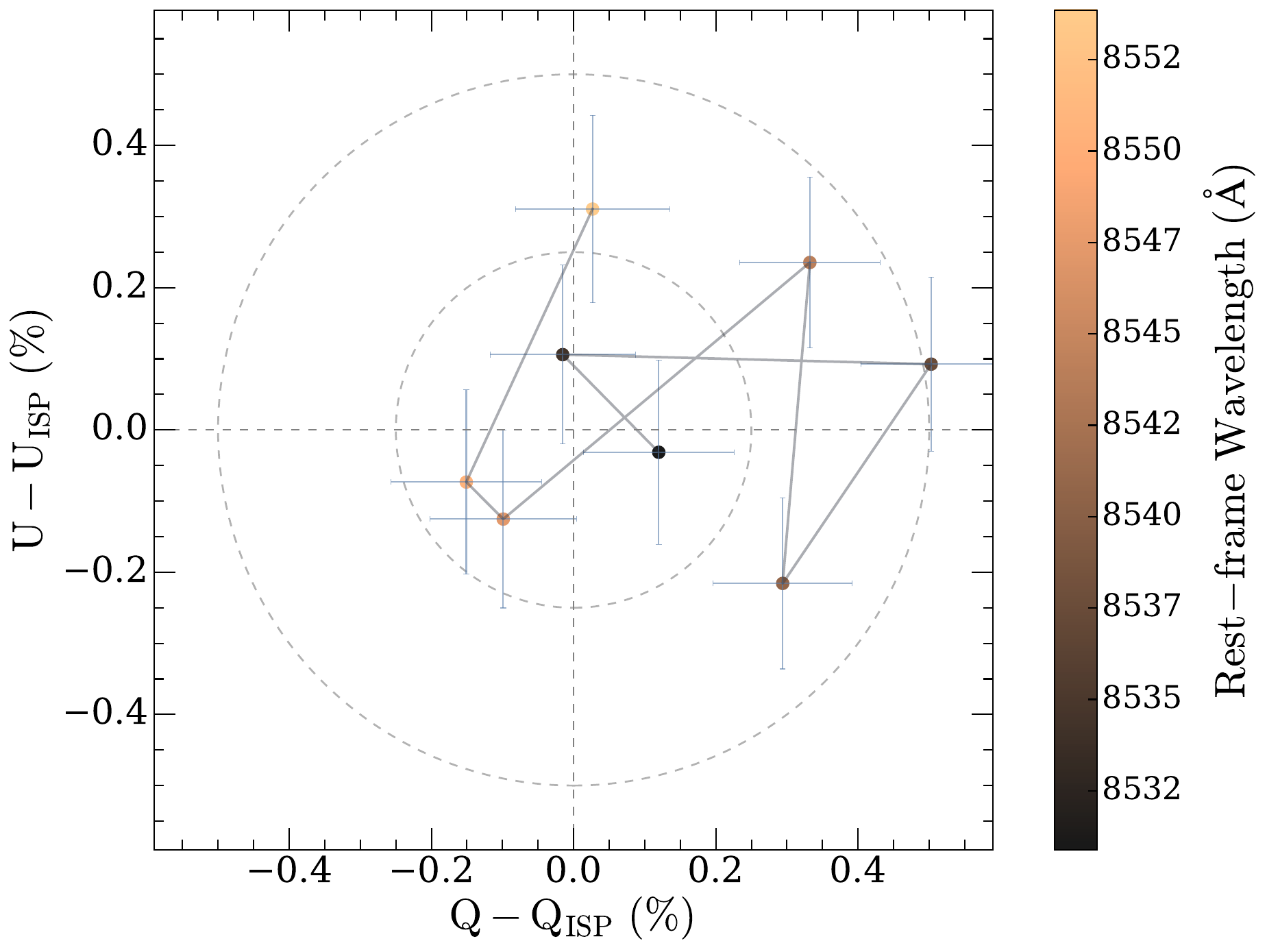}
    \includegraphics[width=0.33\textwidth, trim=0cm 0cm 0cm 0cm, clip]{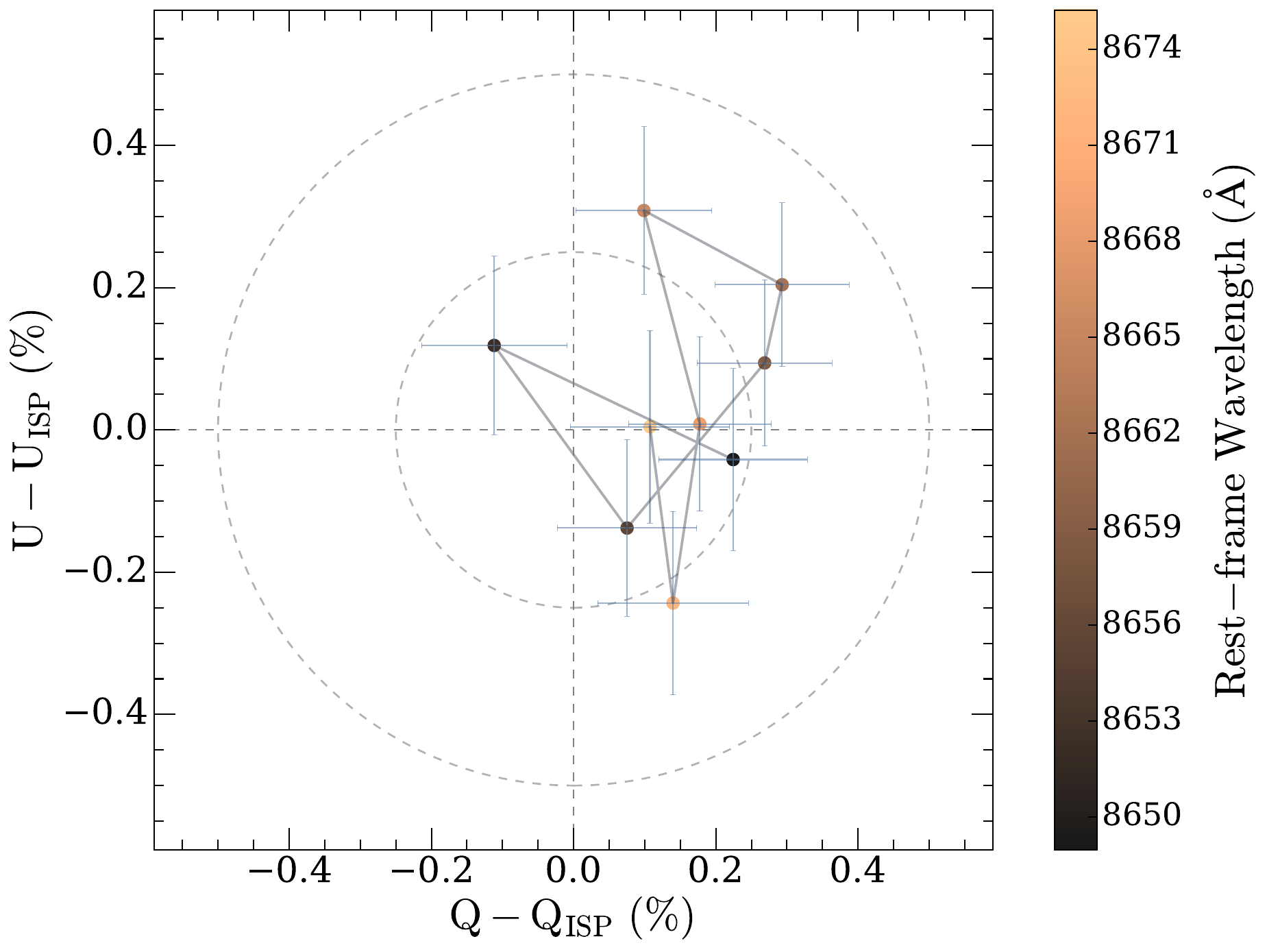}
    \caption{$Q$–$U$ diagrams for the Ca\,{\sc ii} infrared triplet lines in IRAS~17163, obtained on July~4 and~8,~2023. The data are unbinned and continuum-subtracted. The left, centre and right panels correspond to the Ca\,{\sc ii}~8498.018\,\AA, 8542.089\,\AA, and 8662.140\,\AA\ lines, respectively.}
    \label{fig:IRAS17163_specpol_CaII_QU}
\end{figure*}

The $Q$-$U$ diagrams for the Ca\,{\sc ii} lines (Fig.~\ref{fig:IRAS17163_specpol_CaII_QU}) exhibit no clear loops or lines in the Stokes $Q$-$U$ plane. The only possible exception is the 8498.018\,\AA\ line, which shows tentative indications of a loop-like feature.
Loops in the Stokes $Q$-$U$ plane may imply the presence of a clumpy/inhomogeneous envelope whose scattering geometry changes with velocity (i.e., line-of-sight velocity shifts probe different scattering angles/regions).


\footnotetext{Polarisation bias arises because the polarisation degree $P = \sqrt{Q^2 + U^2}$ is always positive, even when the true signal is zero. Noise in the Stokes parameters $Q$ and $U$ therefore leads to a systematic overestimation of $P$. We apply the polarization bias correction following \citet{1997ApJ...476L..27W}:
$P = \left( P_{\mathrm{obs}} - \frac{\sigma_P^2}{P_{\mathrm{obs}}} \right)
\times h\!\left(P_{\mathrm{obs}} - \sigma_P \right)$ ,
where $h$ is the Heaviside function, $P_{\mathrm{obs}}$ is the observed
polarization, and $\sigma_p$ is the $1\sigma$ error on the polarization.
}

\section{Interferometric modelling}


  
To interpret the VLTI/MATISSE observables of IRAS~17163, we adopted a stepwise modelling strategy. As a first step, we applied simple geometric brightness distributions to the continuum and Br$\alpha$ line emission in order to derive basic size estimates and assess the relative extent of the two components. These models provide a useful benchmark and highlight the systematic differences between the compact L-band continuum and the more extended hydrogen line emission. Building on these results, we then employed the parametric spectro-interferometric tool \texttt{PMOIRED}, which allows simultaneous fitting of visibilities, closure phases, differential phases, and flux spectra. This approach enables us to test progressively more sophisticated descriptions of the source geometry, and to explore how additional free parameters improve the fit quality and capture the observed asymmetries.

\subsection{Simple geometric models}
\label{sec:geometric_models}

\begin{figure}[htbp]
    \centering
    \includegraphics[width=\linewidth]{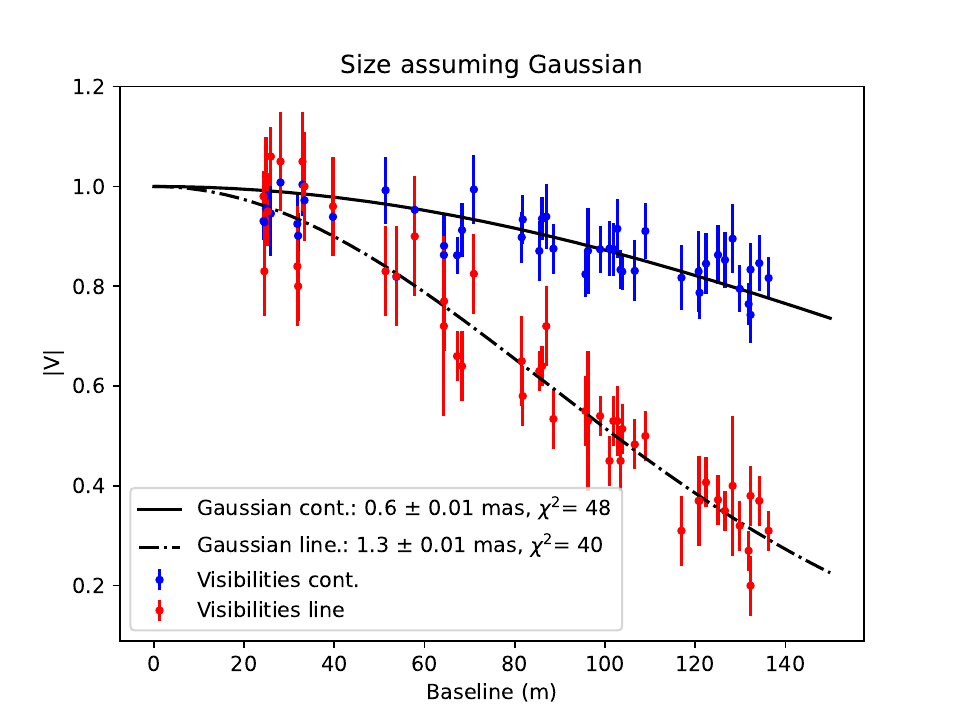}
    \includegraphics[width=\linewidth]{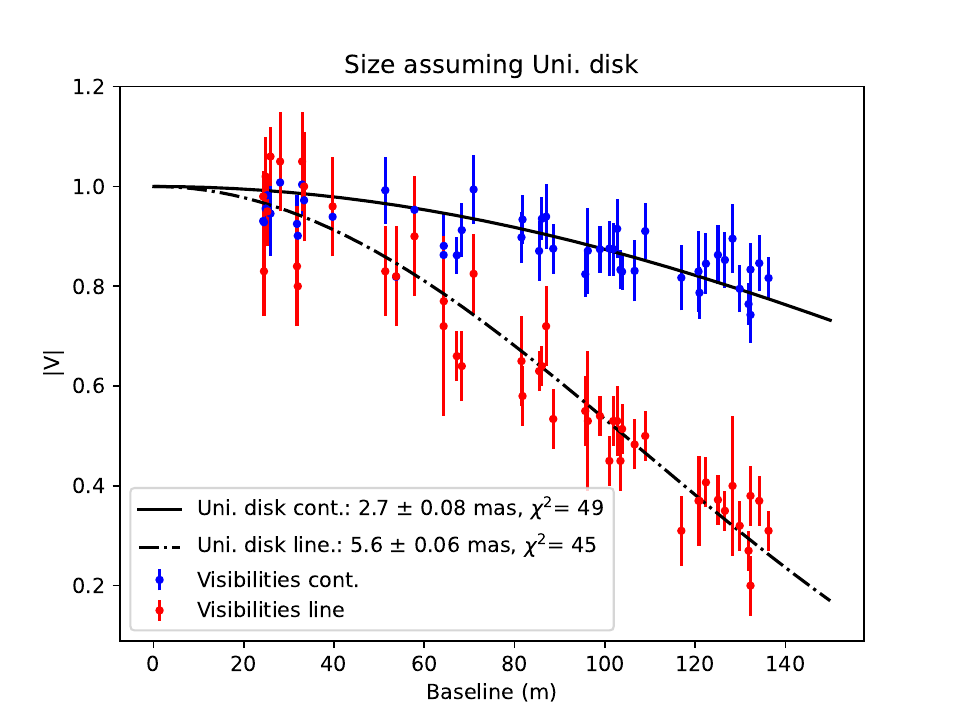}
    \includegraphics[width=\linewidth]{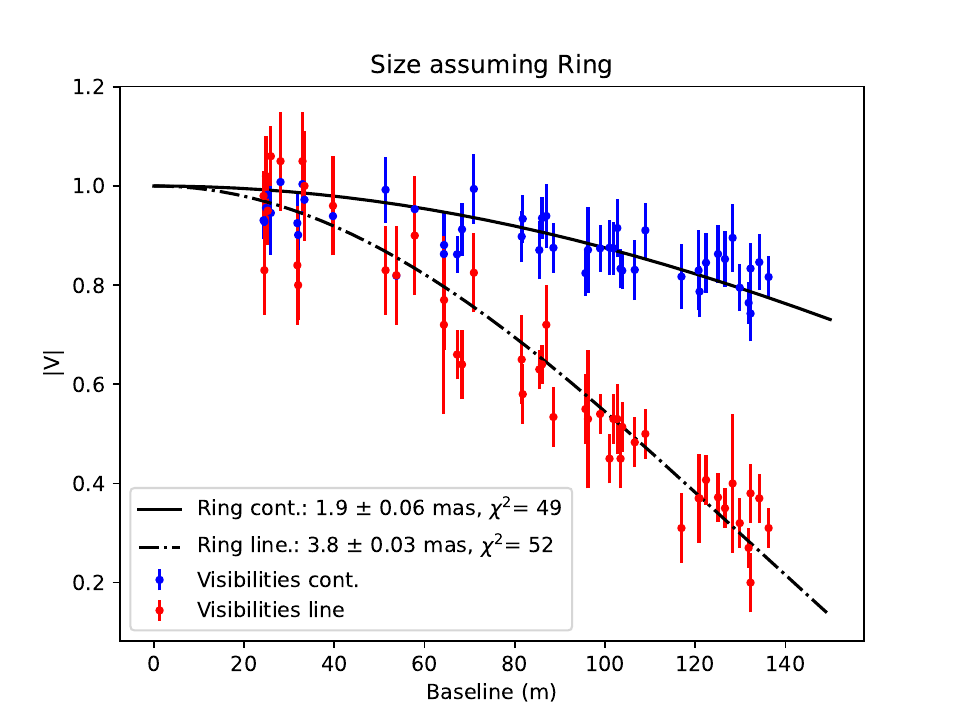}
    \caption{Visibility amplitudes as a function of baseline length, shown for the continuum (averaged for 4.04-4.05 $\mu$m and 4.054-4.06 $\mu$m; blue), and at the peak of the Br$\alpha$ line emission at 4.052 $\mu$m (red). The observed visibilities are fit to three different brightness distributions: a Gaussian (top), a uniform disc (middle), and a ring (bottom). By fitting these models to the observed visibilities, we infer the characteristic size of the emitting region for both the continuum and the Br$\alpha$ emission. The fitted size and reduced $\chi^2$ values for each model are indicated in the corresponding panels, allowing comparison of model performance.}

\label{fig:geom_models_simple}
\end{figure}

Before applying detailed spectro-interferometric modelling with \texttt{PMOIRED}, we performed simple fits of the visibility amplitudes using standard geometric brightness distributions (Gaussian, uniform disc, and ring; Fig~\ref{fig:geom_models_simple}) in Python. This approach provides a first-order estimate of the characteristic size of the continuum and Br$\alpha$--emitting regions. In all cases, the Br$\alpha$ emission is found to be about a factor of two larger than the compact continuum source. 

The Gaussian model yields characteristic sizes of $\sim$0.6\,mas for the continuum and $\sim$1.3\,mas for the Br$\alpha$ emission, the uniform-disc model $\sim$2.7\,mas and $\sim$5.6\,mas, and the ring model $\sim$1.9\,mas and $\sim$3.8\,mas, respectively. These simple fits robustly demonstrate that the hydrogen line emission is substantially more extended than the L-band continuum and motivate the use of a more elaborate parametric description. We, therefore, in the following section, apply \texttt{PMOIRED} to simultaneously model the continuum and Br$\alpha$ emission in order to better constrain their geometry and flux contributions, also taking into account the observed differential and closure phases.

\subsection{PMOIRED}
\label{PMOIRED_sec}

To characterise the spatial morphology of the L-band emission observed with VLTI/MATISSE, we performed parametric modelling using the dedicated tool \texttt{PMOIRED}\footnote{PMOIRED is available at https://github.com/amerand/PMOIRED} \citep{Merand2022}. This software is optimised for spectro-interferometric data and enables polychromatic modelling of various astrophysical components through combinations of standard brightness distributions, such as Gaussians, uniform discs, and rings.

In \texttt{PMOIRED}, the total flux distribution is described by:
\begin{equation}
F(x, y, \lambda) = I(x, y) \times S(\lambda),
\end{equation}
where $(x, y)$ are angular sky coordinates and $\lambda$ is the wavelength. The term $I(x, y)$ denotes the wavelength-independent spatial structure (e.g. geometry and orientation), while $S(\lambda)$ represents the component’s spectral dependence (e.g. flat continuum, emission line profile). This approach allows the user to independently constrain the source geometry and its spectral features, offering flexibility and physical interpretability \citep{Monnier2003}.

\texttt{PMOIRED} computes interferometric observables (visibilities, closure phases, differential phases) for a given model brightness distribution and compares them to the observations using a least-squares minimisation fit. Uncertainties on the best model parameters are estimated via bootstrapping, and the covariance and correlation matrices were checked to assess potential degeneracies. This diagnostic is essential in evaluating the reliability of the model parameters, especially in cases where the data are not sufficient to independently constrain all dimensions of the model.

For this study, we focus on the L-band continuum and the Br$\alpha$ emission. The geometric progression fits and the resulting brightness distribution for the continuum and line emission are described below. 


\subsubsection{L-band continuum modelling with PMOIRED}
\label{sec:pmoi_continuum}

The continuum was modelled in the $L$-band spectral range near the Br$\alpha$ line ($4.067$–$4.080~\mu$m) using a progressive sequence of parametric models. As a first step, the visibilities and phase observables were fitted with a simple uniform disc representing the stellar photosphere. This model already reproduced the data with a reduced $\chi^{2}$ of $\sim$0.9, consistent with the star being only marginally resolved at the available baselines. Allowing for a radial intensity profile ($\mu^{0.5}$ limb darkening) yielded nearly identical results, confirming the robustness of the inferred stellar diameter.  

Overall, the continuum can be well described by a partially resolved continuum emission with an angular size of $\sim$2.42$\pm$0.03 mas (consistent with the size of 2.7~mas obtained in Sec.~\ref{sec:geometric_models} when accounting for the uniform disc distribution) and a small contribution from fully resolved extended flux ($\sim$5\%). To place the L-band continuum size in context, we compare it to previous constraints on the stellar angular diameter. 
Adopting $d=1.2$\,kpc, the best-fitting uniform-disc continuum diameter, $\theta_{\rm UD}=2.42$ mas, corresponds to a linear diameter of $D=\theta d \simeq 2.90$\,au, i.e. a radius of $R \simeq 1.45$\,au ($R\simeq 312\,R_\odot$). 
This is consistent with the expected stellar diameter of $\sim$2.5\,mas inferred from the stellar parameters presented in \citet{Koumpia2020} (corresponding to $D\sim3$\,au, or $\sim$650\,$R_\odot$ in diameter), within the systematic uncertainties associated with simple geometric brightness distributions. A comparison with previous near-infrared interferometric measurements indicates that the continuum size of IRAS~17163 is wavelength dependent.
In particular, the K-band continuum size derived from VLTI/GRAVITY in \citet{Koumpia2020} was smaller (e.g.\ $\theta_{\rm UD}\simeq1.73\pm0.02$\,mas). 

A natural question is whether the larger apparent size in the L-band could arise from a wavelength-dependent photospheric radius, i.e. because the continuum $\tau_\lambda \simeq 1$ surface forms at slightly different depths at 2--4\,$\mu$m. For a star with $T \sim 8000$\,K, the near-infrared opacity varies only weakly with wavelength, so the corresponding shift of the photospheric radius is expected to be at most a few atmospheric scale heights, i.e. $\lesssim 10^{4}$\,km even under very conservative assumptions. At a distance of 1.2\,kpc this translates to $\ll 1~\mu$as, which is entirely negligible compared to the milliarcsecond-scale difference observed between the K- and L-band continuum sizes. The leading cause of the chromatic size increase seems to be due to the growing contribution of spatially extended circumstellar emission at longer wavelengths rather than a change in the stellar radius itself.


The present continuum model provides a stable baseline for the more complex line-emission modelling discussed as follows.

\subsubsection{Br$\alpha$ line emission modelling with PMOIRED}
\label{sec:pmoi_continuum}

To fit the line emission with PMOIRED, we adopted a parametric description of a compact stellar source surrounded by an extended Br$\alpha$--emitting environment. The model included seven free parameters: the environment-to-star continuum flux ratio ($f_{\rm env}$), the angular size assuming a Gaussian spatial distribution (FWHM) of the environment, the flux and Gaussian width of the Br$\alpha$ line emission, the photocenter offsets of the line-emitting region compared to the continuum along right ascension and declination, and a flux contribution from an extended, over-resolved component.


The fitting procedure was carried out in a progressive manner, starting with simple brightness distributions and gradually including additional observables and free parameters until a satisfactory solution was reached. We initially restricted the fit to visibility amplitudes only, in order to constrain the overall size of the emission region, and subsequently added the differential and closure phases, as well as the spectrally dispersed flux, to account for asymmetries and spectral-line behaviour. We explore simple geometries, such as uniform-disc and Gaussian brightness distributions. These models reproduced the continuum visibilities reasonably well and provided reduced $\chi^{2}$ in the range 1.8-3.4, once all epochs and observables were considered. 

To explore more flexibility, we tested radial profiles and Gaussian intensity distributions, which marginally improved the fits but still failed to fully capture the observed phase signals. A more refined model was then introduced by allowing the Br$\alpha$ line-emitting region to be offset from the continuum (free $(x,y)$ parameters). This configuration significantly reduced the residuals ($\chi^{2}_{\nu}\sim 1.8$), and reproduced both the amplitude drops in the line and some signatures, but not fully, in the differential-phases. Finally, we examined more complex geometries, such as azimuthally varied profiles, but these resulted in degrading the fit solutions ($\chi^{2}_{\nu}\sim 5$), confirming that such a degree of complexity is not required for these observations. This stepwise approach shows that while simple symmetric profiles can describe the continuum, the inclusion of spatial offsets in the line emission is essential to reproduce the full set of interferometric observables, paving the way for the disc-like model solution discussed below. 

The best-fit solution reached a reduced $\chi^{2} = 1.03$.  In particular, the environment-to-star flux ratio was found to be $f_{\rm env} = 0.023 \pm 0.009$, with a size of uniform disc of $5.55 \pm 0.12$~mas. The Br$\alpha$ line was reproduced with a flux of $1.101 \pm 0.021$ (in continuum units) and a Gaussian width of $2.09 \pm 0.03$~nm, while the line-emitting region showed photocenter marginal offsets of $\Delta x = 0.194 \pm 0.071$~mas and $\Delta y = 0.089 \pm 0.079$~mas. Although these offsets correspond to sub-resolution spatial scales for MATISSE, they are measurable through differential phase information: spectro-astrometry enables photocenter shifts to be constrained well below the nominal angular resolution by exploiting the differential signal across the spectral line. Such shifts do not necessarily imply a bulk displacement of the line-emitting region with respect to the continuum, but can naturally arise from structural inhomogeneities or asymmetric brightness distributions within the Br$\alpha$-emitting gas. The present modelling, therefore, captures the minimum level of asymmetry required by the interferometric observables, without uniquely prescribing the underlying physical origin.

\begin{figure}[htbp]
    \centering
    \includegraphics[width=0.5\textwidth]{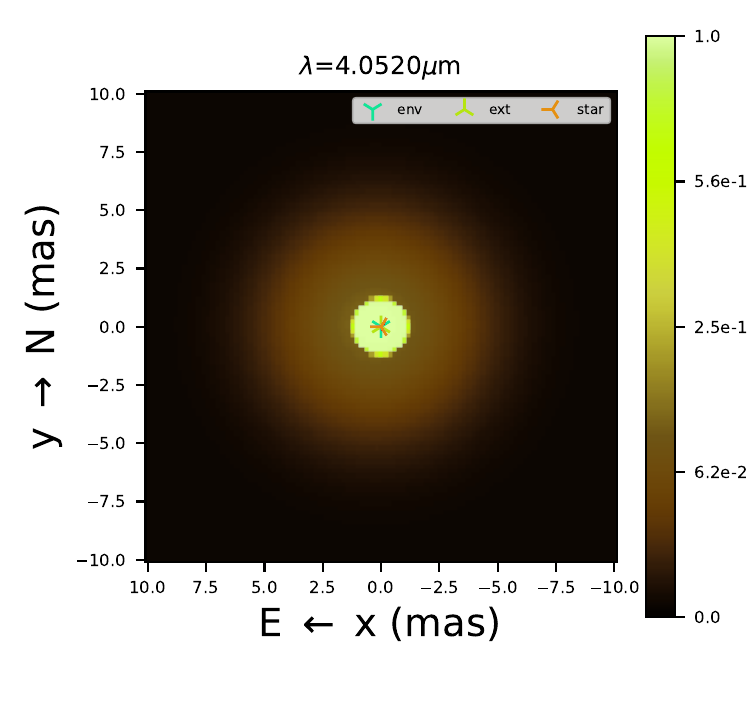}
    \caption{
    Model image corresponding to the best-fit brightness distribution that reproduces the observed MATISSE interferometric observables around the entire line profile of the Br$\alpha$ emission. The image was generated using \texttt{PMOIRED}, adopting the same geometry and parameters that yield the visibility, closure phase, and differential phase fits shown in Fig.~\ref{fig:model_vs_data}. The intensity is normalised and displayed in logarithmic scale to highlight both the compact continuum and more extended Br$\alpha$-emitting structures.
    }
    \label{fig:pmoired_model_image}
\end{figure}

\begin{figure*}[htbp]
    \centering
    \includegraphics[width=0.8\textwidth, trim=0cm 6.5cm 0cm 8.5cm, clip]{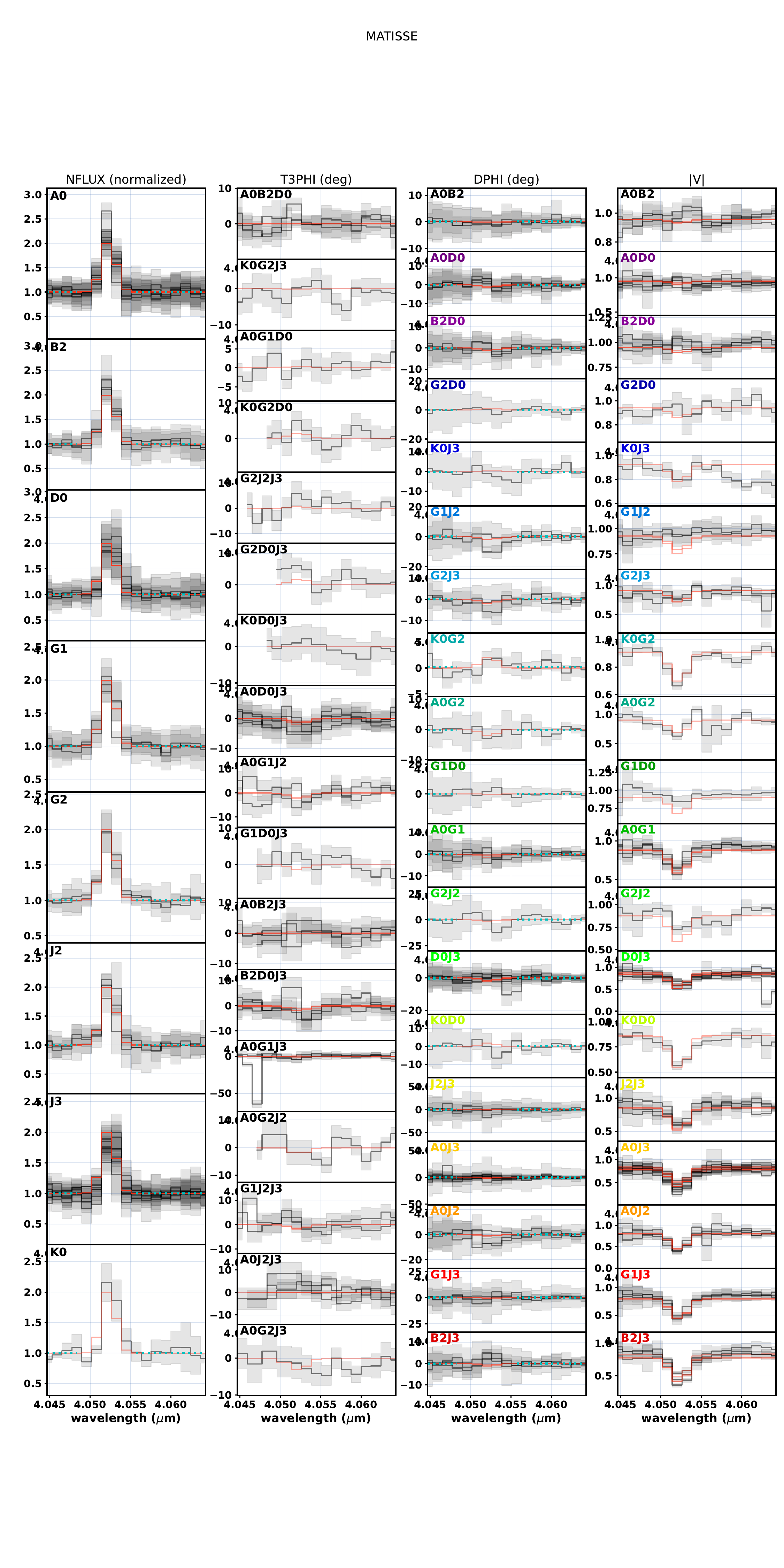}
    \caption{
    Comparison between observed MATISSE interferometric observables (black) and best-fit model (red) predictions across all baseline configurations. From left to right, the panels show flux spectra, closure phases, differential phases, and visibilities. Each baseline is colour-coded to fit the uv-plane shown in Figure~\ref{fig:uvplane}. 
    }
    \label{fig:model_vs_data}
\end{figure*}

An additional extended over-resolved flux contribution of $f_{\rm ext} = 0.041 \pm 0.005$ was also required. The Br$\alpha$ line centroid was measured at $\lambda_{0} = 4.0523~\mu$m, consistent with the expected rest wavelength within the instrumental resolution. The resulting parameters are summarised in Table~\ref{tab:pmoi_red}, the model image in Fig.~\ref{fig:pmoired_model_image}, and the modelled versus observed interferometric observables in Fig.~\ref{fig:model_vs_data}. 


\begin{table}
    \centering
    \caption{Best-fit parameters of the MATISSE interferometric data derived with \texttt{PMOIRED}.}
    \label{tab:pmoi_red}
    \begin{tabular}{lcc}
        \hline\hline
        Parameter & Value & Uncertainty \\
        \hline
        $f_{\rm env}$ (continuum) & 0.93 & $\pm$ 0.01 \\
        $f_{\rm env}$ (Br$\alpha$) & 0.023 & $\pm$ 0.009 \\
        ud size (Br$\alpha$) [mas] & 5.55 & $\pm$ 0.12 \\
        $f_{\rm Br\alpha}$ (line flux) & 1.10 & $\pm$ 0.02 \\
        Br$\alpha$ Gaussian width [nm] & 2.09 & $\pm$ 0.03 \\
        $\Delta x$ [mas] & 0.19 & $\pm$ 0.07 \\
        $\Delta y$ [mas] & 0.09 & $\pm$ 0.08 \\
        $f_{\rm ext}$ (extended flux) & 0.040 & $\pm$ 0.005 \\
        $\lambda_{0}$ (Br$\alpha$ centroid) [$\mu$m] & 4.0523 & (fixed) \\
        $\theta_{\rm UD}$ (stellar disc) [mas] & 2.42 & (fixed for the Br$\alpha$ fit) \\
        \hline
        Reduced $\chi^2$ & \multicolumn{2}{c}{1.03} \\
        \hline
    \end{tabular}
\end{table}

The PMOIRED modelling confirms and refines the results obtained from the uniform-disc fit, with a Br$\alpha$--emitting region of $\sim$5.6\,mas, in excellent agreement with the $\sim$5.6\,mas size derived from the simple model while accounting for a uniform disc brightness distribution. This consistency shows that the hydrogen line emission is systematically more extended than the continuum, while the parametric approach additionally constrains the flux ratio and reveals the presence of an over-resolved component not captured by the uniform-disc approximation.

The sampled channel maps at N equally spaced velocities spanning the maximum symmetric velocity interval permitted by the Br$\alpha$ wavelength window (4.050–4.055 $\mu$m) are shown in Figure~\ref{fig:bralpaha_grid}. The Br$\alpha$ channel maps reveal a pronounced asymmetry in velocity space, with significantly stronger emission on the blue-shifted side of the line. Emission remains detectable up to $v_{\mathrm{LSR}} \sim -120~\mathrm{km\,s^{-1}}$, whereas the red-shifted counterpart is strongly suppressed at comparable offsets from the systemic velocity. At all velocities where Br$\alpha$ emission is detected, the morphology remains compact and centrally peaked, arguing against ordered rotation or a collimated ejection and instead favouring a radially expanding flow. The observed blue--red asymmetry is naturally explained if emission from the receding hemisphere is partially attenuated by line optical depth and/or continuum opacity within the dense inner wind. Taking the maximum detected blue-shift as a lower limit implies projected outflow velocities of at least $\sim140~\mathrm{km\,s^{-1}}$ for the ionised gas, adopting a systemic velocity of $v_{\mathrm{sys}} = +18~\mathrm{km\,s^{-1}}$, as reported previously from optical Fe\,\textsc{ii} in \citet{Wallstrom2017}.

\section{DISCUSSION}
\label{discussion}

\subsection{Distance considerations on FEN}

\label{distance_sec}
The distance of the source has traditionally been a source of debate \citep[e.g.][]{Lagadec2011,Wallstrom2015}.
Here, we discuss our choice of adopting a distance of 1.2\,kpc \citep[DR2;][]{Koumpia2020}. This distance is also supported by a more recent independent distance constraint as provided by the systemic radial velocity of IRAS~17163 ($v_{\mathrm{rad}} \simeq +18\,\mathrm{km\,s^{-1}}$), measured consistently in optical Fe\,\textsc{ii} emission and CO lines \citep{Wallstrom2017}. Comparison with reconstructed three-dimensional Galactic velocity fields based on the HI4PI survey \citep{Soding2025} yields a kinematic distance consistent with $\sim$1.2\,kpc \citep{Kasikov2026,deWitTBS}.

\citet{Oudmaijer2022} used Gaia DR3 measurements and report a distance of $\sim$5.2\,kpc for the source. While the RUWE parameter does not flag IRAS 17163 as problematic, this does not guarantee reliable parallaxes for extended, dusty sources such as IRAS\,17163 due to centroid errors and photocenter shifts. We argue that using the DR3-based distance would introduce substantial inconsistencies in the known physics related to massive stellar evolution.

A strong argumentation against the 5.2\,kpc is the luminosity of the source. At a distance of 1.2\,kpc, \cite{Koumpia2020} derive a luminosity of $L = 5 \times 10^5\,L_\odot$, typical for a yellow hypergiant (YHG). However, if the star were placed at 5.2\,kpc, its luminosity would increase by a factor of $\left(5.2/1.2\right)^2 \approx 18.8$, resulting in $L \approx 9.4 \times 10^6\,L_\odot$. This would make IRAS\,17163 one of the most luminous stars in the Milky Way, comparable to the peak luminosity of $\eta$~Carinae during its Great Eruption ($1-5 \times 10^7\,L_\odot$). Such a luminosity would, in fact, place the star above the empirical Davidson limit ($L \sim (5$--$6)\times10^{5}\,L_\odot$), while exceeding the luminosities of known LBVs, WR stars, and even many supergiants near the Eddington limit. 

For a star to remain stable against radiative acceleration at such extreme luminosity, it must not exceed its Eddington limit,
\[
L_{\rm Edd} = 3.2 \times 10^4 \,(M/M_\odot)\,L_\odot .
\]
Imposing $L \leq L_{\rm Edd}$ for $9.4 \times 10^6\,L_\odot$ yields
$M \gtrsim 294\,M_\odot$, far above the mass of any known Galactic star.
Even if IRAS\,17163 were temporarily in a super-Eddington state (as
$\eta$~Car has been proposed to be), such high luminosity would demand
an extremely massive progenitor, likely exceeding $150\,M_\odot$
(assuming a super-Eddington factor, $\Gamma$\footnote{Defined as
$\Gamma = L / L_{\rm Edd} > 1$, where
$L_{\rm Edd} = 3.2 \times 10^4\,(M/M_\odot)\,L_\odot$; then the required
mass is
$M \gtrsim L / (\Gamma \cdot 3.2 \times 10^4\,L_\odot/M_\odot)$.},
of about 2), formed in a dense, young stellar cluster, yet no such
environment is evident in its vicinity.


Moving on with the argumentation, the 2D radiative transfer modelling in
\cite{Koumpia2020} yields mass-loss rates of
$\dot{M} \approx 6 \times 10^{-7}\,M_\odot\,{\rm yr}^{-1}$ (inner shell),
$9 \times 10^{-4}\,M_\odot\,{\rm yr}^{-1}$ (intermediate shell), and
$5 \times 10^{-5}\,M_\odot\,{\rm yr}^{-1}$ (outer shell), assuming
optically thin dust emission. These inferred mass-loss rates at
1.2\,kpc are moderate and consistent with values observed in yellow
hypergiants and evolved post-RSG stars\footnote{For \object{IRC+10420},
CO modelling suggests two main mass-loss episodes with
$\dot{M}\approx1.2\times10^{-4}$ and
$3\times10^{-4}\,M_\odot\,{\rm yr}^{-1}$ (durations
$\sim3.8\times10^{3}$ and $\sim8\times10^{2}$\,yr, respectively),
separated by a $\sim1.2\times10^{3}$\,yr low-$\dot{M}$ interval
\citep{Castro2007}.}

However, rescaling them to 5.2\,kpc results in unrealistically high
values, particularly for the intermediate shell, which would reach
$\dot{M} \approx 1.7 \times 10^{-2}\,M_\odot\,{\rm yr}^{-1}$, comparable
to eruptive LBV phases. These values are listed in
Table~\ref{tab:mdot_scaling}. Additionally, such an amount of mass would
obscure the star even more. Since the extinction is mainly interstellar
rather than originating from the star itself \citep{Koumpia2020}, this
scenario seems even more unlikely.

The measured angular extent of the Br$\alpha$ emission provides an additional distance-dependent constraint on the ionised wind scale. For a characteristic size of $\theta \simeq 5.6$\,mas, the corresponding physical radius is $r \simeq \theta d \approx 6.7$\,au at $d=1.2$\,kpc, but $r \approx 29.1$\,au at $d=5.2$\,kpc. In a steady, approximately spherical wind with $\rho \propto r^{-2}$ and constant outflow speed $v$, the electron density scales as $n_e(r) \propto \dot{M}/(v r^2)$ and the hydrogen recombination emissivity as $j_{\mathrm{Br}\alpha} \propto \dot{M}^2/(v^2 r^4)$. While a detailed inference of $\dot{M}$ from the Br$\alpha$ size would require dedicated radiative-transfer modelling, the tens of au scale implied at $d\gtrsim 5$\,kpc would push the required ionised wind parameters toward a significantly more extreme regime, reinforcing the tension with the larger distance. 

\begin{table}[ht]
\caption{Shell properties derived at 1.2\,kpc and rescaled to 5.2\,kpc.}
\centering
\begin{tabular}{lcc}
\hline\hline
Shell & $\dot{M}$ at 1.2\,kpc ($M_\odot\,\text{yr}^{-1}$) & $\dot{M}$ at 5.2\,kpc ($M_\odot\,\text{yr}^{-1}$) \\
\hline
Hot inner shell      & $6 \times 10^{-7}$    & $1.13 \times 10^{-5}$ \\
Intermediate shell   & $9 \times 10^{-4}$    & $1.69 \times 10^{-2}$ \\
Outer shell          & $5 \times 10^{-5}$    & $9.38 \times 10^{-4}$ \\
\hline
\end{tabular}
\label{tab:mdot_scaling}
\end{table}


In conclusion, the 1.2\,kpc distance places IRAS\,17163 in the upper-luminosity range of YHGs and evolved massive stars (see, e.g., IRC+10420, $\rho$ Cas, HR\,8752), consistent with its variability, moderate spectroscopic wind signatures, and circumstellar shell properties. It avoids invoking a physically implausible progenitor mass and prevents violating the Eddington limit, while remaining consistent with the observed optical and infrared SED, and mass-loss properties. 


\subsection{Comparison of M-band spectral features in IRAS\,17163 and IRC+10420}
\label{sec:comparison_Mband}

The acquisition of the M-band spectrum of IRAS 17163, utilising high-resolution CRIRES+ spectroscopy, allows for a direct and detailed comparison with the well-studied post-RSG YHG, IRC+10420. This comparative analysis reveals fundamental differences in the kinematic, excitation, and density properties of their innermost circumstellar environments, tracing potential evolutionary divergence between the two objects.

\subsubsection{Contrasting hydrogen recombination and metal lines}

Hydrogen recombination lines, specifically \text{Br}$\alpha$, \text{Pf}$\beta$, \text{Pf}$\gamma$, and various Humphreys series transitions, are clearly detected in both stars, but display significant variations in strength and profile. In IRAS 17163, these lines exhibit strong emission with broad wings and asymmetric profiles. Notably, Pf\,$\beta$ and Pf\,$\gamma$ exhibit broad wings and P~Cygni profiles, indicative of an expanding and sufficiently dense inner wind. Such profiles require significant optical depth in the line-forming region and therefore trace energetic, high-density outflows rather than merely the presence of expansion. These lines are also present in \object{IRC+10420}. The Br\,$\alpha$ line shows strong emission in both stars, with a red-shifted peak in \object{IRC+10420} consistent with its higher systemic velocity relative to IRAS\,17163. Several Humphreys transitions (Hu~$\delta$, Hu~$\epsilon$, Hu~$\theta$), and higher excitation lines (e.g., H\,20--7), are present in \object{IRAS\,17163}. These lines are either absent or significantly weaker in \object{IRC+10420}.

A further distinction lies in the atomic features. IRAS 17163 displays prominent neutral and low-ionisation metal lines in emission or absorption, including Fe I, Fe II, Mg I, and C I. These transitions are permitted lines rather than forbidden ones, indicating formation in a relatively dense inner region where collisional de-excitation suppresses forbidden emission. This differs from the situation in B[e]-type objects, where strong forbidden metal lines trace lower-density extended outflows; the spectrum of IRAS~17163 instead points to a dense, inner wind rather than shock-excited or diffuse circumstellar gas. Conversely, these metal lines are largely absent or significantly weaker in IRC+10420, a signature hinting to a more evacuated inner envelope. This behaviour is fully consistent with the optical comparison presented by \citet{Wallstrom2015}, who directly overplotted the optical spectra of the two objects and found IRAS~17163 to be considerably richer in Fe\,\textsc{ii}, Cr\,\textsc{ii}, and Ca\,\textsc{ii} features than IRC+10420 (their Fig.~B.1).

\subsubsection{Molecular features and wind kinematics}

The $\text{CO}$ first overtone transitions ($v=1 \to 0$), particularly in the R- and P-branches, highlight a stark contrast in the molecular content and kinematics of the two objects (Fig.~\ref{fig:pfbeta_comparison}, Fig.~\ref{fig:crires_overview}). IRC+10420 exhibits clear P Cygni profiles across multiple $\text{CO}$ transitions, characterised by well-defined blue-shifted absorption. This profile is consistent with classical P Cygni line shapes and demonstrates the presence of an outflowing molecular envelope. The detection of $\text{CO}$ in IRC+10420 is indicative of molecular gas at moderate temperatures and may trace a cooler, more extended layer produced during a previous episode of enhanced mass loss. In this interpretation, hydrogen recombination lines would arise predominantly in the denser inner wind, while the CO absorption probes material located farther out along the line of sight, allowing the molecular transitions to display pronounced P~Cygni profiles even if the inner ionised gas does not. The P~Cygni morphology of the CO lines in IRC+10420 also enables an estimate of the molecular outflow speed (e.g.\ from the maximum blue-shift of the absorption trough relative to the systemic velocity). A dedicated analysis of the CO velocity structure and excitation of IRC+10420, including radiative-transfer modelling across multiple transitions, is deferred to future work. 

In contrast, IRAS~17163 shows little to no absorption in the CO rovibrational lines. 
The absence of these hot CO features suggests that molecules are either dissociated in the immediate inner environment or that the molecular gas has been displaced to larger radii. Indeed, millimetre observations detect CO in rotational transitions (e.g.\ $J=2$--1 and $J=3$--2) arising from circumstellar material on scales of $\lesssim20''$, with broad and asymmetric profiles indicative of a structured outflow \citep{Wallstrom2015,Wallstrom2017}. This demonstrates that molecular gas is present, but predominantly in a cooler, more extended component that is not traced by the near-infrared rovibrational lines. As context, IRC+10420 is also known to host an extended molecular envelope traced by CO rotational emission at millimetre wavelengths \citep[e.g.][]{Castro2007,Quintana2016}; the contrast discussed here therefore pertains specifically to the hot, inner regions probed by the M-band rovibrational lines.

The detection of several CO rovibrational transitions toward IRC+10420 naturally raises the question of whether an excitation analysis could be used to constrain the physical conditions of the molecular gas.
In principle, rotational diagrams constructed from the relative strengths of individual CO lines can provide estimates of excitation temperatures and column densities. However, the CO lines detected toward IRC+10420 show complex profiles, including absorption and P Cygni-like signatures, indicating that the gas is part of a dynamic, expanding circumstellar environment rather than a
static slab. Under these conditions, line strengths are affected by
optical depth effects, velocity gradients, and radiative transfer
through the wind, making a simple rotational-diagram analysis
highly model-dependent. Similarly, the absence of statistically
significant CO detections toward IRAS 17163 precludes a meaningful excitation analysis for that source. For these reasons, we do not attempt to derive excitation temperatures from the CO rovibrational ladder in this work.
A dedicated excitation and radiative-transfer analysis, incorporating wind geometry and optical depth effects, is deferred to a future study focused specifically on the molecular component of post-RSG mass loss.


\subsubsection{Implications for evolutionary status}

The observed spectral differences act as a crucial diagnostic tool for probing evolutionary divergence among the two post-RSG YHGs. To summarise, IRAS 17163 shows strong hydrogen and metal emission and the absence of clear molecular outflow signatures. The kinematics indicated by the strong metal lines suggest slower, more quiescent ejecta, consistent with the interpretation that IRAS 17163 likely underwent episodic, fragmented mass-loss events. Conversely, IRC+10420, shows prominent molecular absorption, weaker metal lines, and emission-dominated, yet weaker hydrogen profiles. 

This comparison reveals clear differences in the kinematic and excitation properties of the two yellow hypergiants, but also highlights that their interpretation is not straightforward. In particular, the emission components of the hydrogen recombination lines are systematically narrower in IRAS~17163 than in IRC+10420. For Br$\alpha$, IRAS~17163 exhibits a FWHM of $\sim$45~km\,s$^{-1}$, compared to $\sim$70~km\,s$^{-1}$ in IRC+10420 (Table~\ref{crires_comparison}). Such differences in line width alone can have a strong impact on line opacity: broader velocity fields reduce the optical depth per velocity channel and can flatten line profiles even for comparable column densities. As a result, differences in equivalent width or peak intensity cannot be interpreted uniquely in terms of gas temperature or spatial extent.

In addition, the observed line strengths depend on the gas filling factor relative to the stellar and dust continuum. If the emitting material is distributed in a clumpy or fragmented medium rather than in a smooth wind, optically thick clumps covering only a fraction of the projected area can still produce strong emission while significantly altering the effective opacity. These effects are expected in both pulsation- and wind-driven outflows and introduce further degeneracies between kinematics, density, and geometry. Consequently, while the stronger metal emission and prominent hydrogen lines in IRAS~17163 are consistent with a denser and more ionised inner wind, and the molecular absorption in IRC+10420 points to a cooler, more extended molecular component, the contrasting spectral properties likely reflect a combination of velocity structure, optical-depth effects, and gas filling factors, in addition to intrinsic differences in temperature or size.

The contrasting presence of CO absorption in \object{IRC+10420} and its
absence in \object{IRAS~17163} further suggests that molecular survival
is governed not only by temperature, but also by local shielding and
geometry.
The persistence of CO features in \object{IRC+10420} over several
decades indicates that the molecular gas may reside at relatively small
radii in dense, dust-shielded regions, potentially associated with a
flattened or torus-like structure that is still participating in an
outflow.
In contrast, the strongly ionised inner wind of \object{IRAS~17163}
appears hostile to molecule formation or survival despite evidence for
recent mass loss.

In the context of line-driven winds, relatively small differences in effective temperature are expected to play a first-order role, as they can induce significant opacity changes through ionisation balance shifts, particularly near the second bi-stability jump. Both sources are characterised by effective temperatures close to the temperature known to trigger this instability (8800~K). The overall contrast in line profiles and species, therefore, hint differences in wind properties that are likely rooted in modest shifts in effective temperature and associated opacity changes, with variations in geometry and time-variable ejection behaviour emerging as secondary consequences. This diversity highlights the sensitivity of the mass-loss process in post-RSG yellow hypergiants to subtle differences in stellar parameters, when stars evolve close to the Eddington limit. Small opacity changes can lead to markedly different wind properties.

The observed spectroscopic differences suggest that YHGs may follow diverse evolutionary pathways, influenced by initial mass, binarity, or instabilities, rather than evolving along a uniform sequence. Our results, therefore, emphasise the importance of comparative studies across multiple post-RSGs to capture the full range of mass-loss behaviours at this transitional stage.



\subsection{From ionised winds to dusty shells: Mass-loss across multiple spatial scales}

The combination of high-resolution spectroscopy, long-baseline interferometry, recently published mid-infrared imaging, and optical spectropolarimetry provides a unique, multi-scale view of the circumstellar environment of IRAS 17163. Together, these results demonstrate that the mass-loss process in this post-RSG star is neither smooth nor steady, but instead episodic, structured, and time-variable.

\subsubsection{The innermost wind region}

The CRIRES+ spectra reveal strong and broad hydrogen recombination lines with asymmetric and P-Cygni–like profiles, consistent with the presence of a dense, accelerating outflow. The absence of significant CO absorption, in contrast to IRC+10420, indicates a hotter or more ionised inner environment where molecules are destroyed or absent. The emission in metallic lines such as Fe I, Fe II and C I further supports the presence of dense, excited gas close to the star.

These spectroscopic diagnostics are directly complemented by the MATISSE interferometry, which spatially resolves the Br$\alpha$ line-emitting region. The Br$\alpha$ emission is found to be approximately twice as extended as the L-band continuum, demonstrating that the ionised wind is a distinct component from the compact continuum environment, which is more likely a result of photospheric emission and warm dust emission. The line variability on monthly timescales implies a time-variable mass-loss rate. Together, CRIRES+ and MATISSE show that the current wind of IRAS 17163 is unstable, shaping the innermost circumstellar environment on au scales. 

The Br$\alpha$ channel maps do not show strong velocity-dependent asymmetries that would point to a strongly flattened or rotating structure. This behaviour contrasts with IRC+10420, where interferometric observations at similar spatial scales reveal a clearly asymmetric structure interpreted as an hour-glass or bipolar wind seen close to pole-on \citep[][]{Koumpia2022}. At the spatial scales probed here, the inner wind of IRAS 17163 does not clearly show, but neither can it exclude equatorial structures or bipolar outflows seen near pole-on.

Additional constraints on the geometry of the innermost wind are provided
by the FORS2 spectropolarimetric observations.
Line polarisation effects are detected across H$\alpha$ and are
tentatively present across the Ca\,\textsc{ii} lines, indicating
departures from spherical symmetry on spatial scales well below the
direct imaging resolution. Spectropolarimetric signatures are commonly associated with
asymmetric structures close to the star, including flattened winds,
disc-like geometries, or localised density enhancements.
They may also arise from preferentially directed ejections or compact
clumps propagating within the inner wind, as proposed for luminous blue
variables \citep{Davies2005}.
In this context, the spectropolarimetric results provide an independent
line of evidence that the innermost wind of IRAS~17163 is not strictly
spherically symmetric, complementing the picture inferred from
interferometry and spectroscopy. While these signatures suggest departures from strict spherical symmetry, the detected line polarisation effects are modest. This may indicate that the inner wind of IRAS 17163 was in a relatively quiescent phase during the epoch of our observations, despite the episodic mass-loss history inferred at larger spatial scales.

\subsubsection{Large-scale asymmetry from spectropolarimetry}

The discrepancy between the 1988 polarisation measurements and the stable values seen in 2015 and 2023 may hint at a transient intrinsic contribution. If dust scattering requires material sufficiently close to the star, the intrinsic polarisation detected in 1988 could plausibly be associated with enhanced circumstellar dust during a recent mass-loss episode. In this scenario, the fading of the intrinsic component would be consistent with the dust expanding to larger radii, leaving behind only the interstellar contribution. 

Recent mid-infrared imaging of IRAS 17163 by \citep{deWitTBS} provides an independent and highly complementary view of the large-scale circumstellar geometry inferred from our spectropolarimetry. Their VISIR data reveal that the Fried Egg Nebula is expanding at a measurable angular rate, with shell-specific dynamical ages ranging from $\sim$70 to $\sim$ 400 yr (at 1.2 kpc) for the inner mid-IR rings and up to several thousand years for the outer structures. Importantly, although the shells appear circular in projection, the high-quality PAH1 images show localised brightness enhancements and Kelvin–Helmholtz–type ripples along the inner rims, indicating that the mass-loss is not perfectly isotropic. This morphology is consistent with the possibility, supported by historical polarimetric variability, that IRAS 17163 has exhibited intrinsic asymmetry during past mass-loss episodes, even though the current continuum polarisation is dominated by the interstellar component.

This interpretation is also consistent with the large-scale asymmetries detected at millimetre wavelengths with ACA \citep{Wallstrom2017}, where the CO($2$--$1$) emission revealed a red-shifted "spur" extending from the star toward the south-east (PA $\simeq 112^{\circ}$), suggestive of a unidirectional ejection event. Together, the mid-IR shell inhomogeneities and the ACA spur indicate that departures from spherical symmetry occur on multiple spatial scales in the circumstellar environment of \object{IRAS~17163}.

The VISIR results strengthen the idea that IRAS 17163 undergoes episodic, directionally modulated mass-loss, rather than smooth spherical outflows. The fact that the spectropolarimetric angle remains stable across recombination lines, while the mid-IR images show shell-specific asymmetries, suggests that the preferred axis of mass ejection is long-lived, potentially linked to a persistent large-scale structure in the atmosphere or wind-launching region. The apparent geometric symmetry of the present-day wind therefore does not contradict the asymmetric large-scale shell structure, which reflects older, episodic mass-loss events preserved at larger radii.


\subsubsection{Connecting small and large scales}

By combining these diagnostics, we obtain a coherent multi-scale view of the circumstellar environment of IRAS~17163. The dense, largely symmetric yet time-variable ionised wind revealed by CRIRES+ and MATISSE provides the immediate mass reservoir that ultimately forms the dusty shells now resolved in the mid-infrared. The most recent VISIR imaging demonstrates that these shells expand at a measurable rate and possess dynamical ages of only a few hundred years for the inner structures. This indicates that the present-day wind, although geometrically compact, is fully capable of feeding the larger-scale dusty nebula on short evolutionary timescales.

Despite their near-circular appearance in projection, the VISIR images show localised brightness enhancements and ripple-like substructures along the inner rims of the shells, pointing to inherently inhomogeneous mass ejection. While our FORS2 observations show that the continuum polarisation is dominated by interstellar dust, they do reveal a line effect across H$\alpha$, indicating departures from spherical symmetry in the innermost wind. This small-scale asymmetry does not contradict the larger-scale structures inferred at mid-infrared wavelengths; rather, it suggests that the continuum-forming region is close to spherical on average, with localised or transient deviations that are insufficient to produce a strong intrinsic continuum polarisation.

The model-independent confirmation of a very compact inner emitting region (Sec.~\ref{reconstruction}) complements previous imaging results \citep{Koumpia2020,Lagadec2011}, which reveal multiple extended shells tracing episodic mass-loss over timescales ranging from decades to several millennia. Together, these findings show that the present-day mass loss probed by MATISSE arises from a dense inner wind embedded within a much more extended and multi-episodic circumstellar environment.

At this point, it is important to distinguish between temporal variability and geometric asymmetry, which are, in principle, independent properties of mass loss. A wind may vary in strength over time while remaining largely symmetric, just as a geometrically asymmetric outflow does not necessarily imply strongly time-variable or episodic behaviour. In the case of IRAS~17163, the present data indicate a compact, relatively symmetric ionised wind that exhibits variability in strength, alongside a larger-scale dusty environment that records discrete mass-loss episodes and localised asymmetries. The observed shell structure therefore reflects the time-integrated imprint of mass-loss variability, rather than requiring the instantaneous wind to be strongly asymmetric.

Several physical mechanisms have been proposed to account for such episodic mass loss in yellow hypergiants. In particular, both pulsation-driven instabilities and opacity-driven variations associated with the second bi-stability jump have been discussed in detail for IRAS~17163 in \citet{Koumpia2020}. Pulsation models predict cycles of atmospheric inflation, shocks, and partial envelope ejection as stars approach the cool edge of the yellow-void instability region \citep[e.g.][]{Glatzel2024}, while bi-stability effects in line-driven winds can lead to abrupt changes in wind density and velocity as the star evolves through a narrow range of effective temperatures \citep[e.g.][]{Vink1999,Petrov2016}. 

Either mechanism is capable, in principle, of producing recurrent mass-loss episodes and concentric shell structures. In pulsation-driven scenarios, such episodes can lead to discrete ejection events associated with atmospheric inflation and shock propagation, while both pulsation- and wind-driven outflows may develop stochastic density enhancements or modest latitude-dependent variations. In the case of line-driven winds, intrinsic clumping and, if the star is rotating, latitudinal variations in effective temperature and radiative acceleration can likewise imprint asymmetries on the ejecta. These effects may become observationally apparent only once the material expands and cools, forming the fragmented and locally enhanced structures seen at larger radii.

The new observational constraints presented here provide additional leverage on this picture. The combination of a compact, time-variable ionised wind traced by Br$\alpha$, moderate present-day wind velocities, and largely optically thin dusty shells point toward relatively mild, recurrent enhancements in mass loss operating over months to more eruptive events over decades to centuries timescales. In this context, perhaps a combined mechanism of pulsation-driven instabilities, potentially modulated by changes in wind opacity near the second bi-stability jump, offers a natural explanation for the observed shell sequence and its asymmetric substructure. Taken together, the interferometric, spectroscopic, polarimetric, and mid-infrared diagnostics indicate that IRAS~17163 undergoes mass loss that is neither steady nor strictly spherically symmetric, but governed by instabilities operating as the star evolves through the dynamically unstable region bordering the yellow void.

\subsubsection{In search of a companion}

The centroid of the Br$\alpha$ line remains stable over all available epochs, spanning more than 300 days (Fig.~\ref{fig:bralpha_profiles}).
While the line profile and peak intensity vary significantly, no systematic shift of the line centroid is detected.
From the scatter of the peak positions we derive a conservative upper limit of $\Delta v \lesssim 10$~km~s$^{-1}$ (Table~\ref{tab:bralpha_gauss_fits}), corresponding to a radial-velocity semi-amplitude $K_1 \lesssim 5$~km~s$^{-1}$.

Any orbital reflex motion of the primary would induce a radial-velocity semi-amplitude
\begin{equation}
K_1 = \left(\frac{2\pi G}{P}\right)^{1/3}
      \frac{M_2 \sin i}{(M_1 + M_2)^{2/3}}
      \frac{1}{\sqrt{1 - e^2}},
\end{equation}
where $M_1$ and $M_2$ are the masses of the primary and companion, $P$ is the orbital period, $i$ the inclination, and $e$ the eccentricity.
The orbital period corresponding to a given separation is obtained from Kepler’s third law,
\begin{equation}
P^2 = \frac{4\pi^2 a^3}{G(M_1+M_2)}.
\end{equation}
For $M_2 \ll M_1$ and $M_1 = 30\,M_\odot$, this yields
$P \simeq 0.18\,(a/\mathrm{au})^{3/2}$~yr, so that separations of
1, 3, and 10~au correspond to orbital periods of approximately
0.2, 1, and 5~yr, respectively.

Interpreting the observed limit $K_1 \lesssim 5$~km~s$^{-1}$ as an upper bound on orbital motion and adopting circular orbits and $\sin i = 1$, we constrain the mass of a putative companion as a function of separation.
Under these assumptions, companions more massive than $\sim1\,M_\odot$ are excluded at separations $\lesssim1$--2~au (orbital periods $\lesssim$ a few months), while companions more massive than $\sim2$--3\,$M_\odot$ are excluded out to separations of $\sim10$~au (orbital periods of several years).
For comparison, a $5\,M_\odot$ companion would induce a radial-velocity semi-amplitude of $K_1 \simeq 20$~km~s$^{-1}$ at $\sim1$~au ($P\sim100$~d) and $K_1 \simeq 14$~km~s$^{-1}$ at $\sim3$~au ($P\sim1$~yr), well above our detection threshold. We note that Br$\alpha$ is formed in the ionised wind rather than in the stellar photosphere. As a result, any orbital motion of the star would be further diluted by wind dynamics, making the above limits conservative.

Complementary constraints on binarity are provided by the VLTI/MATISSE interferometric observations.
We performed a systematic search for faint companions using the detection--limit formalism implemented in \textsc{PMOIRED}, which injects a secondary point source at random positions in the sky plane and determines the minimum detectable flux ratio at a given significance level based on the interferometric observables (visibility amplitudes and closure phases).
The search was restricted to angular separations between $\simeq3.2$~mas, corresponding to half the interferometric resolution ($0.5\,\lambda/B$ at 4~$\mu$m, suggested by the documentation, and a maximum baseline of 130~m), and 450~mas, set by the diffraction limit of the ATs (the effective field of view and sensitivity of the MATISSE observations is actually smaller by a factor of about 2). 

No statistically significant companion was detected at the $3\sigma$ level within this separation range.
The resulting detection limits exclude companions brighter than $\Delta m \simeq 3.3$~mag in the L band, corresponding to a flux ratio of $f_2/f_1 = 10^{-0.4\,\Delta m}$, where $\Delta m = m_2 - m_1$ is the magnitude difference between the companion and the primary in the relevant band.
For our $3\sigma$ detection limit of $\Delta m \simeq 3.3$~mag, this corresponds to $f_2/f_1 \simeq 5\%$.
We note that converting an L-band flux ratio constraint into a limit on companion mass or spectral type is non-trivial for evolved massive stars: the primary may exhibit strong infrared excess from hot dust and/or a free-free emitting ionised wind, so the measured contrast refers to the total interferometric continuum rather than the stellar photosphere alone.
Any inference on companion mass, therefore, depends on assumptions about the circumstellar contribution, extinction, and the companion's intrinsic spectral energy distribution.

The quoted detection limit corresponds to the median of the distribution of $3\sigma$ detection thresholds obtained from the random companion injections. The associated confidence intervals indicate that 68\% (90\%) of the trials yield limits between $\Delta m \simeq 3.18$--3.41~mag (3.09--3.48~mag), with the full 99\% range spanning $\Delta m \simeq 2.97$--3.54~mag. An independent grid-based binary model fit converged to a formal $\chi^2$ of 0.9 minimum corresponding to a faint companion with a flux ratio of $f_2/f_1 \simeq 0.015$ at a projected separation of $\sim33$~mas. This flux ratio corresponds to a magnitude difference of $\Delta m \simeq 4.6$~mag, which lies well below the previously determined $3\sigma$ detection limit of $\Delta m \simeq 3.3$~mag (i.e. $f_2/f_1 \simeq 0.05$). As a result, such a signal cannot be robustly distinguished from statistical fluctuations in the interferometric observables, and we therefore do not consider this solution as a real detection.

The interferometric limits probe projected separations of several to a few hundred astronomical units ($\sim$4-500~au) at the distance of the source (1.2~kpc) and are therefore complementary to the radial velocity constraints derived from the Br$\alpha$ spectroscopy, which are primarily sensitive to even closer-in companions on sub-au to few-au orbits. Taken together, the absence of detectable radial velocity shifts and the lack of interferometric evidence for a companion over a wide range of separations seem to disfavour binarity as the origin of the observed variability. Instead, the combined spectroscopic and interferometric constraints support a scenario in which the observed line-profile variability and spatial asymmetries arise from intrinsic, time-variable structures in the ionised wind.


\section{CONCLUSIONS}

\subsection{Inner-wind diagnostics from M-band spectroscopy}

The CRIRES+ data deliver the first M-band spectrum of IRAS~17163, offering a new probe of its innermost circumstellar environment. 
In direct comparison with the well-studied post-RSG IRC+10420, the spectra reveal clear differences that trace distinct physical conditions and possibly evolutionary stages, which can be summarised as follows:

\begin{itemize}
    
    \item IRC+10420 shows the absence of metal lines in the observed wavelength regimes, 
    suggesting a hotter circumstellar environment, likely due to the clearing of outer layers as the star evolves toward a blue supergiant phase. IRC+10420’s mass loss is dominated by a molecular outflow, but it does not currently host a dense, compact, partially ionised inner wind that produces strong Fe I/Fe II/Mg I/C I emission.
    
    \item In contrast, IRAS 17163 displays prominent metal transitions, including Fe I, Fe II, Mg I, and C I, consistent with a cooler, dense circumstellar envelope typical of a yellow hypergiant or late red supergiant phase.

    \item IRAS 17163 appears to be in a transitional regime in which the gas is still sufficiently cool for neutral species such as Fe I, Mg I, and C I to survive, while at the same time being ionised enough to produce Fe II emission and strong H I recombination lines. The absence of CO indicates that molecular material has either been destroyed in the inner regions or is confined to much larger radii, consistent with the detached shells seen at longer wavelengths. These diagnostics point to a dense, partially ionised, and effectively shielded inner wind.

    \item IRC +10420 is dominated by a molecular outflow in which CO survives and exhibits P-Cygni profiles. The inner regions appear to contain a smaller fraction of partially ionised gas, as evidenced by the lack of strong Fe I, Fe II, Mg I, or C I emission. The circumstellar environment points to a less dense and ionised inner wind compared to IRAS 17163.
    
    \item Hydrogen line profiles, such as Br$\alpha$, Pf$\beta$, and Hu series lines, are broader and stronger in IRAS 17163, indicating higher density gas in a more quiescent environment compared to the weaker hydrogen lines in IRC+10420.

    \item The spectroscopic divergence in the M-band between the two sources reflects differences in mass-loss behaviour and physical conditions of their circumstellar environment. 


    

    
\end{itemize}

\subsection{Spatial structure from L-band interferometry}

The VLTI/MATISSE observations provide the spatially resolved view of the ionised environment of IRAS~17163 in the L band. 
Our interferometric analysis constrains both the geometry and variability of the Br$\alpha$--emitting region relative to the compact continuum emission (photospheric and warm dust contributions), yielding the following conclusions:

\begin{itemize}
    \item The VLTI/MATISSE observations resolve the Br$\alpha$--emitting region of IRAS~17163, revealing it to be significantly more extended than the continuum emission. 
    \item Geometric modelling with \texttt{PMOIRED} constrains the Br$\alpha$--emitting region to a uniform-disc size of $\sim$5.6\,mas ($\approx$7\,au at 1.2\,kpc), with a small but significant contribution ($\sim$4\%) from over-resolved flux.     
    \item The observations reveal spectral variability in the Br$\alpha$ line emission. In particular, the observed profiles show broad ($\sim$ 130 km\,s$^{-1}$) emission with stable centroids but variable peak fluxes (up to 30\% over months), indicating intrinsic variability in the ionised wind on a timescale of months. 
    \item No significant asymmetries or diversions from the photocenter were derived from the modelling process. Weak differential phase signals in a set of baselines may hint towards a non-centrosymmetric Br$\alpha$ emission, but the quality of the data does not allow a robust conclusion. The modelling could not reproduce those features. If confirmed, they could be suggestive of a structured wind or disc-like geometry rather than a spherically symmetric outflow. 
    \item Velocity-resolved Br$\alpha$ channel maps show a blue–red asymmetry and a compact morphology across all velocities, consistent with a radially expanding ionised wind. The maximum blue-shift implies projected outflow velocities of at least $\sim$140~km/s.
    \item No spatial variability of either the continuum or the Br$\alpha$--emitting region is detected within the timeframe of the observations at the given spatial resolution.
    \item The long-term stability of the Br$\alpha$ centroid over $\sim$300 days, combined with the absence of robust continuum photocenter shifts, strongly disfavours binarity. 
    \item The interferometric spatial analysis yields no evidence for a binary companion within the sensitivity and separation range probed by the MATISSE observations, excluding companions brighter than $\sim5\%$ of the primary flux over separations from a few to several hundred au.
    \item The reconstructed L-band continuum resolves a compact $\sim$2~mas emission region, in excellent agreement with the parametric models. This inner structure likely traces the current mass-loss phase that ultimately feeds the far more extended episodic shells previously detected with VISIR, linking the sub-au wind morphology to the long-term mass-loss history of the Fried Egg Nebula.

\end{itemize}

\subsection{Large-scale geometry from optical spectropolarimetry}

The optical spectropolarimetric observations provide constraints on the large-scale geometry of the circumstellar environment and allow us to distinguish between interstellar and intrinsic sources of polarisation. In particular, they probe whether the present-day wind exhibits global departures from spherical symmetry and whether recent mass-loss episodes have imprinted detectable polarimetric signatures.

\begin{itemize}
  
  \item The spectropolarimetric data of IRAS~17163 are well described by a Serkowski law with $\lambda_{\max} = 5492 \pm 77$ \AA, $P_{\max} = 11.33 \pm 0.08$\%, and $K = 1.81 \pm 0.14$.

  \item The wavelength of peak polarisation, $\lambda_{\max} = 5492 \pm 77$ \AA, is consistent with normal Milky Way dust with total-to-selective extinction ratio R$_V$ $\sim$ 3.1.

  \item The relatively high maximum polarisation ($P_{\max} \simeq 11.3\%$) lies below the empirical upper limit of interstellar polarisation and reddening, supporting an origin in the Galactic interstellar medium.

  \item The constant wavelength-independent polarisation angle of $25.8^\circ$ indicates dust aligned with a single magnetic field orientation along the line of sight.

\item The contrast between the stable interstellar polarisation measured in 2015 and 2023 and the discrepant 1988 data suggests that IRAS 17163 may have undergone a more recent mass-loss episode that temporarily produced intrinsic polarisation. This possibility reinforces our conclusion that the star’s circumstellar environment is shaped by episodic and potentially asymmetric ejections.

\item The detection of an H$\alpha$ line polarisation effect with FORS2,
consistent with earlier high-resolution measurements, provides
additional evidence for small-scale asymmetries in the innermost wind
of \object{IRAS~17163}, complementing the interferometric and
spectroscopic diagnostics.

\end{itemize}

These results collectively emphasise that the transition from the red supergiant to hotter phases is accompanied by highly structured and time-variable mass loss. The presence of multiple dusty shells, together with the currently variable ionised wind, implies that IRAS~17163 is losing mass in a sequence of episodic events rather than through a continuous, spherically symmetric flow. The absence of observational signatures of a companion in our interferometric and spectroscopic data disfavours binarity as the origin of this behaviour, indicating that the structured and time-variable mass loss is primarily driven by intrinsic stellar instabilities. The observed diversity in spectroscopic diagnostics and wind properties among YHGs cautions against interpreting any individual object as representative of the class and highlights the need for coordinated spectroscopic and interferometric monitoring of multiple sources to map the range of instability-driven mass-loss behaviours that shape the final pre-supernova evolution of massive stars. Such behaviour may critically influence the angular momentum evolution of the star and the circumstellar conditions into which a future supernova will explode.



\section*{Acknowledgements}

We thank the anonymous referee for their time and effort in evaluating this manuscript. We thank Andrea Mehner and Anni Kasikov for fruitful discussions. Based on observations collected at the European Southern Observatory under ESO programmes 109.22VF.001, 109.22VF.002, 109.22VF.003, 109.22VF.005 (MATISSE), 111.24P0.004 (FORS2), and 107.22TZ.001 (CRIRES+). This research has made use of NASA’s Astrophysics Data System (ADS).
%



\bibliographystyle{aa} 
\bibliography{example}

@ARTICLE{Oudmaijer1999,
       author = {{Oudmaijer}, Rene D. and {Drew}, Janet E.},
        title = "{H{\ensuremath{\alpha}} spectropolarimetry of B[e] and Herbig Be stars}",
      journal = {\mnras},
         year = 1999,
        month = may,
       volume = {305},
       number = {1},
        pages = {166-180},
          doi = {10.1046/j.1365-8711.1999.02383.x},
archivePrefix = {arXiv},
       eprint = {astro-ph/9901032},
 primaryClass = {astro-ph},
       adsurl = {https://ui.adsabs.harvard.edu/abs/1999MNRAS.305..166O}
}

@ARTICLE{1997ApJ...487..314M,
       author = {{Messinger}, D.~W. and {Whittet}, D.~C.~B. and {Roberge}, W.~G.},
        title = "{Interstellar Polarization in the Taurus Dark Clouds: Wavelength-dependent Position Angles and Cloud Structure near TMC-1}",
      journal = {\apj},
         year = 1997,
        month = sep,
       volume = {487},
       number = {1},
        pages = {314-319},
          doi = {10.1086/304610},
archivePrefix = {arXiv},
       eprint = {astro-ph/9706032},
 primaryClass = {astro-ph},
       adsurl = {https://ui.adsabs.harvard.edu/abs/1997ApJ...487..314M}
}

@ARTICLE{1997ApJ...476L..27W,
       author = {{Wang}, Lifan and {Wheeler}, J. Craig and {H{\"o}flich}, Peter},
        title = "{Polarimetry of the Type IA Supernova SN 1996X}",
      journal = {\apjl},
         year = 1997,
        month = feb,
       volume = {476},
       number = {1},
        pages = {L27-L30},
          doi = {10.1086/310495},
archivePrefix = {arXiv},
       eprint = {astro-ph/9609178},
 primaryClass = {astro-ph},
       adsurl = {https://ui.adsabs.harvard.edu/abs/1997ApJ...476L..27W}
}

@ARTICLE{Ababakr2016,
       author = {{Ababakr}, K.~M. and {Oudmaijer}, R.~D. and {Vink}, J.~S.},
        title = "{Linear spectropolarimetry across the optical spectrum of Herbig Ae/Be stars}",
      journal = {\mnras},
         year = 2016,
        month = sep,
       volume = {461},
       number = {3},
        pages = {3089-3110},
          doi = {10.1093/mnras/stw1534},
archivePrefix = {arXiv},
       eprint = {1607.02440},
 primaryClass = {astro-ph.SR},
       adsurl = {https://ui.adsabs.harvard.edu/abs/2016MNRAS.461.3089A}
}

@ARTICLE{2019Gordon,
       author = {{Gordon}, Michael S. and {Humphreys}, Roberta M.},
        title = "{Red Supergiants, Yellow Hypergiants, and Post-RSG Evolution}",
      journal = {Galaxies},
         year = 2019,
        month = dec,
       volume = {7},
       number = {4},
          eid = {92},
        pages = {92},
          doi = {10.3390/galaxies7040092},
archivePrefix = {arXiv},
       eprint = {2009.05153},
 primaryClass = {astro-ph.SR},
       adsurl = {https://ui.adsabs.harvard.edu/abs/2019Galax...7...92G}
}

@INPROCEEDINGS{2009Oudmaijer,
       author = {{Oudmaijer}, R.~D. and {Davies}, B. and {de Wit}, W. -J. and {Patel}, M.},
        title = "{Post-Red Supergiants}",
    booktitle = {The Biggest, Baddest, Coolest Stars},
         year = 2009,
       editor = {{Luttermoser}, D.~G. and {Smith}, B.~J. and {Stencel}, R.~E.},
       series = {Astronomical Society of the Pacific Conference Series},
       volume = {412},
        month = sep,
        pages = {17},
          doi = {10.48550/arXiv.0801.2315},
archivePrefix = {arXiv},
       eprint = {0801.2315},
 primaryClass = {astro-ph},
       adsurl = {https://ui.adsabs.harvard.edu/abs/2009ASPC..412...17O}
}

@ARTICLE{1998deJager,
       author = {{de Jager}, Cornelis},
        title = "{The yellow hypergiants}",
      journal = {\aapr},
         year = 1998,
        month = jan,
       volume = {8},
       number = {3},
        pages = {145-180},
          doi = {10.1007/s001590050009},
       adsurl = {https://ui.adsabs.harvard.edu/abs/1998A&ARv...8..145D}
}

@ARTICLE{1998Msngr..94....1A,
       author = {{Appenzeller}, I. and {Fricke}, K. and {F{\"u}rtig}, W. and {G{\"a}ssler}, W. and {H{\"a}fner}, R. and {Harke}, R. and {Hess}, H. -J. and {Hummel}, W. and {J{\"u}rgens}, P. and {Kudritzki}, R. -P. and {Mantel}, K. -H. and {Meisl}, W. and {Muschielok}, B. and {Nicklas}, H. and {Rupprecht}, G. and {Seifert}, W. and {Stahl}, O. and {Szeifert}, T. and {Tarantik}, K.},
        title = "{Successful commissioning of FORS1 - the first optical instrument on the VLT.}",
      journal = {The Messenger},
         year = 1998,
        month = dec,
       volume = {94},
        pages = {1-6},
       adsurl = {https://ui.adsabs.harvard.edu/abs/1998Msngr..94....1A}
}

@ARTICLE{Bagnulo2017,
       author = {{Bagnulo}, Stefano and {Cox}, Nick L.~J. and {Cikota}, Aleksandar and {Siebenmorgen}, Ralf and {Voshchinnikov}, Nikolai V. and {Patat}, Ferdinando and {Smith}, Keith T. and {Smoker}, Jonathan V. and {Taubenberger}, Stefan and {Kaper}, Lex and {Cami}, Jan and {LIPS Collaboration}},
        title = "{Large Interstellar Polarisation Survey (LIPS). I. FORS2 spectropolarimetry in the Southern Hemisphere}",
      journal = {\aap},
         year = 2017,
        month = dec,
       volume = {608},
          eid = {A146},
        pages = {A146},
          doi = {10.1051/0004-6361/201731459},
archivePrefix = {arXiv},
       eprint = {1710.02439},
 primaryClass = {astro-ph.SR},
       adsurl = {https://ui.adsabs.harvard.edu/abs/2017A&A...608A.146B}
}

@ARTICLE{Cikota2017,
       author = {{Cikota}, Aleksandar and {Patat}, Ferdinando and {Cikota}, Stefan and {Faran}, Tamar},
        title = "{Linear spectropolarimetry of polarimetric standard stars with VLT/FORS2}",
      journal = {\mnras},
         year = 2017,
        month = feb,
       volume = {464},
       number = {4},
        pages = {4146-4159},
          doi = {10.1093/mnras/stw2545},
archivePrefix = {arXiv},
       eprint = {1610.00722},
 primaryClass = {astro-ph.IM},
       adsurl = {https://ui.adsabs.harvard.edu/abs/2017MNRAS.464.4146C}
}

@ARTICLE{Schlafly2011,
       author = {{Schlafly}, Edward F. and {Finkbeiner}, Douglas P.},
        title = "{Measuring Reddening with Sloan Digital Sky Survey Stellar Spectra and Recalibrating SFD}",
      journal = {\apj},
         year = 2011,
        month = aug,
       volume = {737},
       number = {2},
          eid = {103},
        pages = {103},
          doi = {10.1088/0004-637X/737/2/103},
archivePrefix = {arXiv},
       eprint = {1012.4804},
 primaryClass = {astro-ph.GA},
       adsurl = {https://ui.adsabs.harvard.edu/abs/2011ApJ...737..103S}
}

@ARTICLE{Whittet1978,
       author = {{Whittet}, D.~C.~B. and {van Breda}, I.~G.},
        title = "{The correlation of the interstellar extinction law with the wavelength of maximum polarization.}",
      journal = {\aap},
         year = 1978,
        month = may,
       volume = {66},
        pages = {57-63},
       adsurl = {https://ui.adsabs.harvard.edu/abs/1978A&A....66...57W}
}

@ARTICLE{2008MNRAS.385..967P,
       author = {{Patel}, M. and {Oudmaijer}, R.~D. and {Vink}, J.~S. and {Bjorkman}, J.~E. and {Davies}, B. and {Groenewegen}, M.~A.~T. and {Miroshnichenko}, A.~S. and {Mottram}, J.~C.},
        title = "{Spectropolarimetry of the massive post-red supergiants IRC +10420 and HD 179821}",
      journal = {\mnras},
         year = 2008,
        month = apr,
       volume = {385},
       number = {2},
        pages = {967-978},
          doi = {10.1111/j.1365-2966.2008.12889.x},
archivePrefix = {arXiv},
       eprint = {0801.0878},
 primaryClass = {astro-ph},
       adsurl = {https://ui.adsabs.harvard.edu/abs/2008MNRAS.385..967P}
}

@ARTICLE{Serkowski1975,
       author = {{Serkowski}, K. and {Mathewson}, D.~S. and {Ford}, V.~L.},
        title = "{Wavelength dependence of interstellar polarization and ratio of total to selective extinction.}",
      journal = {\apj},
         year = 1975,
        month = feb,
       volume = {196},
        pages = {261-290},
          doi = {10.1086/153410},
       adsurl = {https://ui.adsabs.harvard.edu/abs/1975ApJ...196..261S}
}

@ARTICLE{Whittet1992,
       author = {{Whittet}, D.~C.~B. and {Martin}, P.~G. and {Hough}, J.~H. and {Rouse}, M.~F. and {Bailey}, J.~A. and {Axon}, D.~J.},
        title = "{Systematic Variations in the Wavelength Dependence of Interstellar Linear Polarization}",
      journal = {\apj},
         year = 1992,
        month = feb,
       volume = {386},
        pages = {562},
          doi = {10.1086/171039},
       adsurl = {https://ui.adsabs.harvard.edu/abs/1992ApJ...386..562W}
}

@ARTICLE{Cikota2019MNRAS.490..578C,
       author = {{Cikota}, Aleksandar and {Patat}, Ferdinando and {Wang}, Lifan and {Wheeler}, J. Craig and {Bulla}, Mattia and {Baade}, Dietrich and {H{\"o}flich}, Peter and {Cikota}, Stefan and {Clocchiatti}, Alejandro and {Maund}, Justyn R. and {Stevance}, Heloise F. and {Yang}, Yi},
        title = "{Linear spectropolarimetry of 35 Type Ia supernovae with VLT/FORS: an analysis of the Si II line polarization}",
      journal = {\mnras},
         year = 2019,
        month = nov,
       volume = {490},
       number = {1},
        pages = {578-599},
          doi = {10.1093/mnras/stz2322},
archivePrefix = {arXiv},
       eprint = {1908.07526},
 primaryClass = {astro-ph.HE},
       adsurl = {https://ui.adsabs.harvard.edu/abs/2019MNRAS.490..578C}
}

@ARTICLE{2001PASP..113.1420V,
       author = {{van Dokkum}, Pieter G.},
        title = "{Cosmic-Ray Rejection by Laplacian Edge Detection}",
      journal = {\pasp},
         year = 2001,
        month = nov,
       volume = {113},
       number = {789},
        pages = {1420-1427},
          doi = {10.1086/323894},
archivePrefix = {arXiv},
       eprint = {astro-ph/0108003},
 primaryClass = {astro-ph},
       adsurl = {https://ui.adsabs.harvard.edu/abs/2001PASP..113.1420V}
}

@ARTICLE{2006PASP..118..146P,
       author = {{Patat}, Ferdinando and {Romaniello}, Martino},
        title = "{Error Analysis for Dual-Beam Optical Linear Polarimetry}",
      journal = {\pasp},
         year = 2006,
        month = jan,
       volume = {118},
       number = {839},
        pages = {146-161},
          doi = {10.1086/497581},
archivePrefix = {arXiv},
       eprint = {astro-ph/0509153},
 primaryClass = {astro-ph},
       adsurl = {https://ui.adsabs.harvard.edu/abs/2006PASP..118..146P}
}

@ARTICLE{2010A&A...510A.108P,
       author = {{Patat}, F. and {Maund}, J.~R. and {Benetti}, S. and {Botticella}, M.~T. and {Cappellaro}, E. and {Harutyunyan}, A. and {Turatto}, M.},
        title = "{VLT spectropolarimetry of the optical transient in NGC 300. Evidence of asymmetry in the circumstellar dust}",
      journal = {\aap},
         year = 2010,
        month = feb,
       volume = {510},
          eid = {A108},
        pages = {A108},
          doi = {10.1051/0004-6361/200913083},
archivePrefix = {arXiv},
       eprint = {0908.0942},
 primaryClass = {astro-ph.SR},
       adsurl = {https://ui.adsabs.harvard.edu/abs/2010A&A...510A.108P}
}

@INPROCEEDINGS{Merand2022,
       author = {{M{\'e}rand}, Antoine},
        title = "{Flexible spectro-interferometric modelling of OIFITS data with PMOIRED}",
    booktitle = {Optical and Infrared Interferometry and Imaging VIII},
         year = 2022,
       editor = {{M{\'e}rand}, Antoine and {Sallum}, Stephanie and {Sanchez-Bermudez}, Joel},
       series = {Society of Photo-Optical Instrumentation Engineers (SPIE) Conference Series},
       volume = {12183},
        month = aug,
          eid = {121831N},
        pages = {121831N},
          doi = {10.1117/12.2626700},
archivePrefix = {arXiv},
       eprint = {2207.11047},
 primaryClass = {astro-ph.IM},
       adsurl = {https://ui.adsabs.harvard.edu/abs/2022SPIE12183E..1NM}
}

@ARTICLE{Dorn2014,
       author = {{Dorn}, R.~J. and {Anglada-Escude}, G. and {Baade}, D. and {Bristow}, P. and {Follert}, R. and {Gojak}, D. and {Grunhut}, J. and {Hatzes}, A. and {Heiter}, U. and {Hilker}, M. and {Ives}, D.~J. and {Jung}, Y. and {K{\"a}ufl}, H. -U. and {Kerber}, F. and {Klein}, B. and {Lizon}, J. -L. and {Lockhart}, M. and {L{\"o}winger}, T. and {Marquart}, T. and {Oliva}, E. and {Origlia}, L. and {Pasquini}, L. and {Paufique}, J. and {Piskunov}, N. and {Pozna}, E. and {Reiners}, A. and {Smette}, A. and {Smoker}, J. and {Seemann}, U. and {Stempels}, E. and {Valenti}, E.},
        title = "{CRIRES+: Exploring the Cold Universe at High Spectral Resolution}",
      journal = {The Messenger},
         year = 2014,
        month = jun,
       volume = {156},
        pages = {7-11},
       adsurl = {https://ui.adsabs.harvard.edu/abs/2014Msngr.156....7D}
}

@ARTICLE{Dorn2023,
       author = {{Dorn}, R.~J. and {Bristow}, P. and {Smoker}, J.~V. and {Rodler}, F. and {Lavail}, A. and {Accardo}, M. and {van den Ancker}, M. and {Baade}, D. and {Baruffolo}, A. and {Courtney-Barrer}, B. and {Blanco}, L. and {Brucalassi}, A. and {Cumani}, C. and {Follert}, R. and {Haimerl}, A. and {Hatzes}, A. and {Haug}, M. and {Heiter}, U. and {Hinterschuster}, R. and {Hubin}, N. and {Ives}, D.~J. and {Jung}, Y. and {Jones}, M. and {Kaeufl}, H. -U. and {Kirchbauer}, J. -P. and {Klein}, B. and {Kochukhov}, O. and {Korhonen}, H.~H. and {K{\"o}hler}, J. and {Lizon}, J. -L. and {Moins}, C. and {Molina-Conde}, I. and {Marquart}, T. and {Neeser}, M. and {Oliva}, E. and {Pallanca}, L. and {Pasquini}, L. and {Paufique}, J. and {Piskunov}, N. and {Reiners}, A. and {Schneller}, D. and {Schmutzer}, R. and {Seemann}, U. and {Slumstrup}, D. and {Smette}, A. and {Stegmeier}, J. and {Stempels}, E. and {Tordo}, S. and {Valenti}, E. and {Valenzuela}, J.~J. and {Vernet}, J. and {Vinther}, J. and {Wehrhahn}, A.},
        title = "{CRIRES$^{+}$ on sky at the ESO Very Large Telescope. Observing the Universe at infrared wavelengths and high spectral resolution}",
      journal = {\aap},
         year = 2023,
        month = mar,
       volume = {671},
          eid = {A24},
        pages = {A24},
          doi = {10.1051/0004-6361/202245217},
archivePrefix = {arXiv},
       eprint = {2301.08048},
 primaryClass = {astro-ph.IM},
       adsurl = {https://ui.adsabs.harvard.edu/abs/2023A&A...671A..24D}
}

@ARTICLE{Koumpia2020,
       author = {{Koumpia}, E. and {Oudmaijer}, R.~D. and {Graham}, V. and {Banyard}, G. and {Black}, J.~H. and {Wichittanakom}, C. and {Ababakr}, K.~M. and {de Wit}, W. -J. and {Millour}, F. and {Lagadec}, E. and {Muller}, S. and {Cox}, N.~L.~J. and {Zijlstra}, A. and {van Winckel}, H. and {Hillen}, M. and {Szczerba}, R. and {Vink}, J.~S. and {Wallstr{\"o}m}, S.~H.~J.},
        title = "{Optical and near-infrared observations of the Fried Egg Nebula. Multiple shell ejections on a 100 yr timescale from a massive yellow hypergiant}",
      journal = {\aap},
         year = 2020,
        month = mar,
       volume = {635},
          eid = {A183},
        pages = {A183},
          doi = {10.1051/0004-6361/201936177},
archivePrefix = {arXiv},
       eprint = {2002.02499},
 primaryClass = {astro-ph.SR},
       adsurl = {https://ui.adsabs.harvard.edu/abs/2020A&A...635A.183K}
}

@ARTICLE{Oudmaijer2022,
       author = {{Oudmaijer}, Ren{\'e} D. and {Jones}, Emma R.~M. and {Vioque}, Miguel},
        title = "{A census of post-AGB stars in Gaia DR3: evidence for a substantial population of Galactic post-RGB stars}",
      journal = {\mnras},
         year = 2022,
        month = oct,
       volume = {516},
       number = {1},
        pages = {L61-L65},
          doi = {10.1093/mnrasl/slac088},
archivePrefix = {arXiv},
       eprint = {2208.02832},
 primaryClass = {astro-ph.SR},
       adsurl = {https://ui.adsabs.harvard.edu/abs/2022MNRAS.516L..61O}
}

@ARTICLE{Cruzalebes2019,
       author = {{Cruzal{\`e}bes}, P. and {Petrov}, R.~G. and {Robbe-Dubois}, S. and {Varga}, J. and {Burtscher}, L. and {Allouche}, F. and {Berio}, P. and {Hofmann}, K.-H. and {Hron}, J. and {Jaffe}, W. and {Lagarde}, S. and {Lopez}, B. and {Matter}, A. and {Meilland}, A. and {Meisenheimer}, K. and {Millour}, F. and {Schertl}, D.},
        title = "{A catalogue of stellar diameters and fluxes for mid-infrared interferometry}",
      journal = {\mnras},
         year = 2019,
        month = dec,
       volume = {490},
       number = {3},
        pages = {3158-3176},
          doi = {10.1093/mnras/stz2803},
archivePrefix = {arXiv},
       eprint = {1910.00542},
 primaryClass = {astro-ph.SR},
       adsurl = {https://ui.adsabs.harvard.edu/abs/2019MNRAS.490.3158C}
}

@ARTICLE{Monnier2003,
       author = {{Monnier}, John D.},
        title = "{Optical interferometry in astronomy}",
      journal = {Reports on Progress in Physics},
         year = 2003,
        month = may,
       volume = {66},
       number = {5},
        pages = {789-857},
          doi = {10.1088/0034-4885/66/5/203},
archivePrefix = {arXiv},
       eprint = {astro-ph/0307036},
 primaryClass = {astro-ph},
       adsurl = {https://ui.adsabs.harvard.edu/abs/2003RPPh...66..789M}
}

@ARTICLE{Moriya2014,
   author = {{Moriya}, T.~J. and {Maeda}, K. and {Taddia}, F. and {Sollerman}, J. and 
	{Blinnikov}, S.~I. and {Sorokina}, E.~I.},
    title = "{Mass-loss histories of Type IIn supernova progenitors within decades before their explosion}",
  journal = {\mnras},
archivePrefix = "arXiv",
   eprint = {1401.4893},
 primaryClass = "astro-ph.SR",
     year = 2014,
    month = apr,
   volume = 439,
    pages = {2917-2926},
      doi = {10.1093/mnras/stu163},
   adsurl = {https://ui.adsabs.harvard.edu/abs/2014MNRAS.439.2917M}
}

@ARTICLE{Heger1998,
       author = {{Heger}, A. and {Langer}, N.},
        title = "{The spin-up of contracting red supergiants}",
      journal = {\aap},
         year = 1998,
        month = jun,
       volume = {334},
        pages = {210-220},
archivePrefix = {arXiv},
       eprint = {astro-ph/9803005},
 primaryClass = {astro-ph},
       adsurl = {https://ui.adsabs.harvard.edu/abs/1998A&A...334..210H}
}

@ARTICLE{Wallstrom2017,
   author = {{Wallstr{\"o}m}, S.~H.~J. and {Lagadec}, E. and {Muller}, S. and 
	{Black}, J.~H. and {Cox}, N.~L.~J. and {Galv{\'a}n-Madrid}, R. and 
	{Justtanont}, K. and {Longmore}, S. and {Olofsson}, H. and {Oudmaijer}, R.~D. and 
	{Quintana-Lacaci}, G. and {Szczerba}, R. and {Vlemmings}, W. and 
	{van Winckel}, H. and {Zijlstra}, A.},
    title = "{ALMA Compact Array observations of the Fried Egg nebula. Evidence for large-scale asymmetric mass-loss from the yellow hypergiant IRAS 17163-3907}",
  journal = {\aap},
archivePrefix = "arXiv",
   eprint = {1612.02510},
 primaryClass = "astro-ph.SR",
     year = 2017,
    month = jan,
   volume = 597,
      eid = {A99},
    pages = {A99},
      doi = {10.1051/0004-6361/201628416},
   adsurl = {http://adsabs.harvard.edu/abs/2017A%26A...597A..99W}
}

@ARTICLE{Lagadec2011,
   author = {{Lagadec}, E. and {Zijlstra}, A.~A. and {Oudmaijer}, R.~D. and 
	{Verhoelst}, T. and {Cox}, N.~L.~J. and {Szczerba}, R. and {M{\'e}karnia}, D. and 
	{van Winckel}, H.},
    title = "{A double detached shell around a post-red supergiant: IRAS 17163-3907, the Fried Egg nebula}",
  journal = {\aap},
archivePrefix = "arXiv",
   eprint = {1109.5947},
 primaryClass = "astro-ph.SR",
     year = 2011,
    month = oct,
   volume = 534,
      eid = {L10},
    pages = {L10},
      doi = {10.1051/0004-6361/201117521},
   adsurl = {http://adsabs.harvard.edu/abs/2011A%26A...534L..10L}
}

@ARTICLE{Wallstrom2015,
   author = {{Wallstr{\"o}m}, S.~H.~J. and {Muller}, S. and {Lagadec}, E. and 
	{Black}, J.~H. and {Oudmaijer}, R.~D. and {Justtanont}, K. and 
	{van Winckel}, H. and {Zijlstra}, A.~A.},
    title = "{Investigating the nature of the Fried Egg nebula. CO mm-line and optical spectroscopy of IRAS 17163-3907}",
  journal = {\aap},
archivePrefix = "arXiv",
   eprint = {1501.03362},
 primaryClass = "astro-ph.SR",
     year = 2015,
    month = feb,
   volume = 574,
      eid = {A139},
    pages = {A139},
      doi = {10.1051/0004-6361/201321516},
   adsurl = {http://adsabs.harvard.edu/abs/2015A%26A...574A.139W}
}

@ARTICLE{Meynet2015,
   author = {{Meynet}, G. and {Chomienne}, V. and {Ekstr{\"o}m}, S. and {Georgy}, C. and 
	{Granada}, A. and {Groh}, J. and {Maeder}, A. and {Eggenberger}, P. and 
	{Levesque}, E. and {Massey}, P.},
    title = "{Impact of mass-loss on the evolution and pre-supernova properties of red supergiants}",
  journal = {\aap},
archivePrefix = "arXiv",
   eprint = {1410.8721},
 primaryClass = "astro-ph.SR",
     year = 2015,
    month = mar,
   volume = 575,
      eid = {A60},
    pages = {A60},
      doi = {10.1051/0004-6361/201424671},
   adsurl = {http://ukads.nottingham.ac.uk/abs/2015A%26A...575A..60M}
}

@ARTICLE{Hutsemekers2013,
   author = {{Hutsem{\'e}kers}, D. and {Cox}, N.~L.~J. and {Vamvatira-Nakou}, C.
	},
    title = "{A massive parsec-scale dust ring nebula around the yellow hypergiant Hen 3-1379}",
  journal = {\aap},
archivePrefix = "arXiv",
   eprint = {1303.4292},
 primaryClass = "astro-ph.SR",
     year = 2013,
    month = apr,
   volume = 552,
      eid = {L6},
    pages = {L6},
      doi = {10.1051/0004-6361/201321380},
   adsurl = {http://adsabs.harvard.edu/abs/2013A%26A...552L...6H}
}

@ARTICLE{LeBertre89,
   author = {{Lebertre}, T. and {Epchtein}, N. and {Gouiffes}, C. and {Heydari-Malayeri}, M. and 
	{Perrier}, C.},
    title = "{Optical and infrared observations of four suspected proto-planetary objects}",
  journal = {\aap},
     year = 1989,
    month = nov,
   volume = 225,
    pages = {417-431},
   adsurl = {http://ukads.nottingham.ac.uk/abs/1989A%26A...225..417L}
}

@ARTICLE{Oudmaijer1994,
       author = {{Oudmaijer}, Rene D. and {Geballe}, T.~R. and {Waters}, L.~B.~F.~M. and {Sahu}, K.~C.},
        title = "{Discovery of near-infrared hydrogen line emission in the peculiar F8 hypergiant IRC +10420.}",
      journal = {\aap},
         year = 1994,
        month = jan,
       volume = {281},
        pages = {L33-L36},
       adsurl = {https://ui.adsabs.harvard.edu/abs/1994A&A...281L..33O}
}

@ARTICLE{Oudmaijer1996,
       author = {{Oudmaijer}, Rene D. and {Groenewegen}, M.~A.~T. and {Matthews}, H.~E. and {Blommaert}, J.~A.~D.~L. and {Sahu}, K.~C.},
        title = "{The spectral energy distribution and mass-loss history of IRC+10420}",
      journal = {\mnras},
         year = 1996,
        month = jun,
       volume = {280},
       number = {4},
        pages = {1062-1070},
          doi = {10.1093/mnras/280.4.1062},
       adsurl = {https://ui.adsabs.harvard.edu/abs/1996MNRAS.280.1062O}
}

@ARTICLE{Humphreys1997,
       author = {{Humphreys}, Roberta M. and {Smith}, Nathan and {Davidson}, Kris and {Jones}, Terry Jay and {Gehrz}, Robert T. and {Mason}, Christopher G. and {Hayward}, Thomas L. and {Houck}, James R. and {Krautter}, Joachim},
        title = "{HST and Infrared Images of the Circumstellar Environment of the Cool Hypergiant IRC + 10420}",
      journal = {\aj},
         year = 1997,
        month = dec,
       volume = {114},
        pages = {2778},
          doi = {10.1086/118686},
       adsurl = {https://ui.adsabs.harvard.edu/abs/1997AJ....114.2778H}
}

@ARTICLE{Humphreys2002,
       author = {{Humphreys}, Roberta M. and {Davidson}, Kris and {Smith}, Nathan},
        title = "{Crossing the Yellow Void: Spatially Resolved Spectroscopy of the Post-Red Supergiant IRC +10420 and Its Circumstellar Ejecta}",
      journal = {\aj},
         year = 2002,
        month = aug,
       volume = {124},
       number = {2},
        pages = {1026-1044},
          doi = {10.1086/341380},
archivePrefix = {arXiv},
       eprint = {astro-ph/0205247},
 primaryClass = {astro-ph},
       adsurl = {https://ui.adsabs.harvard.edu/abs/2002AJ....124.1026H}
}

@ARTICLE{Koumpia2022,
       author = {{Koumpia}, Evgenia and {Oudmaijer}, R.~D. and {de Wit}, W. -J. and {M{\'e}rand}, A. and {Black}, J.~H. and {Ababakr}, K.~M.},
        title = "{Tracing a decade of activity towards a yellow hypergiant. The spectral and spatial morphology of IRC+10420 at au scales}",
      journal = {\mnras},
         year = 2022,
        month = sep,
       volume = {515},
       number = {2},
        pages = {2766-2777},
          doi = {10.1093/mnras/stac1998},
archivePrefix = {arXiv},
       eprint = {2207.05812},
 primaryClass = {astro-ph.SR},
       adsurl = {https://ui.adsabs.harvard.edu/abs/2022MNRAS.515.2766K}
}

@INPROCEEDINGS{Thibaut2008,
   author = {{Thi{\'e}baut}, E.},
    title = "{MIRA: an effective imaging algorithm for optical interferometry}",
booktitle = {Optical and Infrared Interferometry},
     year = 2008,
   series = {\procspie},
   volume = 7013,
    month = jul,
      eid = {70131I},
    pages = {70131I},
      doi = {10.1117/12.788822},
   adsurl = {http://adsabs.harvard.edu/abs/2008SPIE.7013E..1IT}
}

@ARTICLE{Glatzel2024,
       author = {{Glatzel}, Wolfgang and {Kraus}, Michaela},
        title = "{Instabilities in the yellow hypergiant domain}",
      journal = {\mnras},
         year = 2024,
        month = apr,
       volume = {529},
       number = {4},
        pages = {4947-4957},
          doi = {10.1093/mnras/stae861},
archivePrefix = {arXiv},
       eprint = {2403.14315},
 primaryClass = {astro-ph.SR},
       adsurl = {https://ui.adsabs.harvard.edu/abs/2024MNRAS.529.4947G}
}

@ARTICLE{Jones2025,
       author = {{Jones}, Terry},
        title = "{Red and Yellow Hypergiants}",
      journal = {Galaxies},
         year = 2025,
        month = apr,
       volume = {13},
       number = {2},
          eid = {43},
        pages = {43},
          doi = {10.3390/galaxies13020043},
archivePrefix = {arXiv},
       eprint = {2507.15962},
 primaryClass = {astro-ph.SR},
       adsurl = {https://ui.adsabs.harvard.edu/abs/2025Galax..13...43J}
}

@ARTICLE{Woillez2024,
       author = {{Woillez}, J. and {Petrov}, R. and {Abuter}, R. and {Allouche}, F. and {Berio}, P. and {Dembet}, R. and {Eisenhauer}, F. and {Frahm}, R. and {Gont{\'e}}, F. and {Haubois}, X. and {Houll{\'e}}, M. and {Jaffe}, W. and {Lacour}, S. and {Lagarde}, S. and {Leftley}, J. and {Lopez}, B. and {Matter}, A. and {Meilland}, A. and {Millour}, F. and {Nowak}, M. and {Paladini}, C. and {Rivinius}, T. and {Salabert}, D. and {Schuhler}, N. and {Varga}, J. and {Zins}, G.},
        title = "{GRAVITY for MATISSE. Improving the MATISSE performance with the GRAVITY fringe tracker}",
      journal = {\aap},
         year = 2024,
        month = aug,
       volume = {688},
          eid = {A190},
        pages = {A190},
          doi = {10.1051/0004-6361/202449702},
archivePrefix = {arXiv},
       eprint = {2405.20730},
 primaryClass = {astro-ph.IM},
       adsurl = {https://ui.adsabs.harvard.edu/abs/2024A&A...688A.190W}
}

@ARTICLE{Lopez2022,
       author = {{Lopez}, B. and {Lagarde}, S. and {Petrov}, R.~G. and {Jaffe}, W. and {Antonelli}, P. and {Allouche}, F. and {Berio}, P. and {Matter}, A. and {Meilland}, A. and {Millour}, F. and {Robbe-Dubois}, S. and {Henning}, Th. and {Weigelt}, G. and {Glindemann}, A. and {Agocs}, T. and {Bailet}, Ch. and {Beckmann}, U. and {Bettonvil}, F. and {van Boekel}, R. and {Bourget}, P. and {Bresson}, Y. and {Bristow}, P. and {Cruzal{\`e}bes}, P. and {Eldswijk}, E. and {Fante{\"\i} Caujolle}, Y. and {Gonz{\'a}lez Herrera}, J.~C. and {Graser}, U. and {Guajardo}, P. and {Heininger}, M. and {Hofmann}, K.-H. and {Kroes}, G. and {Laun}, W. and {Lehmitz}, M. and {Leinert}, C. and {Meisenheimer}, K. and {Morel}, S. and {Neumann}, U. and {Paladini}, C. and {Percheron}, I. and {Riquelme}, M. and {Schoeller}, M. and {Stee}, Ph. and {Venema}, L. and {Woillez}, J. and {Zins}, G. and {{\'A}brah{\'a}m}, P. and {Abadie}, S. and {Abuter}, R. and {Accardo}, M. and {Adler}, T. and {Alonso}, J. and {Augereau}, J.-C. and {B{\"o}hm}, A. and {Bazin}, G. and {Beltran}, J. and {Bensberg}, A. and {Boland}, W. and {Brast}, R. and {Burtscher}, L. and {Castillo}, R. and {Chelli}, A. and {Cid}, C. and {Clausse}, J.-M. and {Connot}, C. and {Conzelmann}, R.~D. and {Danchi}, W.-C. and {Delbo}, M. and {Drevon}, J. and {Dominik}, C. and {van Duin}, A. and {Ebert}, M. and {Eisenhauer}, F. and {Flament}, S. and {Frahm}, R. and {G{\'a}mez Rosas}, V. and {Gabasch}, A. and {Gallenne}, A. and {Garces}, E. and {Girard}, P. and {Glazenborg}, A. and {Gont{\'e}}, F.~Y.~J. and {Guitton}, F. and {de Haan}, M. and {Hanenburg}, H. and {Haubois}, X. and {Hocd{\'e}}, V. and {Hogerheijde}, M. and {ter Horst}, R. and {Hron}, J. and {Hummel}, C.~A. and {Hubin}, N. and {Huerta}, R. and {Idserda}, J. and {Isbell}, J.~W. and {Ives}, D. and {Jakob}, G. and {Jask{\'o}}, A. and {Jochum}, L. and {Klarmann}, L. and {Klein}, R. and {Kragt}, J. and {Kuindersma}, S. and {Kokoulina}, E. and {Labadie}, L. and {Lacour}, S. and {Leftley}, J. and {Le Poole}, R. and {Lizon}, J.-L. and {Lopez}, M. and {Lykou}, F. and {M{\'e}rand}, A. and {Marcotto}, A. and {Mauclert}, N. and {Maurer}, T. and {Mehrgan}, L.~H. and {Meisner}, J. and {Meixner}, K. and {Mellein}, M. and {Menut}, J.~L. and {Mohr}, L. and {Mosoni}, L. and {Navarro}, R. and {Nu{\ss}baum}, E. and {Pallanca}, L. and {Pantin}, E. and {Pasquini}, L. and {Phan Duc}, T. and {Pott}, J.-U. and {Pozna}, E. and {Richichi}, A. and {Ridinger}, A. and {Rigal}, F. and {Rivinius}, Th. and {Roelfsema}, R. and {Rohloff}, R.-R. and {Rousseau}, S. and {Salabert}, D. and {Schertl}, D. and {Schuhler}, N. and {Schuil}, M. and {Shabun}, K. and {Soulain}, A. and {Stephan}, C. and {Toledo}, P. and {Tristram}, K. and {Tromp}, N. and {Vakili}, F. and {Varga}, J. and {Vinther}, J. and {Waters}, L.~B.~F.~M. and {Wittkowski}, M. and {Wolf}, S. and {Wrhel}, F. and {Yoffe}, G.},
        title = "{MATISSE, the VLTI mid-infrared imaging spectro-interferometer}",
      journal = {\aap},
         year = 2022,
        month = mar,
       volume = {659},
          eid = {A192},
        pages = {A192},
          doi = {10.1051/0004-6361/202141785},
archivePrefix = {arXiv},
       eprint = {2110.15556},
 primaryClass = {astro-ph.IM},
       adsurl = {https://ui.adsabs.harvard.edu/abs/2022A&A...659A.192L}
}

@ARTICLE{Soding2025,
       author = {{S{\"o}ding}, Laurin and {Edenhofer}, Gordian and {En{\ss}lin}, Torsten A. and {Frank}, Philipp and {Kissmann}, Ralf and {Phan}, Vo Hong Minh and {Ram{\'\i}rez}, Andr{\'e}s and {Zandinejad}, Hanieh and {Mertsch}, Philipp},
        title = "{Spatially coherent 3D distributions of HI and CO in the Milky Way}",
      journal = {\aap},
         year = 2025,
        month = jan,
       volume = {693},
          eid = {A139},
        pages = {A139},
          doi = {10.1051/0004-6361/202451361},
archivePrefix = {arXiv},
       eprint = {2407.02859},
 primaryClass = {astro-ph.GA},
       adsurl = {https://ui.adsabs.harvard.edu/abs/2025A&A...693A.139S}
}

@ARTICLE{Vink1999,
       author = {{Vink}, J.~S. and {de Koter}, A. and {Lamers}, H.~J.~G.~L.~M.},
        title = "{On the nature of the bi-stability jump in the winds of early-type supergiants}",
      journal = {\aap},
         year = "1999",
        month = "Oct",
       volume = {350},
        pages = {181-196},
archivePrefix = {arXiv},
       eprint = {astro-ph/9908196},
 primaryClass = {astro-ph},
       adsurl = {https://ui.adsabs.harvard.edu/abs/1999A&A...350..181V}
}

@ARTICLE{Petrov2016,
       author = {{Petrov}, Blagovest and {Vink}, Jorick S. and {Gr{\"a}fener}, G{\"o}tz},
        title = "{Two bi-stability jumps in theoretical wind models for massive stars and the implications for luminous blue variable supernovae}",
      journal = {\mnras},
         year = "2016",
        month = "May",
       volume = {458},
       number = {2},
        pages = {1999-2011},
          doi = {10.1093/mnras/stw382},
archivePrefix = {arXiv},
       eprint = {1602.05868},
 primaryClass = {astro-ph.SR},
       adsurl = {https://ui.adsabs.harvard.edu/abs/2016MNRAS.458.1999P}
}

@ARTICLE{Vink2000,
       author = {{Vink}, J.~S. and {de Koter}, A. and {Lamers}, H.~J.~G.~L.~M.},
        title = "{New theoretical mass-loss rates of O and B stars}",
      journal = {\aap},
         year = "2000",
        month = "Oct",
       volume = {362},
        pages = {295-309},
archivePrefix = {arXiv},
       eprint = {astro-ph/0008183},
 primaryClass = {astro-ph},
       adsurl = {https://ui.adsabs.harvard.edu/abs/2000A&A...362..295V}
}

@ARTICLE{Vink2001,
       author = {{Vink}, Jorick S. and {de Koter}, A. and {Lamers}, H.~J.~G.~L.~M.},
        title = "{Mass-loss predictions for O and B stars as a function of metallicity}",
      journal = {\aap},
         year = "2001",
        month = "Apr",
       volume = {369},
        pages = {574-588},
          doi = {10.1051/0004-6361:20010127},
archivePrefix = {arXiv},
       eprint = {astro-ph/0101509},
 primaryClass = {astro-ph},
       adsurl = {https://ui.adsabs.harvard.edu/abs/2001A&A...369..574V}
}

@ARTICLE{deJager1998,
       author = {{de Jager}, Cornelis},
        title = "{The yellow hypergiants}",
      journal = {\aapr},
         year = "1998",
        month = "Jan",
       volume = {8},
       number = {3},
        pages = {145-180},
          doi = {10.1007/s001590050009},
       adsurl = {https://ui.adsabs.harvard.edu/abs/1998A&ARv...8..145D}
}

@ARTICLE{Lobel1994,
       author = {{Lobel}, A. and {de Jager}, C. and {Nieuwenhuijzen}, H. and
         {Smolinski}, J. and {Gesicki}, K.},
        title = "{Pulsation of the yellow hypergiant {\ensuremath{\rho}} Cassiopeiae in 1970.}",
      journal = {\aap},
         year = "1994",
        month = "Nov",
       volume = {291},
        pages = {226-238},
       adsurl = {https://ui.adsabs.harvard.edu/abs/1994A&A...291..226L}
}

@ARTICLE{Heger2003,
       author = {{Heger}, A. and {Fryer}, C.~L. and {Woosley}, S.~E. and {Langer}, N. and {Hartmann}, D.~H.},
        title = "{How Massive Single Stars End Their Life}",
      journal = {\apj},
         year = 2003,
        month = jul,
       volume = {591},
       number = {1},
        pages = {288-300},
          doi = {10.1086/375341},
archivePrefix = {arXiv},
       eprint = {astro-ph/0212469},
 primaryClass = {astro-ph},
       adsurl = {https://ui.adsabs.harvard.edu/abs/2003ApJ...591..288H}
}

@ARTICLE{Langer2012,
       author = {{Langer}, N.},
        title = "{Presupernova Evolution of Massive Single and Binary Stars}",
      journal = {\araa},
         year = 2012,
        month = sep,
       volume = {50},
        pages = {107-164},
          doi = {10.1146/annurev-astro-081811-125534},
archivePrefix = {arXiv},
       eprint = {1206.5443},
 primaryClass = {astro-ph.SR},
       adsurl = {https://ui.adsabs.harvard.edu/abs/2012ARA&A..50..107L}
}

@ARTICLE{Smith2014,
       author = {{Smith}, Nathan},
        title = "{Mass Loss: Its Effect on the Evolution and Fate of High-Mass Stars}",
      journal = {\araa},
         year = 2014,
        month = aug,
       volume = {52},
        pages = {487-528},
          doi = {10.1146/annurev-astro-081913-040025},
archivePrefix = {arXiv},
       eprint = {1402.1237},
 primaryClass = {astro-ph.SR},
       adsurl = {https://ui.adsabs.harvard.edu/abs/2014ARA&A..52..487S}
}

@ARTICLE{Davies2005,
       author = {{Davies}, Ben and {Oudmaijer}, Ren{\'e} D. and {Vink}, Jorick S.},
        title = "{Asphericity and clumpiness in the winds of Luminous Blue Variables}",
      journal = {\aap},
         year = 2005,
        month = sep,
       volume = {439},
       number = {3},
        pages = {1107-1125},
          doi = {10.1051/0004-6361:20052781},
archivePrefix = {arXiv},
       eprint = {astro-ph/0505344},
 primaryClass = {astro-ph},
       adsurl = {https://ui.adsabs.harvard.edu/abs/2005A&A...439.1107D}
}

@ARTICLE{Castro2007,
       author = {{Castro-Carrizo}, A. and {Quintana-Lacaci}, G. and {Bujarrabal}, V. and {Neri}, R. and {Alcolea}, J.},
        title = "{Arcsecond-resolution $^{12}$CO mapping of the yellow hypergiants IRC +10420 and AFGL 2343}",
      journal = {\aap},
         year = 2007,
        month = apr,
       volume = {465},
       number = {2},
        pages = {457-467},
          doi = {10.1051/0004-6361:20066169},
archivePrefix = {arXiv},
       eprint = {astro-ph/0702400},
 primaryClass = {astro-ph},
       adsurl = {https://ui.adsabs.harvard.edu/abs/2007A&A...465..457C}
}

@ARTICLE{Kasikov2026,
       author = {{Kasikov}, A. and {Mehner}, A. and {Kolka}, I. and {Aret}, A.},
        title = "{Painting a Family Portrait of the Yellow Super- and Hypergiants in the Milky Way I. Constraining the Distances and Luminosities}",
      journal = {arXiv e-prints},
         year = 2026,
        month = feb,
          eid = {arXiv:2602.02449},
        pages = {arXiv:2602.02449},
          doi = {10.48550/arXiv.2602.02449},
archivePrefix = {arXiv},
       eprint = {2602.02449},
 primaryClass = {astro-ph.SR},
       adsurl = {https://ui.adsabs.harvard.edu/abs/2026arXiv260202449K}
}

@ARTICLE{Tomassini2026,
       author = {{Tomassini}, G. and {Lagadec}, E. and {El Mellah}, I. and {Oudmaijer}, R.~D. and {Chiavassa}, A. and {N'Diaye}, M. and {de Laverny}, P. and {Nardetto}, N. and {Matter}, A.},
        title = "{Characterising the post-red supergiant binary system AFGL 4106 and its complex nebula with SPHERE/VLT}",
      journal = {\aap},
         year = 2026,
        month = jan,
       volume = {706},
          eid = {A5},
        pages = {A5},
          doi = {10.1051/0004-6361/202557705},
archivePrefix = {arXiv},
       eprint = {2512.01543},
 primaryClass = {astro-ph.SR},
       adsurl = {https://ui.adsabs.harvard.edu/abs/2026A&A...706A...5T}
}

@misc{deWitTBS,
  author = {de Wit, W.-J. and Koumpia, E. and Collaborators},
  year   = {to be submitted}
}

@ARTICLE{Quintana2016,
       author = {{Quintana-Lacaci}, G. and {Ag{\'u}ndez}, M. and {Cernicharo}, J. and {Bujarrabal}, V. and {S{\'a}nchez Contreras}, C. and {Castro-Carrizo}, A. and {Alcolea}, J.},
        title = "{A {\ensuremath{\lambda}} 3 mm and 1 mm line survey toward the yellow hypergiant IRC +10420. N-rich chemistry and IR flux variations}",
      journal = {\aap},
         year = 2016,
        month = jul,
       volume = {592},
          eid = {A51},
        pages = {A51},
          doi = {10.1051/0004-6361/201527688},
archivePrefix = {arXiv},
       eprint = {1605.09183},
 primaryClass = {astro-ph.SR},
       adsurl = {https://ui.adsabs.harvard.edu/abs/2016A&A...592A..51Q}
}

\clearpage
\newpage

\begin{appendix}
\section{Complete CRIRES+ spectral coverage}

This appendix presents all spectral windows observed with CRIRES+ with further highlighting the ranges in which line detections were identified (Fig.~\ref{fig:crires_overview}). For each spectral range, we show the extracted spectra of both \object{IRAS\,17163} and \object{IRC+10420} overplotted, to facilitate direct comparison of their line profiles and relative emission strengths. The measurements of the detected lines are presented in Table~\ref{crires_comparison}.


\begin{figure*}[!htbp]
\centering

\begin{subfigure}{\textwidth}
  \centering
  \includegraphics[width=0.9\linewidth]{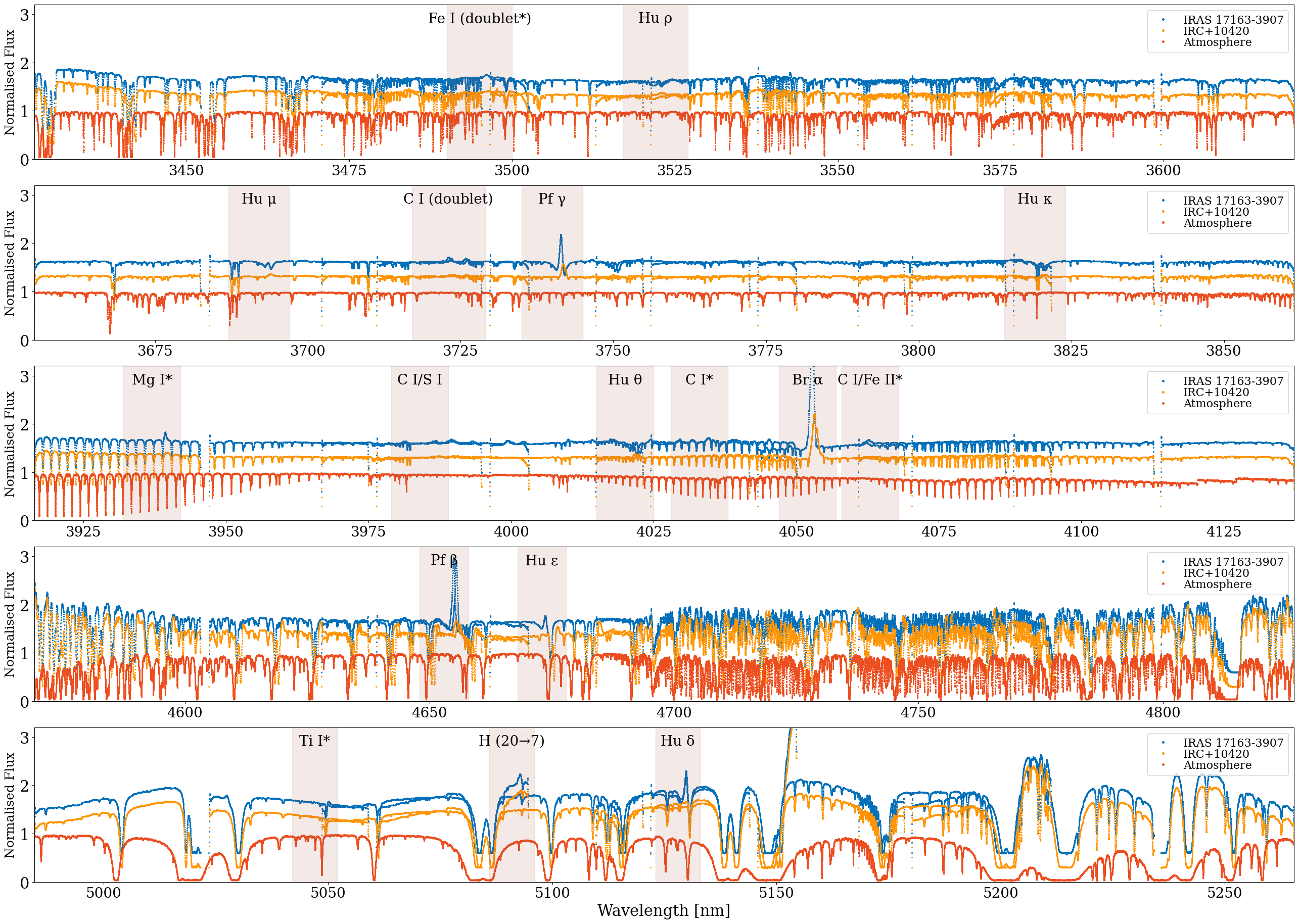}
  \label{fig:crires_combined_full}
\end{subfigure}

\vspace{1.5mm}

\begin{subfigure}{\textwidth}
  \centering
  \includegraphics[width=0.9\linewidth]{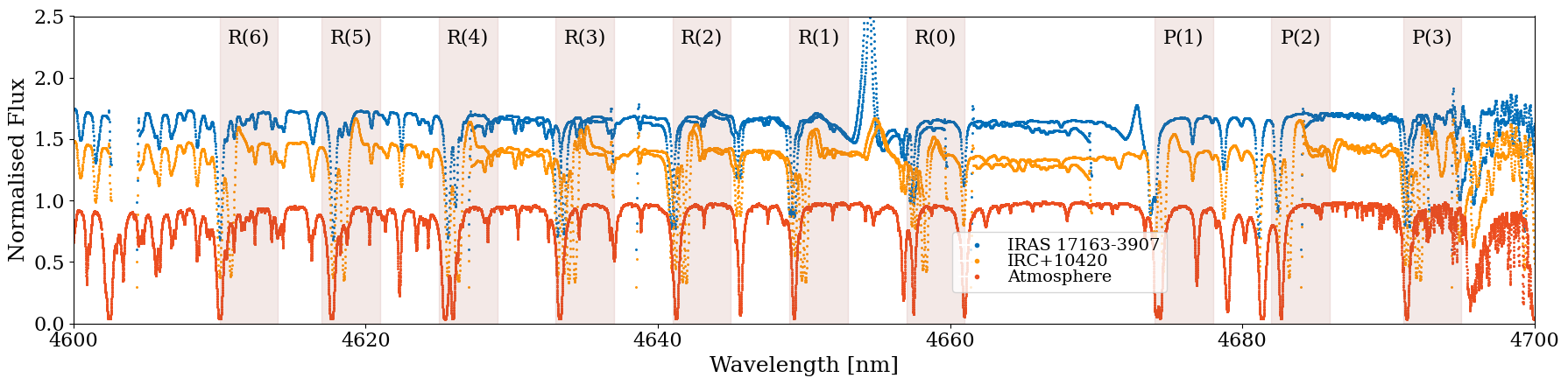}
  \label{fig:crires_combined_CO}
\end{subfigure}

\vspace{1.5mm}

\begin{subfigure}{\textwidth}
  \centering
  \includegraphics[width=0.9\linewidth]{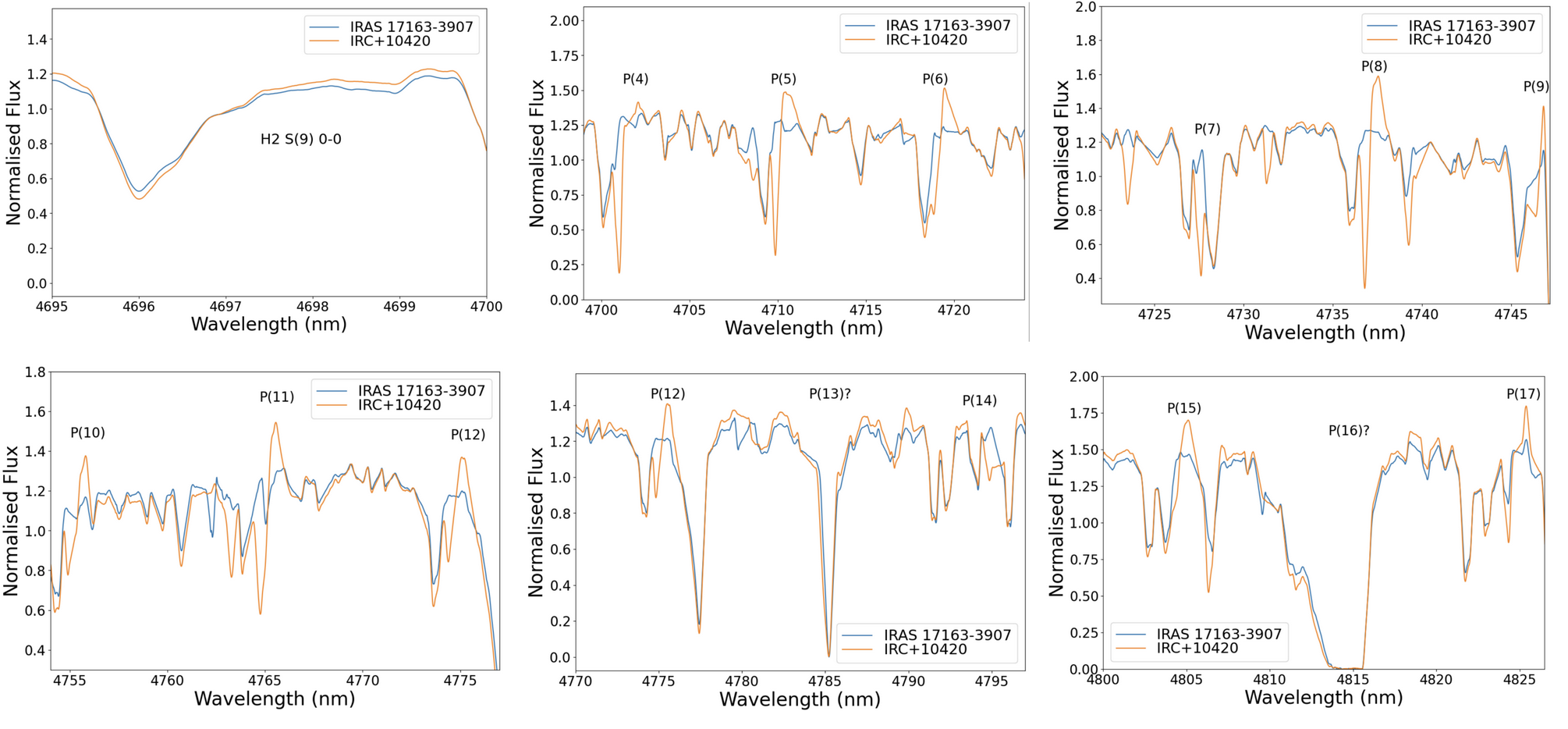}
  \label{fig:crires_combined_highP}
\end{subfigure}

\caption{
Full observed wavelength range of CRIRES+ for \object{IRAS\,17163} (blue) and \object{IRC+10420} (orange), including the telluric reference (red), with line identifications (uncertain identifications marked with $*$; NSO/Kitt Peak FTS atmosphere data produced by NSF/NOAO). The normalised fluxes of the sources have been manually shifted $+0.6$ and $+0.3$ units, respectively. The panels also zoom in the well-resolved fundamental CO rovibrational transitions and covers the higher-order P-branch lines (last 2 panels), shown for completeness despite strong telluric contamination (not included in Table~\ref{crires_comparison}).
}
\label{fig:crires_overview}
\end{figure*}

\begin{figure*}[!htbp]
    \centering
    \includegraphics[width=\textwidth]{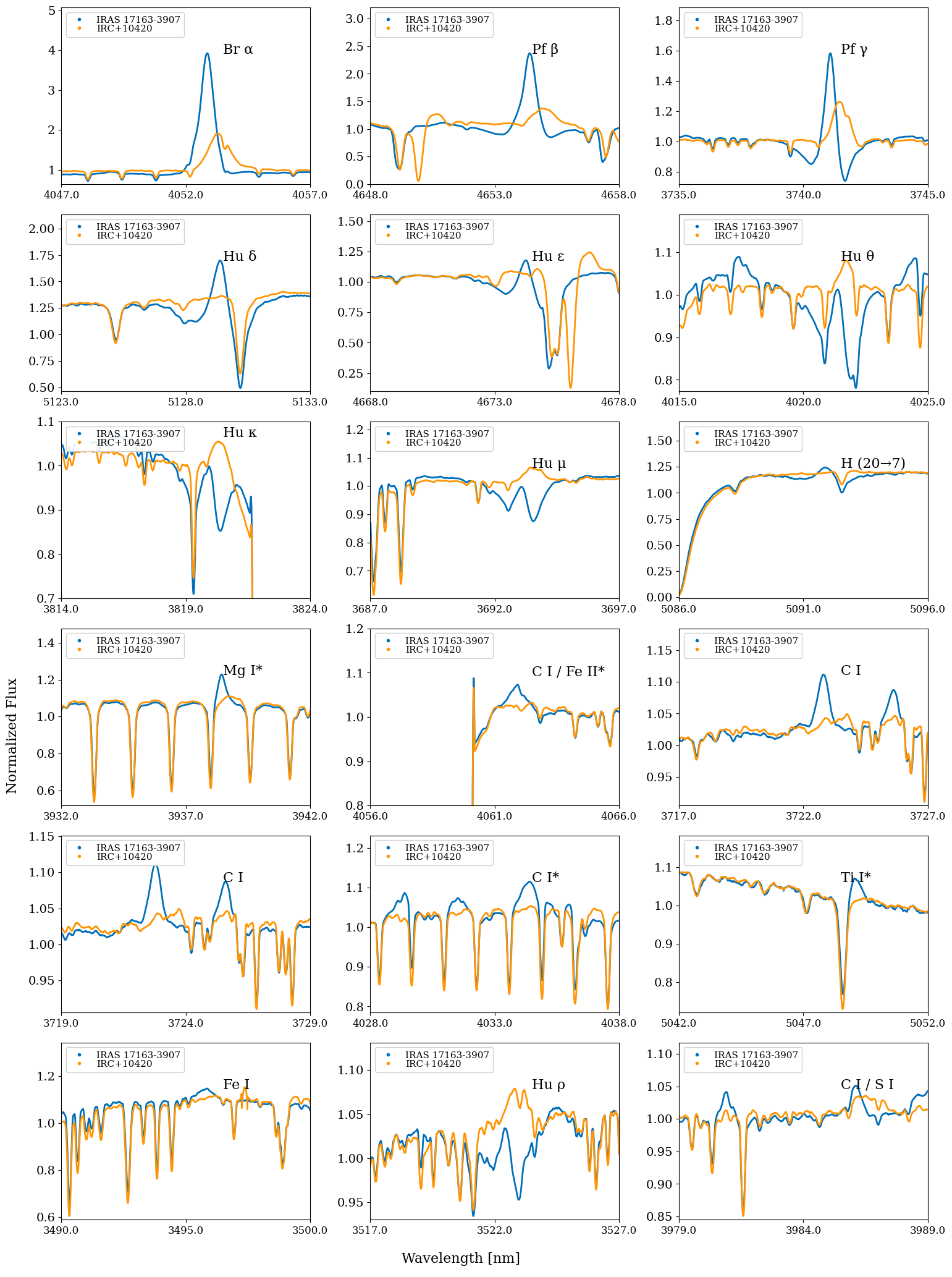}
    \caption{Detected atomic features for \object{IRAS\,17163} and \object{IRC+10420} with CRIRES+. Uncertain identifications are marked with a $*$.}
    \label{fig:crires_select}
\end{figure*}


\begin{table*}[!htbp]
\setlength{\tabcolsep}{14pt}   
\renewcommand{\arraystretch}{1.10} 
\centering
\caption{FWHM, equivalent width (EW), and spectral measurements for CRIRES+ observations. Uncertain measurements are denoted with a $*$. Values that were not calculated due to significant atmospheric interference are denoted as INC.}
\begin{tabular}{cccccc}
\hline
\hline
Spectral Line & Rest Wavelength  & Source & FWHM & EW & Peak $V_{\mathrm{LSR}}$ \\

     & (nm)                  &    & (km/s)        & (Å)         & (km/s)        \\
\hline
\hline
\multicolumn{6}{c}{\textbf{Hydrogen}} \\
\hline
\hline
Br$\alpha$ & 4052.26         & IRAS 17163 & 45.5        & $-22.1$   & 45            \\
           &                               & IRC+10420  & 70.5        & $-9.4$    & 76            \\
Pf$\beta$  & 4653.78              & IRAS 17163 & 38.5        & $-7.0$    & 41          \\
           &                               & IRC+10420  & 61.5        & $-2.3$    & 74.5            \\
Pf$\gamma$ & 3740.56             & IRAS 17163 & 32        & $0.2$    & 41.5            \\
           &                              & IRC+10420  & 60        & $-1.9$    & 75.5            \\
Hu$\delta$ & 5128.65              & IRAS 17163 & 37        & $-3.6$    & 42.5            \\ &
&                 IRC+10420  & INC           & INC         & INC  \\
Hu$\epsilon$ & 4672.51            & IRAS 17163 & 24.5        & $-0.4$    & 37.5            \\
           &                               & IRC+10420  & INC           & INC         & 67           \\
Hu$\theta$   & 4020.87           & IRAS 17163 & 30.5        & $-0.5$    & 37.5            \\
           &                                & IRC+10420  & 41        & $-0.4$    & 62.5            \\

Hu$\kappa$   & 3819.45           & IRAS 17163 &   48.5     &   $-1.0$  &          38.5   \\
           &                                & IRC+10420  &  50.5       &   $-0.3$  &       66.5     \\

Hu$\mu$   & 3692.64           & IRAS 17163 & 43*            &    $-1.2$*  &      36 \\
           &                                & IRC+10420  & 59.5        &   $-0.3$  &         66.5   \\

Hu$\rho$       & 3522.03              & IRAS 17163 &   26.5    &  $-0.1$  & 33            \\
           &                                & IRC+10420  & INC           & INC         & 64.5           \\

$H_{20\rightarrow7}$   & 3740.56         & IRAS 17163 &      33  &   $-0.3$  &          33.5   \\
           &                                & IRC+10420  &      INC   &  INC   &        INC    \\
\hline
\hline
\multicolumn{6}{c}{\textbf{Metals} towards IRAS 17163-3907} \\
\hline
\hline
C I     &      3722.29         & IRAS 17163 &     41   &  $-0.4$ & 41.5            \\
C I &      3725.07         & IRAS 17163 &    43   &  $-0.4$ & 42            \\
C I* &      4033.92        & IRAS 17163 &   57     &  $-0.7$   & 35.5            \\
C I / Fe II* &      4061.60/4061.77         & IRAS 17163 &     65.5    & $-0.7$  & 46/33.5           \\
C I / S I     &  3985.49             & IRAS 17163 &     33   &   $-0.2$  & 47.5            \\
Fe I     & 3495.292/3495.497              & IRAS 17163 &     84.5  &   $-0.6$  & 50/32.5            \\
Mg I* & 3937.95                & IRAS 17163 &     26.5   &   $-0.5$ & 37.5     \\
Ti I*       & 5048.53              & IRAS 17163 &   27     &  $-0.3$  & 30.5            \\
\hline
\hline
\multicolumn{6}{c}{$\mathbf{^{12}\mathrm{C}{}^{16}\mathrm{O}}$ towards IRC+10420} \\
\hline
\hline
CO R(0)    & 4657.5                & Peak &   44      & $-1.2$    &     94.5       \\
& & Dip &19 & $2.9$&N/A\\
CO R(1)    & 4649.3             & Peak &     41   & $1.3$ &     97      \\
& & Dip & 21&$3.5$ &N/A\\
CO R(2)    & 4641.2               & Peak &   43    &  $-1.6$ &   96      \\
& & Dip & 22 & $3.3$&N/A\\
CO R(3)    & 4633.3                & Peak &    INC   &     INC&       99     \\
& & Dip & 21.5&3.3 &N/A\\
CO R(4)    & 4625.4                & Peak &    37.5     &  $-1.1$ &       98     \\
& & Dip & INC & INC &N/A\\
CO R(5)    & 4617.7                & Peak &  34      & $-1.0$  &         97   \\
& & Dip &26.5 & 4.1&N/A\\
CO R(6)    & 4610.0                & Peak &    INC   &   INC &        84.5    \\
& & Dip & 27.5 & 3.6 &N/A\\
CO P(1)    & 4674.2                & Peak &    39    &  $-0.7$  &      93.5    \\
& & Dip & 20 &$2.9$ &N/A\\
CO P(2)    & 4682.6                & Peak &    45.5    &    $-1.7$ &     95.5     \\
& & Dip & 20&$2.6$ &N/A\\
CO P(3)    & 4691.2                & Peak &    INC    &   INC &      81       \\
& & Dip & 21 & 3.0&N/A\\
\hline
\end{tabular}
\begin{flushleft}
\footnotesize
$^{\dagger}$Higher-order CO P-branch transitions are detected toward \object{IRC+10420}
(Fig.~\ref{fig:crires_overview}) but are not quantitatively analysed due to the strong atmospheric
contamination in this wavelength region.
\end{flushleft}

\label{crires_comparison}
\end{table*}


\section{Interferometry}
\subsection{Supporting interferometric measurements and line-emission diagnostics}

This appendix presents the interferometric observables, temporal line diagnostics, and model products that support the analysis described in the main text. For clarity and readability, the corresponding tables and figures are collected here (Fig.~\ref{fig:three_plots_Bralpha}-Fig.~\ref{fig:wray_bralpha_reco_combined}, Table~\ref{tab:visibilities_two_column}, Table~\ref{line_measures}), while their interpretation is discussed in the main text. The material includes the calibrated VLTI/MATISSE interferometric measurements, the time evolution of the Br$\alpha$ line properties derived from spectroscopic fitting, and the best-fit model images and interferometric observables obtained with \textsc{PMOIRED}.


\begin{figure*}[!htbp]
\centering

\begin{minipage}[t]{0.8\linewidth}
    \centering
    \includegraphics[width=\linewidth]{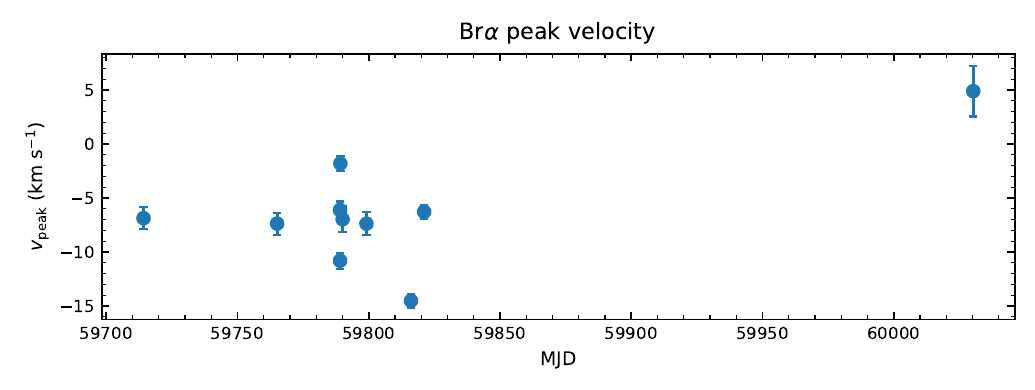}
\end{minipage}

\vspace{2mm}

\begin{minipage}[t]{0.8\linewidth}
    \centering
    \includegraphics[width=\linewidth]{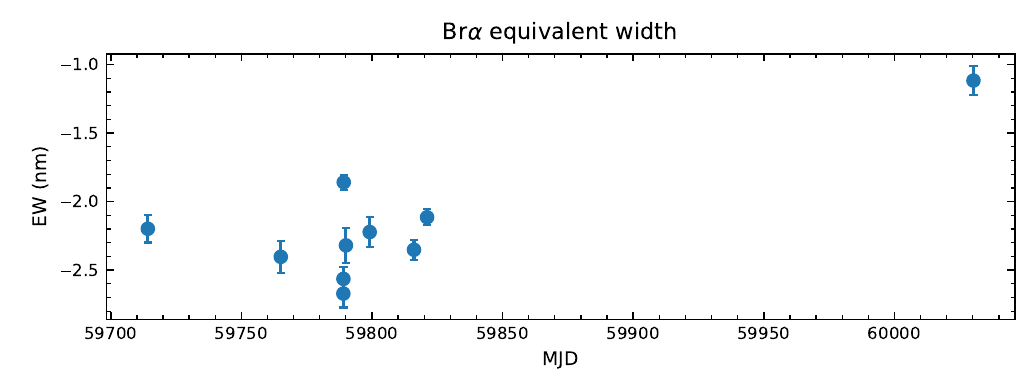}
\end{minipage}

\vspace{2mm}

\begin{minipage}[t]{0.8\linewidth}
    \centering
    \includegraphics[width=\linewidth]{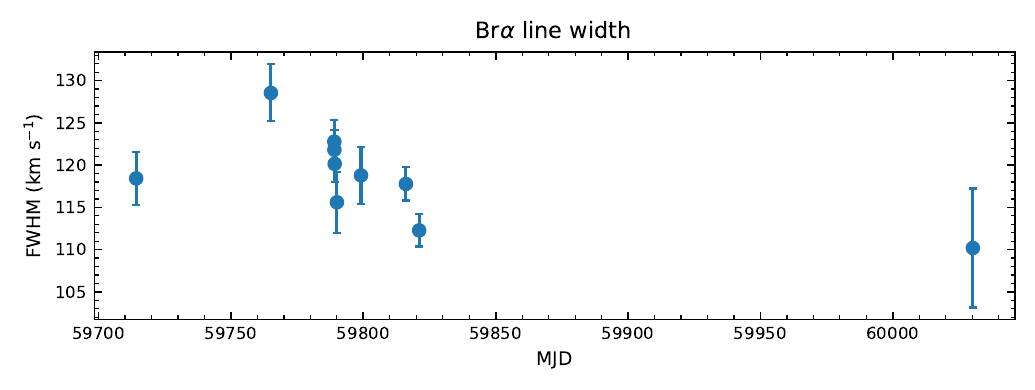}
\end{minipage}

\caption{
From top to bottom, the panels show the line-of-sight velocity with respect to the local standard of rest ($V_{\mathrm{LSR}}$), the equivalent width (EW), and the full width at half maximum (FWHM) of the Br$\alpha$ emission as measured with MATISSE (Table~\ref{line_measures}).
}
\label{fig:three_plots_Bralpha}
\end{figure*}


\subsection{Image reconstruction}
\label{reconstruction}

To complement the geometric modelling and obtain a model-independent view of the spatial distribution of the 4~$\mu$m emission, we performed image reconstruction of the VLTI/MATISSE data using the \texttt{OImaging} package developed by the Jean-Marie Mariotti Center (JMMC\footnote{\url{http://www.jmmc.fr/oimaging}}). The software provides access to several reconstruction algorithms (MiRA \citep{Thibaut2008}, BSMEM, SQUEEZE, WISARD) and regularisation schemes tailored to optical interferometric data.

We started with the reconstruction of selected continuum channels where the emission is stable and unaffected by line variability. The general approach adopted by the imaging algorithms is to minimise a global cost function, $\chi^2 = \chi^2_{\mathrm{data}} + \mu \cdot R(I)$, where $\chi^2_{\mathrm{data}}$ quantifies the fit to the interferometric observables, $R(I)$ encodes the chosen regularisation (e.g.\ smoothness or compactness), and $\mu$ controls the balance between data fidelity and prior information. We tested several regularisation schemes, including quadratic smoothness and total variation, and explored a range of hyperparameters to ensure that the resulting images were stable while limiting reconstruction artefacts. The algorithm converged, resulting in a reduced $\chi^{2}$ of approximately 4.

The continuum reconstruction (Fig.~\ref{fig:wray_bralpha_reco_combined}) reveals a compact morphology with a characteristic radius of $\sim2$~mas, consistent with the continuum size inferred from the geometric modelling (Sect.~\ref{sec:geometric_models}). An integrated Br$\alpha$ image (Fig.~\ref{fig:wray_bralpha_reco_combined}) was obtained by summing the reconstructed intensity across all velocity channels spanning the line. While this integrated map does not correspond to a strictly self-consistent interferometric reconstruction, it provides a convenient visual summary of the overall spatial extent of the Br$\alpha$ emission relative to the continuum. Figure~\ref{fig:wray_bralpha_reco_combined} presents velocity-resolved image reconstructions across the Br$\alpha$ line at eight representative radial velocities spanning the full extent of the observed line profile. The selected velocity slices range from $v_{\rm LSR}\simeq-170$~km~s$^{-1}$ to $+200$~km~s$^{-1}$ and are symmetrically distributed around the systemic velocity ($v_{\rm sys}=18$~km~s$^{-1}$), ensuring that both the blue- and red-shifted wings as well as the line core are sampled. The chosen velocities are not intended to provide a uniformly spaced spectral cube, but rather to trace the gross kinematic structure of the Br$\alpha$ emission at characteristic points across the line profile.

Each channel map was reconstructed independently using the WISARD algorithm, adopting the same field of view and regularisation scheme as for the continuum reconstruction. For visualisation purposes, the images are displayed using a per-channel intensity stretch, adjusted to the local dynamic range of each slice in order to enhance low-level emission. The continuum morphology, reconstructed from line-free spectral channels, is overplotted as contours to provide a spatial reference for the compact dust-emitting region with respect to the line-emitting region. Given the intrinsic variability of the Br$\alpha$ line between epochs, the limited $(u,v)$ coverage and SNR available per spectral channel, and the strong chromatic structure of the line, the velocity-channel reconstructions should be interpreted qualitatively. They are primarily intended to highlight systematic changes in the spatial distribution of the ionised gas across the line profile, rather than to provide a quantitative tomographic view of the velocity field.

\begin{table*}[p]
\centering
\caption{Interferometric baseline information and measured visibilities in the continuum and Br$\alpha$ line.}
\label{tab:visibilities_two_column}
\resizebox{\textwidth}{!}{%
\begin{tabular}{p{2.6cm} p{1.8cm} p{1.8cm} p{1.4cm} p{1.4cm} p{3.5cm} p{3.5cm}}

\toprule
Date & MJD & Baseline & Length [m] & PA [deg] & $V_{\mathrm{cont}} \pm \sigma$ & $V_{\mathrm{Br\alpha}} \pm \sigma$ \\
\midrule
2022-05-15 & 59714.19815 & J2-J3 & 96.2 & 119.9 & 0.87 ± 0.09 & 0.5 ± 0.1 \\
2022-05-15 & 59714.19815 & A0-G1 & 81.8 & 169.3 & 0.93 ± 0.05 & 0.58 ± 0.06 \\
2022-05-15 & 59714.19815 & G1-J2 & 57.8 & 50.2 & 0.95 ± 0.06 & 0.9 ± 0.1 \\
2022-05-15 & 59714.19815 & G1-J3 & 128.3 & 94.9 & 0.90 ± 0.07 & 0.4 ± 0.1 \\
2022-05-15 & 59714.19815 & A0-J2 & 120.9 & 14.0 & 0.79 ± 0.05 & 0.37 ± 0.09 \\
2022-05-15 & 59714.19815 & A0-J3 & 132.2 & 58.4 & 0.83 ± 0.05 & 0.38 ± 0.06 \\
2022-07-05 & 59765.09167 & J2-J3 & 98.9 & 115.4 & 0.87 ± 0.05 & 0.54 ± 0.04 \\
2022-07-05 & 59765.09167 & A0-G2 & 68.3 & 46.7 & 0.91 ± 0.05 & 0.64 ± 0.07 \\
2022-07-05 & 59765.09167 & G2-J2 & 87.0 & 155.3 & 0.94 ± 0.06 & 0.72 ± 0.08 \\
2022-07-05 & 59765.09167 & G2-J3 & 64.3 & 55.4 & 0.88 ± 0.07 & 0.8 ± 0.1 \\
2022-07-05 & 59765.09167 & A0-J2 & 126.6 & 6.0 & 0.85 ± 0.06 & 0.35 ± 0.04 \\
2022-07-05 & 59765.09167 & A0-J3 & 132.3 & 50.9 & 0.74 ± 0.06 & 0.20 ± 0.06 \\
2022-07-29 & 59789.07077 & D0-J3 & 101.9 & 49.3 & 0.87 ± 0.05 & 0.53 ± 0.05 \\
2022-07-29 & 59789.07077 & A0-B2 & 24.9 & 127.1 & 0.96 ± 0.05 & 1.02 ± 0.08 \\
2022-07-29 & 59789.07077 & B2-D0 & 32.9 & 62.1 & 1.00 ± 0.06 & 1.1 ± 0.1 \\
2022-07-29 & 59789.07077 & B2-J3 & 134.2 & 52.4 & 0.85 ± 0.06 & 0.37 ± 0.05 \\
2022-07-29 & 59789.07077 & A0-D0 & 31.8 & 17.0 & 0.92 ± 0.06 & 0.8 ± 0.1 \\
2022-07-29 & 59789.07077 & A0-J3 & 129.9 & 41.8 & 0.79 ± 0.05 & 0.32 ± 0.05 \\
2022-07-30 & 59790.02846 & D0-J3 & 103.8 & 57.4 & 0.83 ± 0.04 & 0.51 ± 0.05 \\
2022-07-30 & 59790.02846 & K0-G2 & 67.2 & 6.6 & 0.86 ± 0.04 & 0.66 ± 0.05 \\
2022-07-30 & 59790.02846 & G2-D0 & 39.7 & 61.6 & 0.94 ± 0.06 & 1.0 ± 0.1 \\
2022-07-30 & 59790.02846 & G2-J3 & 64.3 & 54.8 & 0.86 ± 0.07 & 0.7 ± 0.2 \\
2022-07-30 & 59790.02846 & K0-D0 & 95.7 & 26.4 & 0.82 ± 0.05 & 0.55 ± 0.07 \\
2022-07-30 & 59790.02846 & K0-J3 & 53.8 & 123.5 & 0.82 ± 0.04 & 0.8 ± 0.1 \\
2022-08-08 & 59799.11692 & J2-J3 & 102.7 & 95.5 & 0.92 ± 0.06 & 0.53 ± 0.07 \\
2022-08-08 & 59799.11692 & A0-G1 & 88.6 & 135.6 & 0.88 ± 0.05 & 0.53 ± 0.06 \\
2022-08-08 & 59799.11692 & G1-J2 & 51.4 & 18.9 & 0.99 ± 0.07 & 0.83 ± 0.09 \\
2022-08-08 & 59799.11692 & G1-J3 & 125.0 & 71.9 & 0.86 ± 0.06 & 0.37 ± 0.05 \\
2022-08-08 & 59799.11692 & A0-J2 & 120.7 & 157.9 & 0.83 ± 0.08 & 0.37 ± 0.09 \\
2022-08-08 & 59799.11692 & A0-J3 & 116.9 & 29.1 & 0.82 ± 0.07 & 0.31 ± 0.07 \\
2022-08-25 & 59816.07621 & D0-C1 & 20.4 & 50.7 & 0.79 ± 0.05 & 0.51 ± 0.06 \\
2022-08-25 & 59816.07621 & A0-B2 & 25.3 & 110.9 & 0.98 ± 0.03 & 1.24 ± 0.09 \\
2022-08-25 & 59816.07621 & B2-D0 & 30.5 & 50.7 & 0.98 ± 0.03 & 0.6 ± 0.1 \\
2022-08-25 & 59816.07621 & B2-C1 & 10.2 & 50.8 & 0.83 ± 0.07 & 1.3 ± 0.2 \\
2022-08-25 & 59816.07621 & A0-D0 & 28.4 & 0.0 & 0.95 ± 0.03 & 2.1 ± 0.2 \\
2022-08-25 & 59816.07621 & A0-C1 & 22.1 & 134.5 & 0.80 ± 0.06 & 0.36 ± 0.04 \\
2022-08-30 & 59821.09058 & D0-J3 & 85.4 & 33.1 & 0.87 ± 0.06 & 0.63 ± 0.04 \\
2022-08-30 & 59821.09058 & A0-G1 & 86.0 & 126.0 & 0.94 ± 0.04 & 0.64 ± 0.04 \\
2022-08-30 & 59821.09058 & G1-D0 & 70.9 & 110.5 & 0.99 ± 0.07 & 0.82 ± 0.08 \\
2022-08-30 & 59821.09058 & G1-J3 & 122.4 & 67.6 & 0.85 ± 0.06 & 0.41 ± 0.05 \\
2022-08-30 & 59821.09058 & A0-D0 & 25.9 & 173.0 & 0.95 ± 0.08 & 1.06 ± 0.06 \\
2022-08-30 & 59821.09058 & A0-J3 & 106.6 & 24.1 & 0.83 ± 0.06 & 0.48 ± 0.05 \\
2023-03-27 & 60030.25157 & D0-C1 & 22.6 & 93.2 & 1.01 ± 0.06 & 1.1 ± 0.1 \\
2023-03-27 & 60030.25157 & A0-B2 & 18.1 & 149.5 & 1.00 ± 0.05 & 0.93 ± 0.09 \\
2023-03-27 & 60030.25157 & B2-D0 & 33.8 & 93.3 & 1.07 ± 0.07 & 1.3 ± 0.1 \\
2023-03-27 & 60030.25157 & B2-C1 & 11.3 & 93.3 & 0.98 ± 0.06 & 1.1 ± 0.1 \\
2023-03-27 & 60030.25157 & A0-D0 & 28.1 & 60.9 & 1.00 ± 0.06 & 1.2 ± 0.1 \\
2023-03-27 & 60030.25157 & A0-C1 & 15.1 & 7.8 & 0.94 ± 0.06 & 0.88 ± 0.08 \\
\bottomrule
\end{tabular}%
}
\end{table*}

\begin{table*}[p]
\centering
\caption{Gaussian fit parameters for the Br$\alpha$ line as observed with MATISSE.}
\label{tab:bralpha_gauss_fits}
\begin{tabular}{lcccc}
\hline\hline
Epoch & $\Delta t$ & $v_{\rm peak}$ & FWHM & EW \\
(MJD) & (d) & (km\,s$^{-1}$) & (km\,s$^{-1}$) & (nm) \\
\hline
59714.19815 & 0.00 & $-6.9\pm1.0$ & $118.4\pm3.1$ & $-2.20\pm0.10$ \\
59765.09167 & 50.89 & $-7.4\pm1.0$ & $128.5\pm3.4$ & $-2.40\pm0.12$ \\
59789.04079 & 74.84 & $-6.1\pm0.8$ & $122.8\pm2.5$ & $-2.67\pm0.10$ \\
59789.07077 & 74.87 & $-10.8\pm0.7$ & $121.8\pm2.3$ & $-2.56\pm0.09$ \\
59789.19311 & 74.99 & $-1.8\pm0.7$ & $120.2\pm2.1$ & $-1.86\pm0.05$ \\
59790.02846 & 75.83 & $-7.0\pm1.2$ & $115.6\pm3.6$ & $-2.32\pm0.13$ \\
59799.11692 & 84.92 & $-7.4\pm1.1$ & $118.8\pm3.4$ & $-2.22\pm0.11$ \\
59816.07621 & 101.88 & $-14.5\pm0.6$ & $117.8\pm2.0$ & $-2.35\pm0.07$ \\
59821.09058 & 106.89 & $-6.3\pm0.6$ & $112.3\pm1.9$ & $-2.11\pm0.06$ \\
60030.25157 & 316.05 & $+4.9\pm2.3$ & $110.2\pm7.0$ & $-1.12\pm0.11$ \\
\hline
\label{line_measures}
\end{tabular}
\end{table*}


\begin{figure*}[p]
\centering

\begin{subfigure}{0.32\textwidth}
  \centering
  \includegraphics[width=\linewidth]{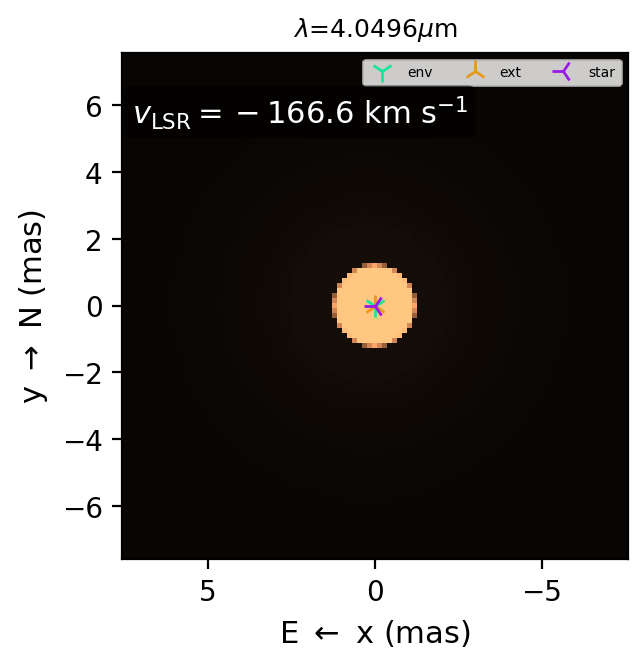}
\end{subfigure}\hfill
\begin{subfigure}{0.32\textwidth}
  \centering
  \includegraphics[width=\linewidth]{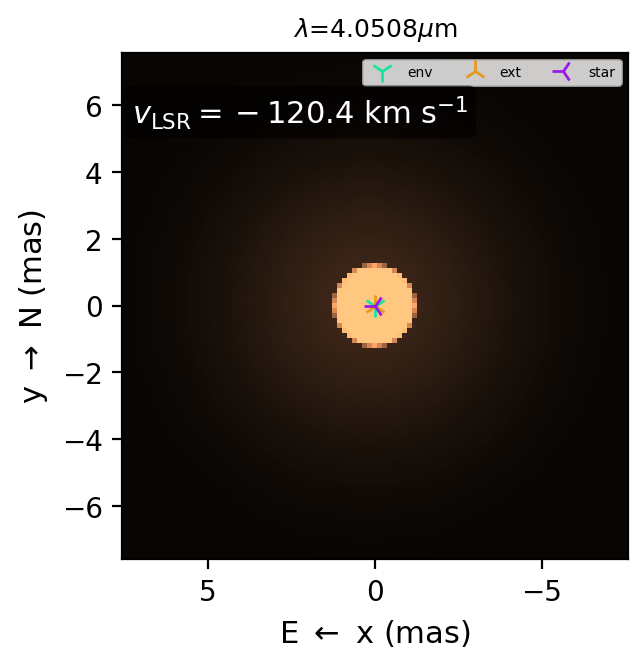}
\end{subfigure}\hfill
\begin{subfigure}{0.32\textwidth}
  \centering
  \includegraphics[width=\linewidth]{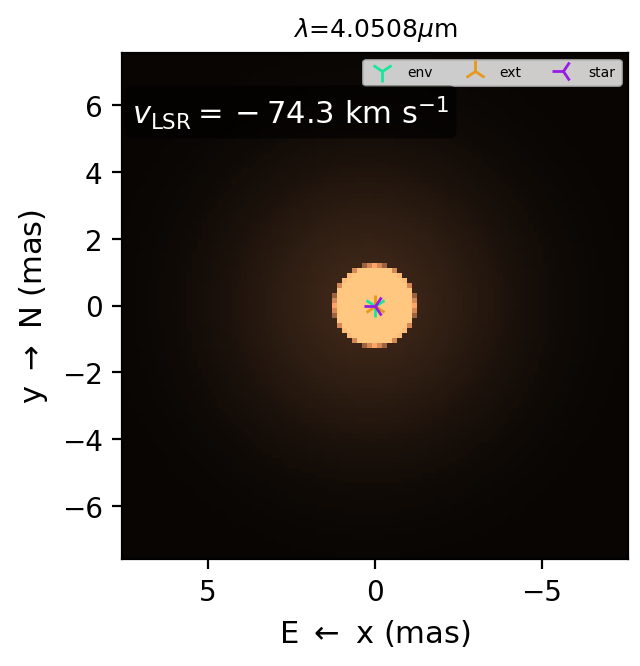}
\end{subfigure}

\vspace{2mm}

\begin{subfigure}{0.32\textwidth}
  \centering
  \includegraphics[width=\linewidth]{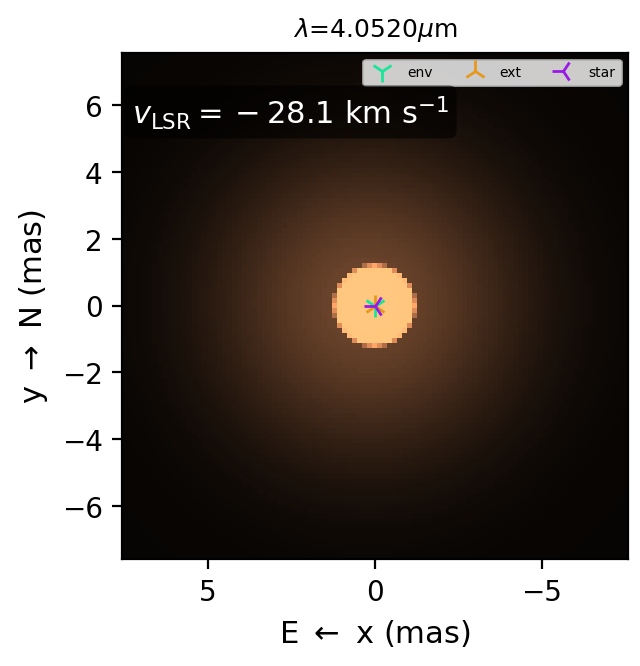}
\end{subfigure}\hfill
\begin{subfigure}{0.32\textwidth}
  \centering
  \includegraphics[width=\linewidth]{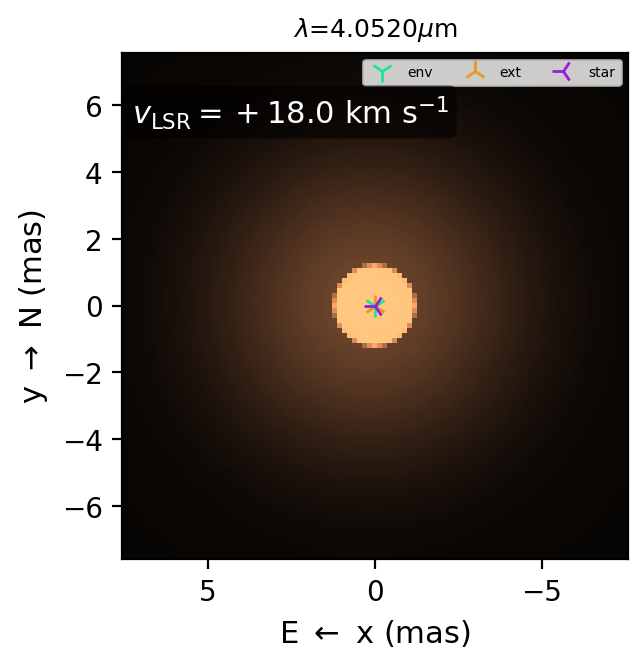}
\end{subfigure}\hfill
\begin{subfigure}{0.32\textwidth}
  \centering
  \includegraphics[width=\linewidth]{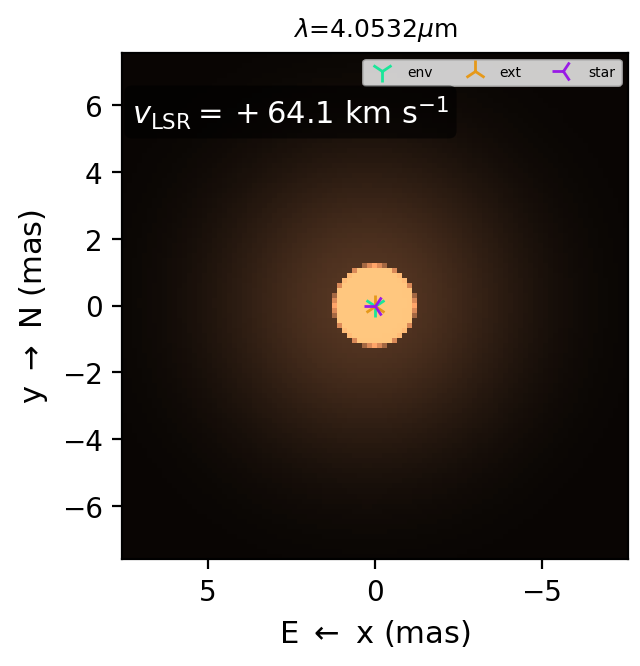}
\end{subfigure}

\vspace{2mm}

\begin{subfigure}{0.32\textwidth}
  \centering
  \includegraphics[width=\linewidth]{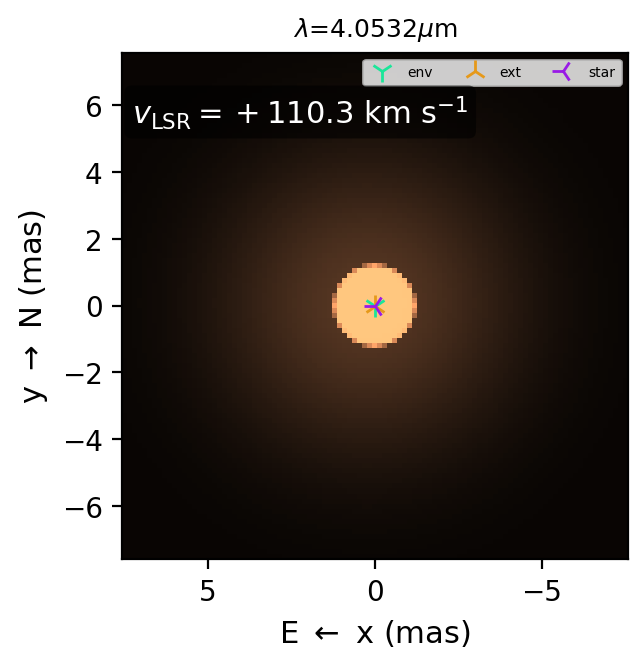}
\end{subfigure}\hfill
\begin{subfigure}{0.32\textwidth}
  \centering
  \includegraphics[width=\linewidth]{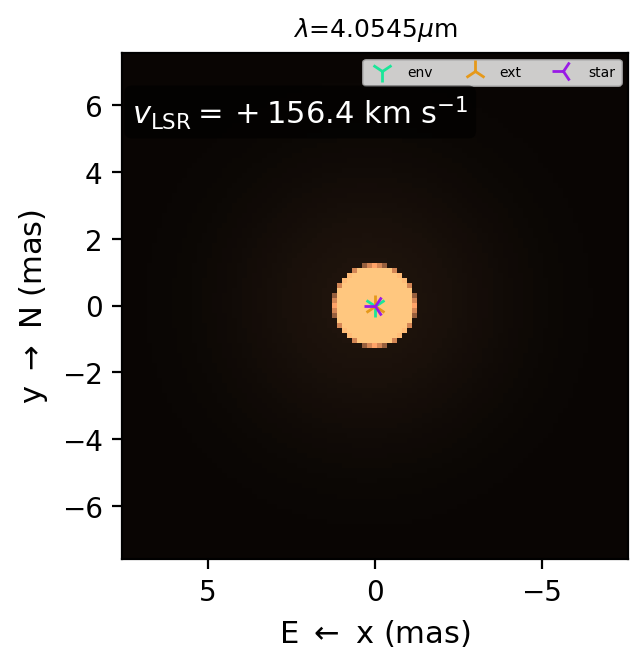}
\end{subfigure}\hfill
\begin{subfigure}{0.32\textwidth}
  \centering
  \includegraphics[width=\linewidth]{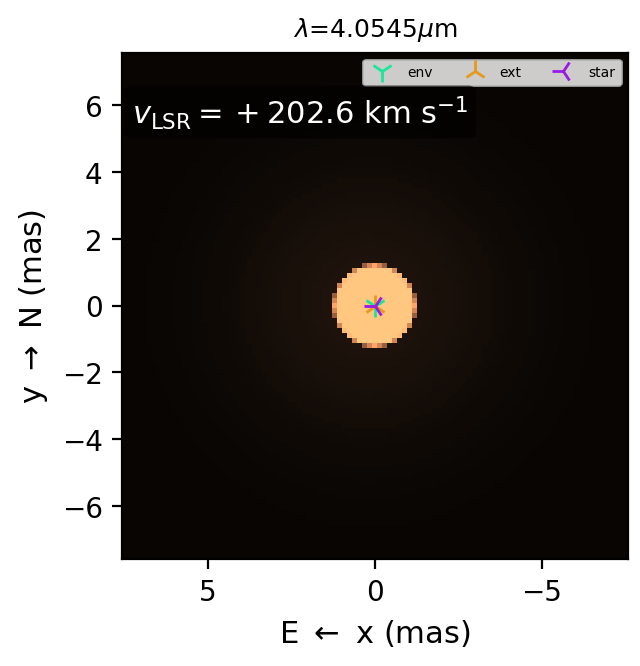}
\end{subfigure}

\caption{Channel maps from the fitting process of the MATISSE data using PMOIRED across the wavelength interval of 4.050--4.056~$\mu$m (Br$\alpha$ emission). 
Each panel corresponds to a single sampled wavelength slice; the associated Doppler velocity with respect to the local standard of rest, $v_{\mathrm{LSR}}$, is indicated in the upper-left corner of each frame (computed relative to the Br$\alpha$ rest wavelength $\lambda_0 = 4.0523~\mu$m). The systemic velocity of 18 km/sec is based on Fe II lines in the optical \citep{Wallstrom2017}.
}
\label{fig:bralpaha_grid}
\end{figure*}


\begin{figure*}[p]
\centering

\begin{subfigure}{0.48\textwidth}
  \centering
  \includegraphics[width=0.95\linewidth]{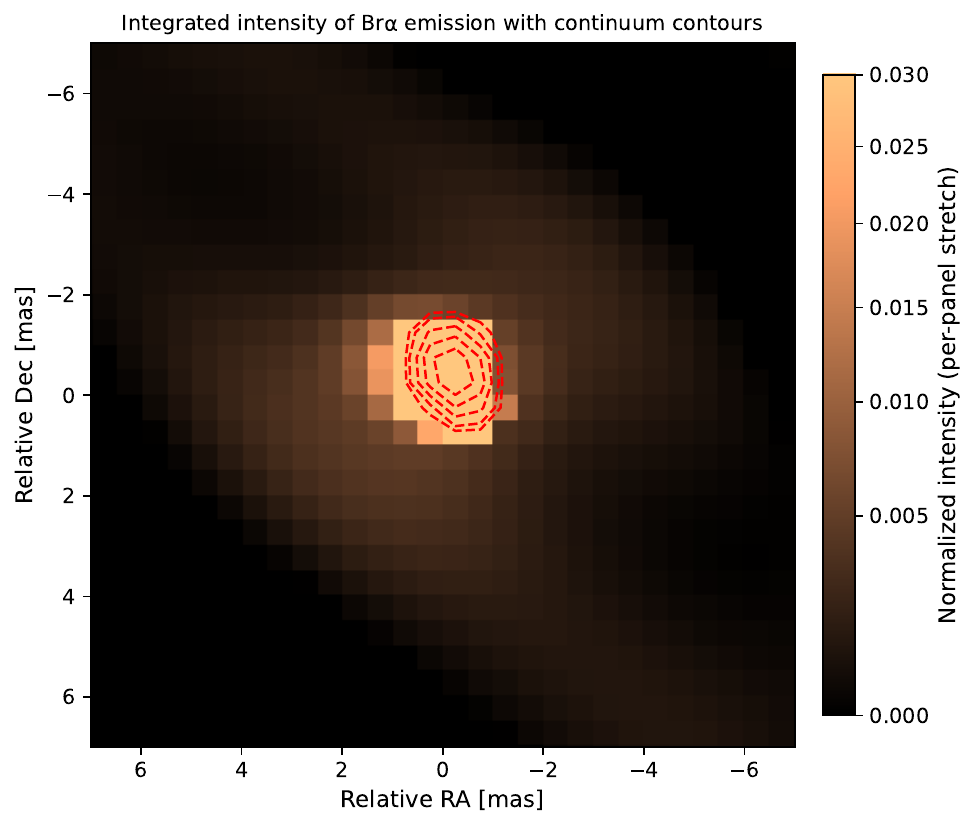}
\end{subfigure}

\vspace{2mm}

\begin{subfigure}{0.85\textwidth}
  \centering
  \includegraphics[width=\linewidth]{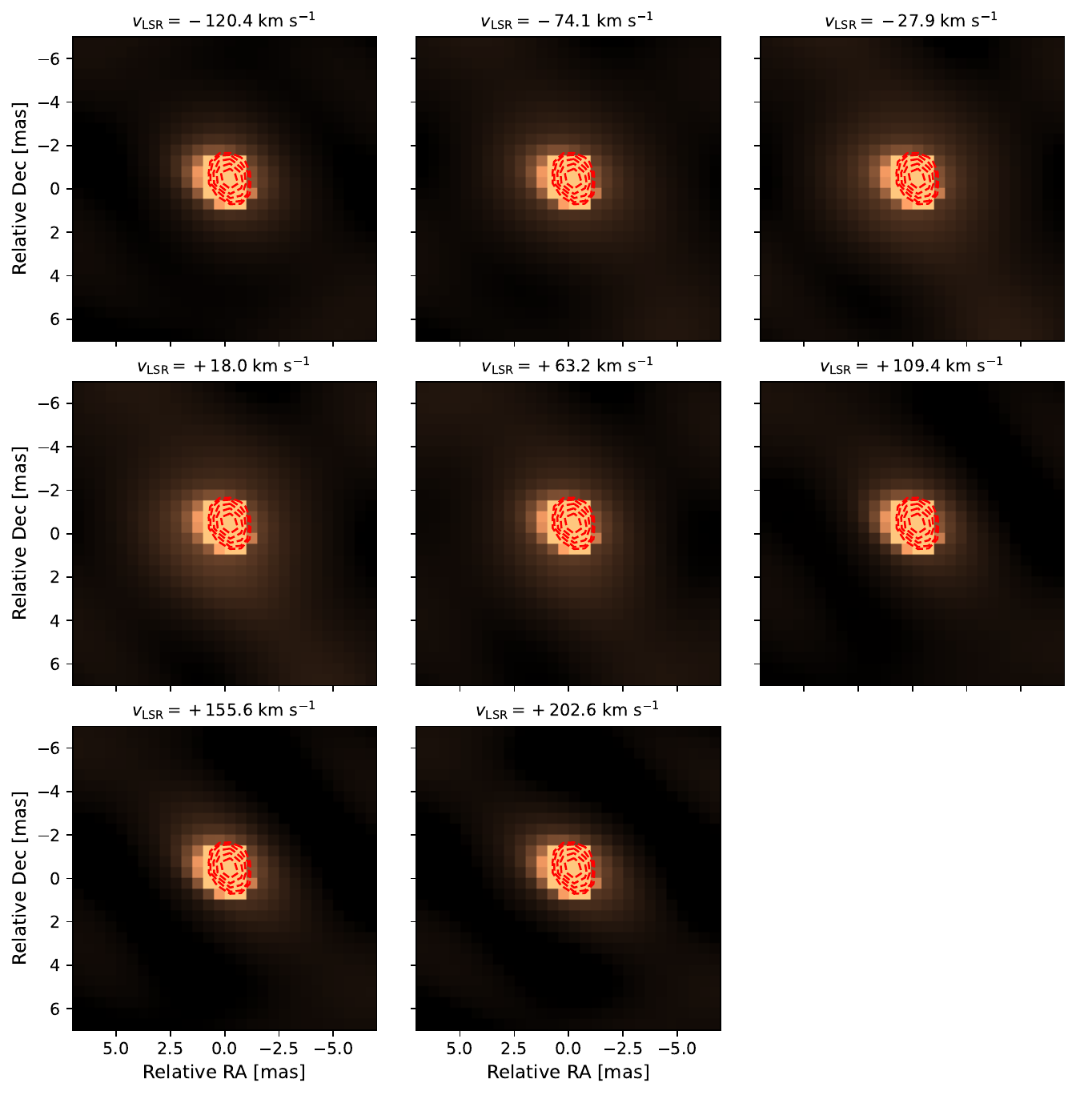}
\end{subfigure}

\caption{
Br$\alpha$ image reconstructions of IRAS~17163 using WISARD.
(Top) Integrated intensity around the Br$\alpha$ emission obtained by summing all velocity channels.
(Grid) Velocity-channel reconstructed maps across the Br$\alpha$ line profile (constant velocity step of $\Delta v$~km~s$^{-1}$), progressing from blue-shifted (top left) to red-shifted (bottom right) emission.
In both panels, the reconstructed L-band continuum emission is overplotted as red contours to highlight the spatial extent of the ionised emission relative to the continuum.
}
\label{fig:wray_bralpha_reco_combined}
\end{figure*}

\end{appendix}

\end{document}